\documentclass[11pt,fleqn]{article}
\usepackage{color}
\usepackage{epsfig}
\usepackage{rotating}
\usepackage{cite}
\usepackage{amsmath}
\usepackage{psfrag}
\allowdisplaybreaks[1]
\usepackage{subfigure}

\newcommand{\alS}{\alpha_s}

\newcommand{\muLow}{{\mu_\text{had}}}
\newcommand{\muEW}{{\mu_\text{ew}}}
\newcommand{\muNP}{{\Lambda}}

\newcommand{\vcb}{|V_{cb}|}

\newcommand{\vub}{|V_{ub}|}
\newcommand{\vts}{|V_{ts}|}

\newcommand{\mvs}{\vbox{\vskip 8mm}}

\def\R1{\varepsilon_1}
\def\E8{\varepsilon_8}

\def\nnb{\nonumber}

\def\epe{\varepsilon'/\varepsilon}
\def\as{\alpha_s}
\def\aspi{\frac{\as}{4\pi}}

\def\Heff{{\cal H}_{\rm eff}}

\newcommand{\mt}{m_{\rm t}}

\newcommand{\mc}{m_{\rm c}}

\newcommand{\mb}{m_{\rm b}}
\newcommand{\mw}{M_{\rm W}}

\newcommand{\gev}{\, {\rm GeV}}

\newcommand{\mev}{\, {\rm MeV}}
\newcommand{\bsi}{B_6^{(1/2)}}
\newcommand{\bei}{B_8^{(3/2)}}

\newcommand{\bea}{\begin{eqnarray}}
\newcommand{\eea}{\end{eqnarray}}
\newcommand{\bd}{\begin{displaymath}}
\newcommand{\ed}{\end{displaymath}}

\newcommand{\beq}{\begin{equation}}
\newcommand{\eeq}{\end{equation}}
\newcommand{\be}{\begin{equation}}
\newcommand{\ee}{\end{equation}}
\newcommand{\bi}{\begin{itemize}}
\newcommand{\ei}{\end{itemize}}
\newcommand{\ord}{{\mathcal{O}}}

\newcommand{\f}{\frac}

\def\kpn{K^+\rightarrow\pi^+\nu\bar\nu}

\def\klpn{K_{\rm L}\rightarrow\pi^0\nu\bar\nu}
\newcommand{\kmm}{K_{\rm L} \to \mu^+ \mu^-}

\def\aspi{\frac{\as}{4\pi}}

\begin{document}
\thispagestyle{empty}
\phantom{xxx}
\vskip1truecm
\begin{flushright}
 AJB-22-10 \\
\end{flushright}
\vskip1.0truecm
\begin{center}
{\LARGE\bf Climbing  NLO and NNLO Summits  }\\
   \vskip0.3truecm
{\LARGE\bf  of Weak Decays: 1988 - 2023}
   \vskip1truecm
{\Large\bf Andrzej J. Buras}
\vskip0.2truecm
{
TUM Institute for Advanced Study,
    Lichtenbergstr. 2a, D-85748 Garching, Germany \\[0.2cm]
Physik Department, TUM School of Natural Sciences, TU M\"unchen,\\ James-Franck-Stra{\ss}e, D-85748 Garching, Germany
}
\end{center}

\vskip1truecm
\centerline{\bf Abstract}
\vskip0.2truecm
I describe the history of the calculations of NLO and NNLO QCD corrections
to weak decays of mesons, particle-antiparticle mixing and electric dipole moments (EDMs)
 in the period 1988--2023. Also existing calculations of  electroweak
and QED corrections to these processes are included in this presentation. 
These efforts bear some analogies to the climbing of Himalayas and various 
expeditions by several teams of strongly motivated ``climbers''allowed to move this field 
from the LO  through the NLO to the NNLO level. We also summarize the
most recent calculations within the Standard Model Effective Field Theory. The material is meant to be an up to date review of this very advanced field in non-technical terms as much as possible and a
guide to the rich literature on NLO and NNLO corrections in question. In
particular we stress for which processes these calculations are crucial
 for the tests of the Standard Model and to be able to differentiate between
numerous New Physics models. 
It includes also several anecdotes related to the climbs that I was involved in.
I hope that some of the comments made in the course of the presentation could turn out to be not only amusing but also instructive.

\thispagestyle{empty}

\mbox{}

\newpage

\pagenumbering{roman}

\tableofcontents

\newpage

\pagenumbering{arabic}

\setcounter{page}{1}

\section{Introduction}
\setcounter{equation}{0}

In April 1988 the workshop on ``Hadronic Matrix Elements and Weak 
Decays" took place in the Ringberg castle at the Tegernsee lake near Munich. 
This workshop, organized by Jean-Marc G\'erard and myself, with a great help 
of Willy Huber, the last secretary of Werner Heisenberg, gathered a large 
portion of experts that were making the first steps towards the calculation of 
hadronic matrix elements relevant for $K^0-\bar K^0$ mixing, 
$B^0_{s,d}-\bar B^0_{s,d}$ mixings and non-leptonic 
decays of $K$ mesons, in particular $K\to \pi\pi$. Also non--perturbative 
aspects of semi-leptonic 
$K$--meson decays belonged to the important topics of this workshop. The 
representatives of lattice calculations, large N approach, QCD sum rules, 
hadronic sum rules and chiral perturbation theory were presenting their views
on the subject. In particular Bill Bardeen 
summarized the large N approach to weak non-leptonic $K$ meson decays 
developed by 
him, Jean-Marc G\'erard and myself in 1986-1988 
\cite{Bardeen:1986vp,Bardeen:1986uz,Bardeen:1986vz,Bardeen:1987vg} which these days is known as the Dual QCD (DQCD) approach. 
Using this approach we were 
able to obtain the first {quantitative}, even if approximate, results in QCD for 
the matrix 
elements relevant for the $\Delta I=1/2$ rule, $\epe$ and $K^0-\bar K^0$
mixing ($\hat B_K$). These results were 
certainly not appreciated by other groups during this workshop that were 
confident to obtain much better results in the following years.
 In particular 
Luciano Maiani and Guido Martinelli promised us to provide in a few 
years a much 
better explanation of the observed $\Delta I=1/2$ rule in $K\to\pi\pi$ 
decays within the lattice 
framework as well as the value of $\hat B_K$ in QCD that in the Standard Model (SM) enters the evaluation of the parameter $\varepsilon_K$ and of the $K^0-\bar K^0$ mass difference $\Delta M_K$.

Also 
representatives of chiral perturbation theory and hadronic sum rules were 
rather critical about our work. But in 1988 only very few understood our
approach, the whole field was in its infancy and it is not surprising that all 
competing groups had rather differing views on the subject. In fact 
as I will summarize at the end of this writing, only 32  years later, lattice QCD was able to confirm  our 1986 claim that  the $\Delta I=1/2$ rule is governed by the QCD dynamics in the hadronic matrix elements of current-current operators. But the physical picture related to this rule,
evident in the analytical DQCD approach, cannot be provided by Lattice QCD
and this also applies to chiral perturbation theory, where this dynamics
is hidden in the values of low energy constants. 

Also the value of  the parameter $\hat B_K$, that was {controversial} in 1988, obtained in the 2010s by various lattice groups 
is known by now with high precision and in a perfect agreement with our result
of 1988. But here we will concentrate on perturbative aspects of QCD
and describe briefly the status of non-perturbative calculations at the end of this writing. More details on the present status of the $\Delta I=1/2$ rule,
the parameter  $\hat B_K$  and in particular the ratio $\epe$ can be found in
 \cite{Buras:2020wyv,Buras:2022cyc}.

The 1988 Ringberg workshop was certainly a great success with hot discussions in 
essentially all rooms of the castle and several participants not leaving it for 
the five days of the workshop. In spite of this it was rather clear to me that 
I do not want to take part in this enterprise any longer. I was sceptical 
that one 
could improve the accuracy of our calculations and I certainly did not want to 
be a member of a big lattice group or joining QCD sum rule and chiral 
perturbation groups that were active already for many years. Looking back 
this was certainly the right decision even if my scepticism on the improvement 
of the accuracy of our calculations was not justified from the present perspective.  Indeed in 2014 Bardeen,  G\'erard 
and myself \cite{Buras:2014maa}, using various advances made separately by us over many years, could significantly improve on our work in the 1980s. More about it at the end of 
this writing.

During the last supper of the Ringberg workshop Guido Martinelli and  me 
realized that 
it would be important to calculate NLO QCD corrections to the Wilson 
coefficients of penguin operators relevant for $K\to \pi\pi$ decays. 
In 1981 Guido
 took
part in the 
pioneering calculation of the two loop anomalous dimensions of 
the current-current operators. 
This calculation done in collaboration with Guido 
Altarelli, Curci and Petrarca \cite{Altarelli:1980te,Altarelli:1980fi}
has been unfortunately performed 
in the dimensional reduction scheme (DRED) \cite{Siegel:1979wq} that 
was not familiar to most phenomenologists and its complicated structure 
discussed in detail by these authors most probably scared many from 
checking their results. Moreover it was known that the treatment of $\gamma_5$ 
in the DRED 
scheme, similarly to the dimensional regularization scheme with 
anticommunicating $\gamma_5$ (known presently as the NDR scheme), may lead to 
mathematically inconsistent results. Consequently it was not clear in 1988 
whether the result of Altarelli et al. was really correct. Being a member of 
the 
 theory group of the Max-Planck Institute for Physics in Munich (MPI) 
 for six years I was exposed very much to 
this problematic. 
Peter Breitenlohner and Dieter Maison \cite{Breitenlohner:1977hr} were rather critical about the DRED 
and NDR schemes. According to them and other field theorists only the
 't Hooft-Veltman (HV) scheme \cite{'tHooft:1972fi} for $\gamma_5$ was mathematically 
self-consistent. 
However, 
this scheme was also unfamiliar before 1988 to many phenomenologists.

While visiting Technical University of Athens in 1984 I learned about the 
second two-loop calculation of anomalous dimensions of current-current 
operators. Two young greek physicists, Tracas and Vlachos \cite{Tracas:1982gp}, performed in 1982
the Altarelli et al. calculation of 1981 in the NDR scheme, obtaining the 
result that differed considerably from the one of the Italian group. They 
could not clarify the reason for this discrepancy and in 1984 being involved 
heavily in other projects I simply did not have time to have a closer look 
at this problem.

At this last supper of the Ringberg 1988 workshop Guido told me that he 
will put 
some of his PhD students to look into NLO QCD corrections to Wilson coefficients of QCD penguin
operators relevant for $K\to \pi\pi$ decays and I told 
him that I will look at this problem as well. However, in April 1988 I was
 still at the MPI and did not have any PhD students who could join 
me in this enterprise. Moreover, due to heavy involvement in the 
organization of the ICHEP 1988 in Munich and other time consuming matters 
like the proceedings of the 1988 Ringberg workshop \cite{Buras:1989er}, lectures on our large N 
approach to weak decays at the Summer School in Jaca (Spain) \cite{Buras:1988ky}
 and most 
importantly because of my moving from MPI to the Technical University 
Munich (TUM), I did not have time to start this new project until October
of the same year.

During my summer vacation 1988 I read several books about the Himalaya 
expeditions. Among them the one by colonel Hunt, in which he described in 
detail the well known 1953 Mount Everest expedition that he organized. From 
these books I learned also about the competition between Reinhold Messner 
and polish climber Janusz Kukuczka to conquer the fourteen highest
Himalayan summits, the ones over 8000 m \footnote{Messner won this competition 
but Kukuczka was the second to reach all these highest summits by October 
1988. Unfortunately he died on October 24th, 1989 in an avalanche on the South 
Face of Lhotse.}. 
These were truly fantastic achievements but I wondered whether the difficulty 
of climbing a 8000 m high mountain by an experienced mountainer could be 
comparable in 1988 to the difficulty of a NLO calculation of weak decays 
performed by 
an experienced physicist like me. This comparison is not fully idiotic. 
After all the  difficult and often pioneering NLO and NNLO calculations in the last 
35 years required not only high technical skills
but also certain {planning} in advance and first of all psychological and 
physical strength. I mean here the ability to be involved in a calculation 
that results sometimes in a single number but lasts at least six months and 
often a year or longer. The air during  NLO and NNLO 
calculations can be very thin 
indeed.

These thoughts prompted me to generalize my plan for the NLO analysis of 
$K\to \pi\pi$ decays  to all relevant $K$ and $B$ decays 
including rare, 
radiative and in particular CP-violating decays. In 1988 this field was, with 
respect to NLO QCD corrections, essentially unexplored and all NLO summits were 
still waiting to be conquered. Being the first to complete all these 
calculations 
would certainly be an achievement with a lasting impact on the 
phenomenology of weak decays.

These were my dreams of 1988. Feeling like colonel Hunt before the Mount 
Everest expedition I made a list of 
most interesting decays and the corresponding operators. This list is given 
in Section~\ref{sec:2.2} and in various tables below. 
As mentioned before, only the current-current 
operators $Q_1$ and $Q_2$ {had}  been studied at 
NLO before 1988, but the status of these calculations was unclear as described
above.

The next step was to find a team of physicists with whom I would perform 
these calculations. To do them alone in 1988 would be  pure madness. 
 Like Guido 
Martinelli I could in principle count on PhD students, but in October 1988 
I had none who could be put on this project and even if I was hoping to get 
some students at TUM soon, it was not certain that it would 
happen. 
Moreover, the knowledge of gauge theories at TUM in 1988 was very limited 
(Hans Peter Nilles and {I} were hired to change this situation) and without a 
series of lectures on renormalization, renormalization group methods and loop 
calculations, sending young students to do NLO calculations in QCD would be 
impractical and certainly irresponsible. I estimated that before the fall of 
1989 I could not count on any help from my future PhD students and/or 
post-docs that were supposed to arrive in October 1989.

However, I certainly could not wait until the fall of 1989. Afterall I was 
convinced that Guido already worked on this project with his students. 
Therfore I told Jean-Marc G\'erard, who was at the MPI at that time, about
 my 
plans. Between 1984 and 1988 we have written 11 papers together and I was 
convinced that he was the right person for this grand project. Unfortunately 
Jean-Marc did not want to join me in this expedition. He basically told me 
that I was crazy to think about calculating 
NLO corrections to weak decays that were 
polluted by hadronic uncertainties. In principle I could also ask Bill 
Bardeen with whom I 
did my first NLO calculations for deep inelastic structure functions and 
photon structure functions ten years earlier \cite{Bardeen:1978yd,Bardeen:1978hg}. But we were seperated by 
the Atlantic which, in contrast to now,  was a real obstacle in a collaboration.
 Moreover, I had some doubts that Bill would be interested in this project.

On  my last day at MPI, the members of the MPI theory group were giving 
5 min talks about their research, in the spirit of similar talks in the Theory 
Group at CERN. 
At this 
meeting I informed my MPI colleagues about my project and that I was 
looking for collaborators. There was no reaction. I left MPI 
rather frustrated.

Few days later, sitting already in my office at TUM, I got a phone call from 
Peter Weisz, who joined MPI few months earlier. I knew Peter from his work on
Lattice QCD
with Martin L\"uscher, but as his field was rather different from mine I had 
only a few conversations with him until then. To my great surprise, Peter was
very much interested in my project and asked me whether he could join me in 
this enterprise. I told him that I was delighted. On this day the Munich 
NLO Club (MNLC) was born. The club consisted only of two members but our 
team could start the first climb.

In 2014, 26 years later I could report in the second version of this writing \cite{Buras:2011we} 
that the MNLC consists of 31
members, all with the exception of Gerhard Buchalla, Ulrich Haisch and Peter Weisz  working  now outside Munich or being not any more active in research. In addition to Peter Weisz and myself they
are:
\paragraph{PhD Students}
          {\sl Matth\"aus Bartsch}, {\sl Guido Bell}, Christoph Bobeth, {\sl Stefan Bosch}, {\sl Joachim Brod },  Gerhard Buchalla, Thorsten Ewerth,  Robert Fleischer, Martin Gorbahn,  Ulrich Haisch, Stefan Herrlich, Sebastian J\"ager, Markus E. Lautenbacher,  Alex Lenz, Manfred M\"unz, Ulrich Nierste,  Gaby Ostermaier, {\sl Volker Pilipp},  Nikolas Pott,  Emanuel Stamou, where in ''{\sl slanted}'' PhD students of my PhD students are indicated.        
\paragraph{Postdocs and Visitors} Patricia Ball, Kostja Chetyrkin, Andrzej Czarnecki
Paolo Gambino, Jennifer Girrbach-Noe,  Matthias Jamin,  Axel Kwiatkowski,
Mikolaj Misiak,  J\"org Urban.
\\

The grand project that I outlined in 1988 and started with
Peter Weisz soon after, turned out to be a great success. Based on the number
of citations collected for our NLO and NNLO papers, that I will describe in this review, it had a clear impact on the field of flavour physics. Peter was an active member of the MNLC only in the period
$1988-1992$ but these were very important years and his participation had an 
invaluable impact on the full project.
As far as QCD calculations are concerned, {the} project
has been completed within the Standard Model (SM) in the first years
of this millennium. Subsequently some NLO electroweak contributions have been computed. Also some calculations beyond the SM, in particular
within the MSSM, have been done. We were not always the first to climb a 
given NLO summit, but MNLC is  basically the only group that
 calculated NLO corrections to all relevant decays within the SM.
 In the last  two decades  several NNLO calculations have
also been performed by the members of our club, mostly outside Munich. From my point of view the task of 
calculating these higher order corrections within the SM has been completed 
even if in some corners still some improvements could be made. Beyond the
SM several NLO calculations have to be done still, but also here significant
progress in the last two decades has been made. I will mention them briefly
here and there and more systematically in Section~\ref{sec:SMEFT}.

Three generations of physicists took part in this enterprise with my PhD 
students  listed in roman above  and the PhD students of 
my PhD students in ``slanted''. Bartsch, Bell, Bosch and Pilipp were PhD 
students of Gerhard Buchalla at the Ludwig-Maximilian University in 
Munich and  Joachim Brod graduated in 
Karlsruhe under the supervision of Uli Nierste, becoming the member of 
MNLC as a PostDoc in the group of Martin Gorbahn. Emanuel 
Stamou got his Diploma under the supervision of Martin Gorbahn and PhD under my supervision. Finally, Jennifer 
Girrbach-Noe, also a PhD student of Nierste, made her  first NLO steps at IAS in Munich. The remaining 
members except for Peter Weisz and myself were assistants, post docs or visitors in my group or at MPI.

In the days of INSPIRE it is easy to verify that our work 
had an impact on particle physics. Roughly 100 papers on NLO and NNLO 
corrections have been published by the members of the MNLC, where mainly 
papers are counted in which NLO and NNLO calculations have been performed 
and not papers in which only the phenomenological implications of these 
calculations have been analyzed. As you will see at the end of this writing, 
the papers from our club amount to roughly 2/3 of all 
papers in the field of weak decays in which such calculations have been 
performed. 

The purpose of the following presentation is the recollection of these
activities and a summary of the present status of NLO and NNLO calculations
in weak decays.
I have organized the material in the following manner. In the next section 
I will 
summarize the theoretical framework for weak decays. This will be a  
compact presentation to which I will refer frequently in subsequent sections. 
The full exposition of the technicalities that I will try to avoid as much as 
possible can be found in the Rev.~Mod.~Phys. article written in collaboration 
with Gerhard Buchalla and Markus Lautenbacher \cite{Buchalla:1995vs}, my Les Houches 
lectures \cite{Buras:1998raa}, my book on weak decays \cite{Buras:2020xsm}
 and of course in the original papers. Section~\ref{sec:3} is devoted to
NLO  QCD corrections to 
$\Delta S=1$ and $\Delta B=1$ non-leptonic decays. I will be rather detailed 
about the  structure of QCD corrections to these decays because the operators involved there 
play an important role, 
even if indirectly, in essentially all weak decays. Also existing NNLO 
calculations for these decays will be summarized.
Section~\ref{sec:4} describes $\Delta S=2$ and 
$\Delta B=2$ transitions in some detail and very briefly $\Delta B=0$ 
transitions. Section~\ref{sec:5} is devoted to rare $K$ and $B$ decays, 
in particular $\kpn$, $\klpn$ and $B_{s,d}\to\mu^+\mu^-$. Section~\ref{sec:6}  is 
devoted to the K2 of weak decays, the inclusive decay 
 $B\to X_s\gamma$, with a few comments on the 
$B\to X_s$~gluon decay. In addition to a detailed description of the 
history of NLO calculations we will summarize the present status of 
NNLO calculations. Finally, we will list the literature for the 
corresponding exclusive decays $B\to K^*(\rho)\gamma$. In Section~\ref{sec:7}  
we discuss the NLO corrections to $K_L\to \pi^0 l^+l^-$ and in Section~\ref{sec:8} the NLO and NNLO {corrections} to
$B\to X_s\ell^+\ell^-$. Here also the decays $B\to K^*(\rho)l^+l^-$  will 
be mentioned.

A very special Section is Section~\ref{sec:qcdf} because it is not written by me but
 by one of 
the prominent members of MNLC, Gerhard Buchalla, who is also one of 
the fathers of the QCD Factorization approach to non-leptonic decays. 
I thought that the QCD calculations in this approach should also have 
a place in this presentation and I asked Gerhard to help me in this matter.
In Section~\ref{sec:EDMs} electric dipole moments are discussed and in Section~\ref{sec:SMEFT} we summarize the present status of QCD corrections in the Weak Effective Theory (WET) and in the Standard Model Effective Field Theory (SMEFT).
We conclude in Section~\ref{sec:12}.

While in certain parts of our review we will enter some details, the material
is meant to be a guide to the rich literature 
on NLO and NNLO corrections to weak decays. It is certainly not as technical
as the reviews and lectures in  
\cite{Buchalla:1995vs,Buras:1998raa,Buras:2020xsm}, although it includes a lot of information 
after the first two long treatises on this subject have been published. It also includes new information relative to my book \cite{Buras:2020xsm}. I just wanted to summarize what has been 
done during the last 35 years, listing in particular the first climbs 
of the existing NLO and NNLO summits and few subsequent climbs that used 
different methods or routes to reach a given summit. Thus the 
full material 
can be considered as a non-technical chronicle of NLO and NNLO calculations (1988-2023) 
in weak decays with 
several anecdotes behind the scene related to the climbs that I was involved in and several, 
hopefully, instructive comments for non-experts 
that probably are hard to find in the most recent very technical
literature on NNLO corrections.

Before describing my NLO-story in more explicit terms I will make an express 
review of the 
theoretical framework for weak decays summarizing at the end the present 
status of NLO and NNLO calculations that are discussed in detail in the 
subsequent sections. I should stress that several NLO and NNLO QCD calculations 
in the framework of  SCET, QCD sum rules, light-cone sum rules, for non-leading terms in heavy quark expansions, heavy quark 
effective theory, charmonia are left out from 
this presentation because I did not participate in these studies. With the 
help of INSPIRE clicking the names of Ball, Bauer, Bell, Beneke, Brambilla, Chetyrkin, Feldmann, Greub, 
Hoang, Jamin, K\"uhn, Lenz, Neubert, Steinhauser, Stewart and several 
other masters of these fields 
one can easily find all papers on these topics. 
The stories behind these calculations are unknown to me. In this context one 
should mention in particular numerous papers of Matthias Jamin on QCD 
corrections relevant for QCD sum rules, numerous papers by the Karlsruhe 
QCD club lead by Konstatin Chetyrkin, Hans K\"uhn and Matthias Steinhauser, in particular their results on quark masses, 
and the work of Andre Hoang  and his collaborators, now in Vienna,  among  others.

The following section is rather heavy but I hope that it will facilitate
the reading of the subsequent sections for non-experts. Experts, knowing this 
technology, can skip this section to go directly to Sections~\ref{sec:3}-\ref{sec:12} in order 
to check whether I cited them properly. On the other hand the classification of QCD corrections into eight  classes, presented in Section~\ref{sec:2.6}, could 
be of interest  to them.

{Finally, I would like to emphasize that the  Rev.~Mod.~Phys. article \cite{Buchalla:1995vs} discusses only NLO QCD corrections to weak decays as known in 1995 and
  the era of NNLO calculations begins first in the year 2000 with the
  paper in collaboration with Paolo Gambino and Ulrich Haisch \cite{Buras:1999st}.}

\section{Theoretical Framework for Weak Decays}\label{sec:2}
\setcounter{equation}{0}
\subsection{OPE and Renormalization Group}
The basis for any serious phenomenology of weak decays of
hadrons is the {\it Operator Product Expansion} (OPE) \cite{Wilson:1969zs,Zimmermann:1972tv},
which allows to write
the effective weak Hamiltonian simply as follows
\be\label{b1}
{\cal H}_{eff}=\frac{G_F}{\sqrt{2}}\sum_i V^i_{\rm CKM}C_i(\mu)Q_i~.
\ee
Here $G_F$ is the Fermi constant and $Q_i$ are the relevant local
operators which govern the decays in question. 
As we will see below they are built out of quark and lepton fields.
The Cabibbo-Kobayashi-Maskawa
factors $V^i_{\rm CKM}$ \cite{Cabibbo:1963yz,Kobayashi:1973fv} 
and the Wilson coefficients $C_i(\mu)$  describe the 
strength with which a given operator enters the Hamiltonian.
The latter coefficients can be considered as scale dependent
``couplings'' related to ``vertices'' $Q_i$ and 
can be calculated using perturbative methods as long as the scale $\mu$ is
not too small.

An amplitude for a decay of a given meson 
$M= K, B,..$ into a final state $F=\mu^+\mu^-,~\pi\nu\bar\nu$,
$\pi\pi,~DK$ is then
simply given by
\be\label{amp5}
A(M\to F)=\langle F|{\cal H}_{eff}|M\rangle
=\frac{G_F}{\sqrt{2}}\sum_i V^i_{CKM}C_i(\mu)\langle F|Q_i(\mu)|M\rangle,
\ee
where $\langle F|Q_i(\mu)|M\rangle$ 
are the matrix elements of $Q_i$ between $M$ and $F$, evaluated at the
renormalization scale $\mu$. 

The essential virtue of the OPE is this one. It allows to separate the problem
of calculating the amplitude
$A(M\to F)$ into two distinct parts: the {\it short distance}
(perturbative) calculation of the coefficients $C_i(\mu)$ and 
the {\it long-distance} (generally non-perturbative) calculation of 
the matrix elements $\langle Q_i(\mu)\rangle$. The scale $\mu$
separates, roughly speaking, the physics contributions into short
distance contributions contained in $C_i(\mu)$ and the long distance 
contributions
contained in $\langle Q_i(\mu)\rangle$. Our presentation is mainly devoted 
to the calculations of the coefficients $C_i(\mu)$ within the SM and within 
some of its extensions.  The status of the hadronic matrix elements $\langle Q_i(\mu)\rangle$ will be briefly summarized at the end of this writing.

It should be stressed at this point that our presentation would not 
exist without the asymptotic freedom in QCD \cite{Gross:1973id,Politzer:1973fx}
that allows the
calculations of Wilson coefficients by means of ordinary or 
renormalization group improved perturbation  theory. 
The precision of these calculations increased in the last thirty years 
not only because of NLO and NNLO QCD calculations but also because of  
the more accurate determination of the strong coupling
$\alpha_s$ for which the most recent result from PDG22 reads
\cite{Workman:2022ynf}: 
\be
\alpha^{\overline{\rm MS}}_s(M_Z)=0.1179\pm0.0009~.
\ee
Now, the coefficients $C_i$ include in addition to tree level 
contributions from the $W$-exchange,
virtual top quark contributions and
contributions from other heavy particles such as W, Z bosons, charged
Higgs particles, supersymmetric particles in the supersymmetric extensions
of the SM and other heavy objects in numerous extensions of this model.
Consequently $C_i(\mu)$ depend generally 
on $m_t$ and also on the masses of new particles if extensions of the 
SM are considered. This dependence can be found by evaluating 
so-called {\it box} and {\it penguin} diagrams with full W, Z, top quark and 
new particle exchanges and {\it properly} including short distance QCD 
effects. The latter govern the $\mu$-dependence of $C_i(\mu)$. In models 
in which the GIM mechanism \cite{Glashow:1970gm} is absent, also {\it tree} diagrams can contribute to flavour changing neutral current (FCNC) processes. The 
point is that a given $C_i$ receives  generally 
contributions from all these three classes of diagrams.

The value of $\mu$ can be chosen arbitrarily but the final result
must be $\mu$-independent.
Therefore 
the $\mu$-dependence of $C_i(\mu)$ has to cancel the 
$\mu$-dependence of $\langle Q_i(\mu)\rangle$. In other words as far as heavy 
mass independent terms are concerned 
 it is a
matter of choice what exactly belongs to $C_i(\mu)$ and what to 
$\langle Q_i(\mu)\rangle$. This cancellation
of the $\mu$-dependence involves generally several terms in the expansion 
in (\ref{amp5}).
The  coefficients $C_i(\mu)$ 
depend also
on the renormalization scheme.
This scheme dependence must also be canceled
by the one of $\langle Q_i(\mu)\rangle$ so that the physical amplitudes are 
renormalization scheme independent. Again, as in the case of the 
$\mu$-dependence, the cancellation of
the renormalization scheme dependence involves generally several 
terms in the expansion (\ref{amp5}). One of the type of scheme dependences 
is the manner in which $\gamma_5$ is defined in $D=4-2\varepsilon$ dimensions 
implying for instance 
the three schemes NDR, HV and DRED mentioned earlier.

Although $\mu$ is in principle arbitrary,
it is customary 
to choose
$\mu$ to be of the order of the mass of the decaying hadron. 
This is $\ord (\mb)$ and $\ord(\mc)$ for $B$ decays and
$D$ decays respectively. In the case of $K$ decays the typical choice is
 $\mu=\ord(1-2~\gev)$
instead of $\ord(m_K)$, which is much too low for any perturbative 
calculation of the couplings $C_i$.
Now due to the fact that $\mu\ll  M_{W,Z},~ m_t$, large logarithms 
$\ln\mw/\mu$ compensate in the evaluation of
$C_i(\mu)$ the smallness of the QCD coupling constant $\alpha_s$ and 
terms $\alpha^n_s (\ln\mw/\mu)^n$, $\alpha^n_s (\ln\mw/\mu)^{n-1}$ 
etc. have to be resummed to all orders in $\alpha_s$ before a reliable 
result for $C_i$ can be obtained.
This can be done very efficiently by means of the renormalization group
methods. 
The resulting {\it renormalization group improved} perturbative
expansion for $C_i(\mu)$ in terms of the effective coupling constant 
$\alpha_s(\mu)$ does not involve large logarithms and is more reliable.
The related technical issues are discussed in detail in \cite{Buchalla:1995vs,Buras:1998raa,Buras:2020xsm} 
and we will recall here only those that are essential for 
our presentation.

All this looks rather formal but in fact should be familiar.
Indeed,
in the simplest case of the $\beta$-decay, ${\cal H}_{eff}$ takes 
the familiar form
\be\label{beta}
{\cal H}^{(\beta)}_{eff}=\frac{G_F}{\sqrt{2}}
\cos\theta_c(\bar u\gamma_\mu(1-\gamma_5)d)
(\bar e \gamma^\mu (1-\gamma_5)\nu_e)~,
\ee
where $V_{ud}$ has been expressed in terms of the Cabibbo angle 
\cite{Cabibbo:1963yz}. In this
particular case the Wilson coefficient is equal to unity and the local
operator, the object between the square brackets, is given by a product 
of two $V-A$ currents. 
Equation (\ref{beta}) represents the Fermi theory for $\beta$-decays 
as formulated by Sudarshan and
Marshak \cite{Sudarshan:1958vf} and Feynman and Gell-Mann \cite{Feynman:1958ty} more than 
sixty years ago, 
except that in (\ref{beta})
the quark language has been used and following Cabibbo a small departure of
$V_{ud}$ from unity has been incorporated. In this context the basic 
formula (\ref{b1})
can be regarded as a generalization of the Fermi Theory to include all known
quarks and leptons as well as their strong and electroweak interactions as
summarized by the SM. 

Due to the interplay of electroweak 
and strong interactions the structure of the local operators is 
much richer than in the case of the $\beta$-decay. They can be classified 
with respect to  Lorentz structure,  Dirac structure, the 
colour structure and the type of quarks 
and leptons relevant for a given decay. Some of these operators are
unimportant in the SM but could be relevant in some extensions of the SM. We will now list all the operators 
{whose} Wilson coefficients will be mentioned in subsequent sections.

\subsection{Local Operators in the SM}\label{sec:2.2}
\begin{figure}[hbt]
\centerline{
\epsfysize=4.3in
\epsffile{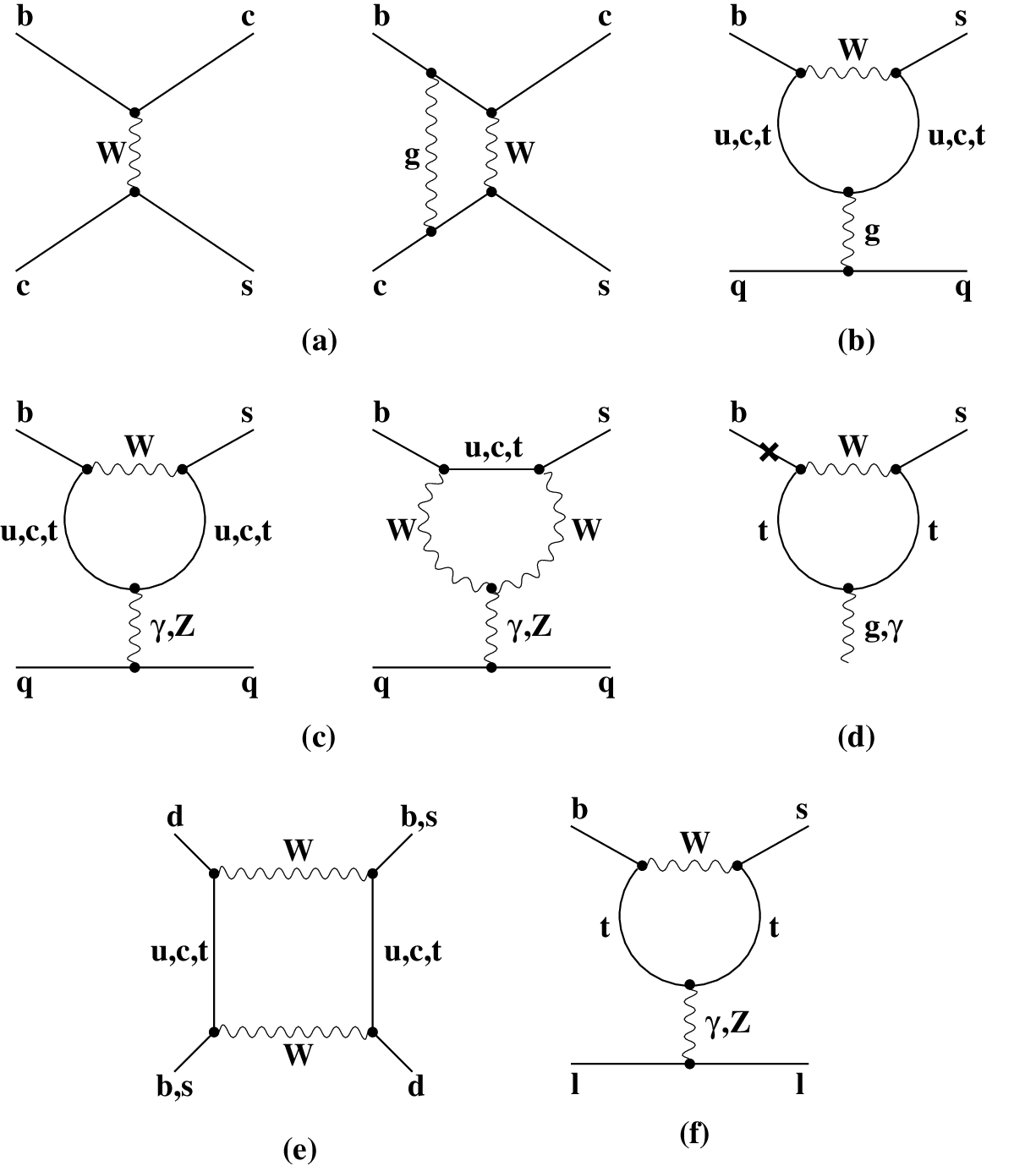}
}
\caption{Typical Tree, Penguin and Box Diagrams in the SM.}
\label{fig:fdia}
\end{figure}

We give below first a list of operators that play the role 
in weak $B$ decays.
Typical diagrams in
the full theory from which these operators originate are 
 shown in Fig.~\ref{fig:fdia}. The cross in the diagram 1d indicates  
that dipole penguins originate from the mass-term on the external
line in the usual QCD or QED penguin diagrams. The operators relevant 
for $K$ decays are discussed subsequently.

\subsubsection{Nonleptonic Operators}
Of particular interest are the 
operators involving quarks only. In the case of the $\Delta B=1$
transitions the relevant set of operators is given as follows:

{\bf Current--Current (Fig.~1a):}
\begin{equation}\label{O1} 
Q_1 = (\bar c_{\alpha} b_{\beta})_{V-A}\;(\bar s_{\beta} c_{\alpha})_{V-A}
\,,~~~~~Q_2 = (\bar c b)_{V-A}\;(\bar s c)_{V-A} 
\end{equation}

{\bf QCD--Penguins (Fig.~1b):}
\begin{equation}\label{O2}
Q_3 = (\bar s b)_{V-A}\sum_{q=u,d,s,c,b}(\bar qq)_{V-A}\,,~~~~~   
 Q_4 = (\bar s_{\alpha} b_{\beta})_{V-A}\sum_{q=u,d,s,c,b}(\bar q_{\beta} 
       q_{\alpha})_{V-A}\,,
\end{equation}
\begin{equation}\label{O3}
 Q_5 = (\bar s b)_{V-A} \sum_{q=u,d,s,c,b}(\bar qq)_{V+A}\,,~~~~  
 Q_6 = (\bar s_{\alpha} b_{\beta})_{V-A}\sum_{q=u,d,s,c,b}
       (\bar q_{\beta} q_{\alpha})_{V+A}\,.
\end{equation}

{\bf Electroweak Penguins (Fig.~1c):}
\begin{equation}\label{O4} 
Q_7 = \frac{3}{2}\;(\bar s b)_{V-A}\sum_{q=u,d,s,c,b}e_q\;(\bar qq)_{V+A} 
\,,~~~~ Q_8 = \frac{3}{2}\;(\bar s_{\alpha} b_{\beta})_{V-A}\sum_{q=u,d,s,c,b}e_q
        (\bar q_{\beta} q_{\alpha})_{V+A}
\end{equation}
\begin{equation}\label{O5} 
 Q_9 = \frac{3}{2}\;(\bar s b)_{V-A}\sum_{q=u,d,s,c,b}e_q(\bar q q)_{V-A}
\,,~~~~Q_{10} =\frac{3}{2}\;
(\bar s_{\alpha} b_{\beta})_{V-A}\sum_{q=u,d,s,c,b}e_q\;
       (\bar q_{\beta}q_{\alpha})_{V-A}\,. 
\end{equation}
Here, $\alpha,\beta$ denote colours and $e_q$ denotes the electric quark charges reflecting the
electroweak origin of $Q_7,\ldots,Q_{10}$. Finally,
$(\bar c b)_{V-A}\equiv \bar c_\alpha\gamma_\mu(1-\gamma_5) b_\alpha$. 

These operators play a crucial role in non-leptonic decays of $B^\pm$, $B^0_s$ and $B^0_d$ 
mesons and have through mixing under renormalization also an impact on other 
processes as we will see below. In this context let me also 
 make one useful remark. In the literature 
operators in $B$ physics appear sometimes with $(\bar bs)_{V-A}$ or 
$(\bar s b)_{V-A}$, dependent on whether $B_s^0$ or $\bar B_s^0$ are studied, 
respectively. When using their Wilson coefficients given in the 
literature it is crucial to remember that they are complex conjugates of 
each other. This  
distinction is crucial for obtaining correct CP asymmetries.

For non-leptonic $K$ decays the flavours have to be changed appropriately. 
Explicit expressions can be found in \cite{Buchalla:1995vs,Buras:1998raa,Buras:2020xsm}.
In particular the analogues of 
$Q_1$ and $Q_2$ govern the $\Delta I=1/2$ rule in $K_L\to\pi\pi$ decays, while 
the corresponding QCD penguins and electroweak penguins enter directly 
the ratio $\epe$.

\subsubsection{Dipole Operators}
In the case of $B\to X_s\gamma$ decay and  $B\to X_sl^+l^-$ decays 
as well as corresponding exclusive decays the crucial role is played by

{\bf Dipole Penguins (Fig.~1d):}
\begin{equation}\label{O6}
Q_{7\gamma}  =  \frac{e}{8\pi^2} m_b \bar{s}_\alpha \sigma^{\mu\nu}
          (1+\gamma_5) b_\alpha F_{\mu\nu}\,,\qquad            
Q_{8G}     =  \frac{g}{8\pi^2} m_b \bar{s}_\alpha \sigma^{\mu\nu}
   (1+\gamma_5)T^a_{\alpha\beta} b_\beta G^a_{\mu\nu}\,.  
\end{equation}
Again, when using the results in the literature care must be taken 
whether $b$ or $\bar b$ is present in the operator and what are the 
factors multiplying the Dirac structures. The operator $Q_{8G}$ can 
also be relevant in nonleptonic decays. Also dipole penguins with 
$(1+\gamma_5)$ replaced by $(1-\gamma_5)$ are present but they are 
suppressed within the SM with respect to the operators in (\ref{O6}) 
by $m_s/m_b$.

\boldmath
\subsubsection{$\Delta F= 2$ Operators}
\unboldmath
In the case of  $K^0-\bar K^0$ mixing and  $B_d^0-\bar B^0_d$ mixing
the relevant operators within the SM are

{\bf \boldmath{$\Delta S= 2$} and \boldmath{$\Delta B=2$} Operators (Fig.~1e):}
\begin{equation}\label{O7}
Q(\Delta S = 2)  = (\bar s d)_{V-A} (\bar s d)_{V-A}\,,~~~~
 Q(\Delta B = 2)  = (\bar b d)_{V-A} (\bar b d)_{V-A} \,.
\end{equation}
For $B_s^0-\bar B^0_s$ mixing one has to replace $d$ by $s$ in 
the last operator.

\boldmath
\subsubsection{Semileptonic Operators}
\unboldmath
In the case of $b\to s\ell^+\ell^-$ transitions that govern
decays like $B\to K(K^*)\ell^+\ell^-$, $B\to X_sl^+l^-$ also the following operators originating in 
{\bf Fig.~1f} on top of the
dipole penguins contribute
\begin{equation}\label{9V}
Q_{9V}  = (\bar s b  )_{V-A} (\bar \mu\mu)_{V}\,,~~~~~
Q_{10A}  = (\bar s b )_{V-A} (\bar \mu\mu)_{A},
\end{equation}
where we set $\ell^{\pm}=\mu^{\pm}$.
Changing appropriately flavours one obtains the corresponding 
operators relevant for 
$B\to X_d\ell^+\ell^-$ and $K_L\to\pi^0\ell^+\ell^-$.

The rare decays $B\to X_s\nu\bar\nu$, $B\to K^*\nu\bar\nu$,  
$B\to K\nu\bar\nu$ and  $B_s\to\bar\mu\mu$ are governed by
\begin{equation}\label{10V}
Q_{\nu\bar\nu}(B)  = (\bar s b  )_{V-A} (\bar \nu\nu)_{V-A}\,,~~~~~
Q_{\mu\bar\mu}(B)  = (\bar s b )_{V-A} (\bar \mu\mu)_A~.
\end{equation}

The rare decays $K\to\pi\nu\bar\nu$ and $K_L\to\bar\mu\mu$ are governed on the 
other hand by
\begin{equation}\label{11V}
Q_{\nu\bar\nu}(K)  = (\bar s d  )_{V-A} (\bar \nu\nu)_{V-A}\,,~~~~~
Q_{\mu\bar\mu}(K)  = (\bar s d )_{V-A} (\bar \mu\mu)_A~.
\end{equation}

\subsection{Local Operators in Extensions of the SM}\label{BMU}

New physics (NP) can generate new operators. Typically new operators 
are generated through the presence of right-handed (RH) currents and 
{\it scalar} currents with the latter strongly suppressed within the SM.
New gauge bosons and scalar exchanges are at the origin of these operators 
that can have an important impact on phenomenology. Below we give examples of 
new operators being aware that this list is incomplete. A much more extensive 
discussion can be found in \cite{Buchmuller:1985jz,Grzadkowski:2010es}. We
will return to it when discussing the SMEFT in Section~\ref{sec:SMEFT}.

\boldmath
\subsubsection{$\Delta F=2$ Non-leptonic Operators}
\unboldmath

For definiteness, we shall consider here operators responsible for the
$K^0$--$\bar{K}^0$ mixing. There are 8 such operators of dimension 6.
They can be split into 5 separate sectors, according to the chirality
of the quark fields they contain. The operators belonging to the first
three sectors (VLL, LR and SLL) read \cite{Buras:2000if} (our competition 
in Rome \cite{Ciuchini:1997bw} uses a different basis):
\bea 
Q_1^{\rm VLL} &=& (\bar{s}^{\alpha} \gamma_{\mu}    P_L d^{\alpha})
              (\bar{s}^{ \beta} \gamma^{\mu}    P_L d^{ \beta}),
\nnb\\[4mm] 
Q_1^{\rm LR} &=&  (\bar{s}^{\alpha} \gamma_{\mu}    P_L d^{\alpha})
              (\bar{s}^{ \beta} \gamma^{\mu}    P_R d^{ \beta}),
\nnb\\
Q_2^{\rm LR} &=&  (\bar{s}^{\alpha}                 P_L d^{\alpha})
              (\bar{s}^{ \beta}                 P_R d^{ \beta}),
\nnb\\[4mm]
Q_1^{\rm SLL} &=& (\bar{s}^{\alpha}                 P_L d^{\alpha})
              (\bar{s}^{ \beta}                 P_L d^{ \beta}),
\nnb\\
Q_2^{\rm SLL} &=& (\bar{s}^{\alpha} \sigma_{\mu\nu} P_L d^{\alpha})
              (\bar{s}^{ \beta} \sigma^{\mu\nu} P_L d^{ \beta}),
\label{normal}
\eea
where $\sigma_{\mu\nu} = \f{1}{2} [\gamma_{\mu}, \gamma_{\nu}]$ and
$P_{L,R} =\f{1}{2} (1\mp \gamma_5)$. The operators belonging to the
two remaining sectors (VRR and SRR) are obtained from $Q_1^{\rm VLL}$ and
$Q_i^{\rm SLL}$ by interchanging $P_L$ and $P_R$. In the SM only the 
operator $Q_1^{\rm VLL}=Q(\Delta S=2)/4$ is present.

\boldmath
\subsubsection{$\Delta F=1$ Non-leptonic Current-Current Operators}
\unboldmath
In the present section, we list the
current--current four-quark $\Delta F=1$ operators.  For
this purpose, we choose the operators in such a manner that all the
four flavours they contain are different: $\bar{s}$, $d$, $\bar{u}$,
$c$. In such a case, the only possible diagrams are the
current--current ones. Penguin diagrams are discussed subsequently.

Twenty linearly independent operators can be built out of four
different quark fields.  They can be split into 8 separate sectors,
between which there is no mixing. The operators belonging to the first
four sectors (VLL, VLR, SLR and SLL) read \cite{Buras:2000if}
\bea 
Q_1^{\rm VLL} &=& (\bar{s}^{\alpha} \gamma_{\mu}    P_L d^{ \beta})
              (\bar{u}^{ \beta} \gamma^{\mu}    P_L c^{\alpha}) ~=~ 
\tilde{Q}_{V_L V_L},
\nnb\\
Q_2^{\rm VLL} &=& (\bar{s}^{\alpha} \gamma_{\mu}    P_L d^{\alpha})
              (\bar{u}^{ \beta} \gamma^{\mu}    P_L c^{ \beta}) ~=~ 
Q_{V_L V_L},
\nnb\\[4mm]
Q_1^{\rm VLR} &=& (\bar{s}^{\alpha} \gamma_{\mu}    P_L d^{ \beta})
              (\bar{u}^{ \beta} \gamma^{\mu}    P_R c^{\alpha}) ~=~ 
\tilde{Q}_{V_L V_R},
\nnb\\
Q_2^{\rm VLR} &=& (\bar{s}^{\alpha} \gamma_{\mu}    P_L d^{\alpha})
              (\bar{u}^{ \beta} \gamma^{\mu}    P_R c^{ \beta}) ~=~ 
Q_{V_L V_R},
\nnb\\[4mm]
Q_1^{\rm SLR} &=& (\bar{s}^{\alpha}                 P_L d^{ \beta})
              (\bar{u}^{ \beta}                 P_R c^{\alpha}) ~=~ 
\tilde{Q}_{LR},
\nnb\\
Q_2^{\rm SLR} &=& (\bar{s}^{\alpha}                 P_L d^{\alpha})
              (\bar{u}^{ \beta}                 P_R c^{ \beta}) ~=~ 
Q_{LR},
\nnb\\[4mm]
Q_1^{\rm SLL} &=& (\bar{s}^{\alpha}                 P_L d^{ \beta})
              (\bar{u}^{ \beta}                 P_L c^{\alpha}) ~=~ 
\tilde{Q}_{LL},
\nnb\\
Q_2^{\rm SLL} &=& (\bar{s}^{\alpha}                 P_L d^{\alpha})
              (\bar{u}^{ \beta}                 P_L c^{\beta}) ~=~ 
Q_{LL},
\nnb\\
Q_3^{\rm SLL} &=& (\bar{s}^{\alpha} \sigma_{\mu\nu} P_L d^{ \beta})
              (\bar{u}^{ \beta} \sigma^{\mu\nu} P_L c^{\alpha}) ~=~ 
\tilde{Q}_{T_L T_L},
\nnb\\
Q_4^{\rm SLL} &=& (\bar{s}^{\alpha} \sigma_{\mu\nu} P_L d^{\alpha})
              (\bar{u}^{ \beta} \sigma^{\mu\nu} P_L c^{ \beta}) ~=~ 
Q_{T_L T_L},
\label{normal1}
\eea
where on the r.h.s. we have shown the notation of the Rome group 
\cite{Ciuchini:1997bw}.

The operators belonging to the four remaining sectors (VRR, VRL, SRL
and SRR) are obtained from the above by interchanging $P_L$ and $P_R$.
Obviously, it is sufficient to calculate the anomalous dimensions 
(ADMs) only for the VLL,
VLR, SLR and SLL sectors.  The ``mirror'' operators in the VRR, VRL,
SRL and SRR sectors will have exactly the same properties under QCD
renormalization. On the other hand their Wilson coefficients being governed 
by some new weak interactions can be different. In the SM only the operators $Q_1^{\rm VLL}$ and $Q_2^{\rm VLL}$ are present.

\boldmath
\subsubsection{$\Delta F=1$ Non-leptonic Penguin Operators}
\unboldmath
The operators in (\ref{normal}) and (\ref{normal1}) do not constitute the full set of six-dimensional four 
quark operators contributing to $\Delta F=1$ processes. In addition to 
QCD penguins and electroweak penguins of the SM there are other penguin 
operators.
In our paper \cite{Buras:2000if} we have therefore generalized our  analysis of 
two-loop anomalous dimensions to the full set of $\Delta F=1$ four-quark
 operators. These 
results are much less known but should be useful in the extensions of the SM 
one day. The list of these operators can be found in \cite{Buras:2000if}.
Their resurrection took place recently in the context of the NLO analysis 
within the general WET \cite{Aebischer:2021raf}.

\subsubsection{Dipole Operators}
In the presence of right-handed (RH) currents, mediated for instance by a 
very heavy $W_R$ in left-right symmetric models the dipole penguins 
\begin{equation}\label{O6R}
\tilde Q_{7\gamma}  =  \frac{e}{8\pi^2} m_b \bar{s}_\alpha \sigma^{\mu\nu}
          (1-\gamma_5) b_\alpha F_{\mu\nu}\qquad            
\tilde Q_{8G}     =  \frac{g}{8\pi^2} m_b \bar{s}_\alpha \sigma^{\mu\nu}
   (1-\gamma_5)T^a_{\alpha\beta} b_\beta G^a_{\mu\nu}  
\end{equation}
could be important. Note that in contrast to dipole operators in (\ref{O6})
the operators with $(1-\gamma_5)$ include now the factor  $m_b$ and not $m_s$.

\boldmath
\subsubsection{$\Delta F=1$ Semi-leptonic Operators}
\unboldmath
Concerning the semi-leptonic operators in the extensions of the SM the 
typical examples  of operators related to the presence of RH {\it currents} are

\begin{equation}\label{9VR}
\tilde Q_{9V}  = (\bar s b  )_{V+A} (\bar \mu\mu)_{V}, \qquad
\tilde Q_{10A}  = (\bar s b )_{V+A} (\bar \mu\mu)_{A}.
\end{equation}
\begin{equation}\label{10VR}
\tilde Q_{\nu\bar\nu}(B)  = (\bar s b  )_{V+A} (\bar \nu\nu)_{V-A}, \qquad
\tilde Q_{\mu\bar\mu}(B)  = (\bar s b )_{V+A} (\bar \mu\mu)_A~.
\end{equation}
\begin{equation}\label{11VR}
\tilde Q_{\nu\bar\nu}(K)  = (\bar s d  )_{V+A} (\bar \nu\nu)_{V-A}, \qquad
\tilde Q_{\mu\bar\mu}(K)  = (\bar s d )_{V+A} (\bar \mu\mu)_A~.
\end{equation}

If {\it scalar currents} resulting from scalar exchanges like the heavy 
Higgs in the 2HDM models or sparticles in the MSSM are present, scalar operators enter 
the game. The most prominent are the ones that govern the 
$B^0_s\to \mu^+\mu^-$ decay in 2HDMs and the MSSM at large $\tan\beta$:
\begin{equation}\label{scalarL}
Q_S  = (\bar s P_L b  ) (\bar \mu\mu),~~~~~
Q_P  = (\bar s P_L b ) (\bar \mu\gamma_5 \mu),
\end{equation}
\begin{equation}\label{scalarR}
\tilde Q_S  = (\bar s P_R b  ) (\bar \mu\mu),~~~~~
\tilde Q_P  = (\bar s P_R b ) (\bar \mu\gamma_5 \mu).
\end{equation}

\subsection{Wilson Coefficients}
\subsubsection{General Structure}
The main objects of interest in this review are the QCD and electroweak 
corrections to the Wilson coefficients 
of the operators listed above. Once these coefficients have been  calculated 
at a high energy scale like $M_W$ \footnote{$C_i(M_W)$ are often called 
{\it matching conditions} as they are found through matching of the full 
theory with heavy fields as dynamical degrees of freedom to the effective 
theory where only light fields are dynamical.}, the renormalization group methods 
allow to calculate them at low energy scales at which the 
matrix elements are evaluated by means of non-perturbative methods. Denoting 
this lower scale simply by $\mu$ 
the general expression for $ C_i(\mu) $ is given by
\begin{equation}\label{CV}
 \vec C(\mu) = \hat U(\mu,M_W) \vec C(M_W)~,   
\end{equation}
where $ \vec C $ is a column vector built out of the coefficients $C_i$.
$\vec C(M_W)$ are the initial conditions which depend on the 
short distance physics at high energy scales.
In particular they depend on $m_t$ and the masses and couplings of new heavy particles 
in the extensions of the SM.  
We set the high energy scale
at $\mw$, but other choices are clearly possible.
$ \hat U(\mu,M_W) $, the evolution matrix from $M_W$ down to $\mu$,
is given as follows:
\begin{equation}\label{UM}
 \hat U(\mu,M_W) = T_g \exp \left[ 
   \int_{g(M_W)}^{g(\mu)}{dg' \frac{\hat\gamma^T(g')}{\beta(g')}}\right] 
\end{equation}
with $g$ denoting the QCD effective coupling constant and $T_g$ an
ordering operation defined in \cite{Buras:1998raa}. $ \beta(g) $
governs the evolution of $g$ and $ \hat\gamma $ is the anomalous dimension
matrix of the operators involved. The structure of this equation
makes it clear that the renormalization group approach goes
 beyond the usual perturbation theory.
Indeed $ \hat  U(\mu,M_W) $ sums automatically large logarithms
$ \log M_W/\mu $ which appear for $ \mu\ll M_W $. In the so-called leading
logarithmic approximation (LO) terms $ (g^2\log M_W/\mu)^n $ are summed.
The next-to-leading logarithmic correction (NLO) to this result involves
summation of terms $ (g^2)^n (\log M_W/\mu)^{n-1} $ and so on.
This hierarchic structure gives the renormalization group improved
perturbation theory.

As an example let us consider only QCD effects and the case of a single
operator so that (\ref{CV}) reduces to
\begin{equation}\label{CC}
  C(\mu) =  U(\mu,M_W)  C(M_W)   
\end{equation}
with $C(\mu)$ denoting the coefficient of the operator in question.

 Keeping the first three terms in the expansions of
 $\gamma(g)$ and $\beta(g)$ in powers of $g$:
\begin{equation}\label{gammaexp}
\gamma (g) = \gamma^{(0)} \frac{\alpha_s}{4\pi} + \gamma^{(1)}
\left(\frac{\alpha_s}{4\pi}\right)^2 +  \gamma^{(2)}
\left(\frac{\alpha_s}{4\pi}\right)^3~,\qquad \alpha_s=\frac{g^2}{4\pi}
\ee
\be\label{betaexp}
 \beta (g) = - \beta_0 \frac{g^3}{16\pi^2} - \beta_1 
\frac{g^5}{(16\pi^2)^2}- \beta_2\frac{g^7}{(16\pi^2)^3}
\end{equation}
and inserting these expansions into (\ref{UM}) gives:
\begin{equation}\label{UMNLO}
 U (\mu, M_W) = \Biggl\lbrack 1 + \frac{\alpha_s (\mu)}{4\pi} J_1
+ \left(\frac{\alpha_s (\mu)}{4\pi}\right)^2 J_2
\Biggl\rbrack \Biggl\lbrack \frac{\alpha_s (M_W)}{\alpha_s (\mu)}
\Biggl\rbrack^P \Biggl\lbrack 1 - \frac{\alpha_s (M_W)}{4\pi} J_1-
\left(\frac{\alpha_s (M_W)}{4\pi}\right)^2(J_2-J_1^2)
\Biggl\rbrack 
\end{equation}
where
\begin{equation}
P = \frac{\gamma^{(0)}}{2\beta_0},\qquad J_1 = \frac{P}{\beta_0}
\beta_1 - \frac{\gamma^{(1)}}{2\beta_0},  
\end{equation} 
\be
J_2=\frac{P}{2\beta_0}\beta_2 +
\frac{1}{2}\left(J_1^2-\frac{\beta_1}{\beta_0}J_1\right)
-\frac{\gamma^{(2)}}{4\beta_0}.
\ee
General
formulae for the evolution 
matrix $ \hat U (\mu, M_W) $ in the case of operator mixing and
valid also for electroweak effects at the NLO level
 can be found in \cite{Buchalla:1995vs}. The corresponding
 NNLO formulae are rather complicated  
and given in \cite{Buras:1999st,Buras:2006gb}.
The leading
logarithmic approximation corresponds to setting $ J_1=J_2 = 0 $ in (\ref{UMNLO}). In the NLO $J_2=0$ and the last term in (\ref{UMNLO}) has to be 
removed. 

The coefficients $\beta_i$  are given as follows
\begin{equation}\label{b0b1}
\beta_0=\frac{33-2f}{3},\qquad
\beta_1=\frac{306-38 f}{3}, 
\ee
\be
\beta_2=\frac{2857}{2}-\frac{5033}{18}f+\frac{325}{54}f^2
\ee
where $f$ is the number of quark flavours.
By now three-loop \cite{Tarasov:1980au,Larin:1993tp}, four-loop \cite{vanRitbergen:1997va,Czakon:2004bu}  and five-loop \cite{Baikov:2016tgj,Herzog:2017ohr}   contributions are known.

The expansion for $C(\mw)$ is given by
\be\label{init}
C(\mw)=C_0+\frac{\as(\mw)}{4\pi} C_1+\left(\frac{\alpha_s (M_W)}{4\pi}\right)^2C_2
\ee
where $C_0$, $C_1$ and $C_2$ depend generally on $m_t$, $M_W$, the masses
of the new particles and the new parameters in the extentions of the SM. 
It should be 
stressed that the renormalization scheme dependence of $C_1$ and $C_2$ 
is canceled
by the one of $J_1$ and $J_2$ in the last square bracket in (\ref{UMNLO})
although at the NNLO level this cancellation is rather involved.
 The
scheme dependence of $J_1$ and $J_2$ in the first square bracket in (\ref{UMNLO})
is canceled by the scheme dependence of $\langle Q(\mu)\rangle$. The
power $P$ is scheme independent.
The methods for the calculation of $ \hat U (\mu, M_W) $ and the
discussion of the cancellation of the $\mu$- and renormalization 
scheme dependences
are presented in detail in \cite{Buras:1998raa,Buras:2020xsm} and in the original papers 
where such calculations have been done.

When talking about  the $\mu$-dependence one should distinguish two types of 
dependences. First we have  the 
$\mu$ dependence related to the renormalization of operators and 
present in the evolution matrix. This dependence arises in the presence of 
non-vanishing anomalous dimensions of the operators responsible for weak 
decays.
But in the coefficients $C_i(M_W)$ in (\ref{init})
also heavy quark masses are present that are running 
masses with their scale dependence governed by the anomalous dimension 
of the mass operator 
\begin{equation}\label{gama}
\gamma_m(\as)=\gamma^{(0)}_{m}\aspi + \gamma^{(1)}_{m}\left(\aspi\right)^2
+ \gamma^{(2)}_{m}\left(\aspi\right)^3
\end{equation}
with the coefficients $\gamma^{(i)}_{m}$  given 
as follows \cite{Nanopoulos:1975kd,Nanopoulos:1978hh,Larin:1993tq,Tarasov:2019rwk}
\begin{equation}\label{gm01} 
\gamma^{(0)}_{m}
=8, \qquad \gamma^{(1)}_{m}=\frac{404-40 f}{3}  
\end{equation}
\be
 \gamma^{(2)}_{m}= 2\left[1249 -\left(\frac{2216}{27}+\frac{160}{3}\zeta(3)\right)f
-\frac{140}{81}f^2\right]
\ee
where $\zeta(3)\approx 1.202057$.
These results are valid in the $\overline{\rm MS}$ scheme.

By now also four-loop \cite{Chetyrkin:1997dh,Vermaseren:1997fq} and five loop 
\cite{Baikov:2014qja} contributions to $\gamma_m(\as)$ are 
known but they are  too complicated to be presented here. Moreover, it has been demonstrated how five-loop fermion anomalous dimensions for a general gauge group can be obtained from four-loop massless propagators  \cite{Baikov:2017ujl}.

Restricting the expansions for $\gamma_m(g)$ and $\beta(g)$ to LO and NLO terms 
and expanding in $\as$ gives:
\begin{equation}\label{mmu}
{m(\mu)=
m(\mu_0)
\left[{\as(\mu)\over\as(\mu_0)}\right]^{\gamma^{(0)}_m\over 2\beta_0}
\left[1+\left({\gamma^{(1)}_{m}\over 2\beta_0}-{\beta_1\gamma^{(0)}_{m}\over
  2\beta^2_0}\right){\as(\mu)-\as(\mu_0)\over 4\pi}\right].}
\end{equation}

The value of $m(\mu_0)$ for a given quark is determined using the data. For the top quark, which is 
not confined, this can be done through high energy processes but in the case 
of the remaining five quarks, which are confined in hadrons non-perturbative 
methods are required to find $m(\mu_0)$. The most accurate values are presently 
obtained from lattice simulations. One finds then in FLAG and PDG reports.
\be\label{qm1}
m_u(2\gev)=(2.16\pm0.11)\mev,\qquad m_d(2\gev)=(4.68\pm0.15)\mev,
\ee
\be\label{qm2}
m_c(m_c) = (1.279\pm 0.013) \gev, \qquad m_s(2\gev)=(93.8\pm2.4) \mev,
\ee
\be\label{qm3}
{m_b(m_b)=4.18^{+0.03}_{-0.02}\gev,\qquad m_t(m_t) = 162.83(67)\gev\,,}
\ee
\noindent
where we also list the value of $m_t(m_t)$ which { has been obtained
  in \cite{Brod:2021hsj} using first the so-called pole masses extracted from collider 
experiments.}

If only the leading term $C_0$ is present the choice of the 
$\mu$ for the masses matters. This unphysical scale dependence is cancelled 
by the non-leading terms in (\ref{init}). A detailed discussion of this issue 
can be found in \cite{Buras:1998raa,Buras:2020xsm}  and we will return to it briefly below.

For later purposes it will be useful to generalize the formula (\ref{gammaexp}) 
to include mixing between operators and $\ord(\alpha)$ effects, 
where $\alpha$ is the QED coupling constant. 
This formula is relevant whenever also electroweak effects are present and 
electroweak penguin operators contribute. At the NNLO level in QCD but to 
leading order in $\alpha$ one has:
\begin{equation}\label{ggew}
\hat\gamma(\as,\alpha)=\hat\gamma_s^{(0)}\aspi + 
\hat\gamma_e^{(0)}\frac{\alpha}{4\pi}+ 
\hat\gamma_s^{(1)}\left(\aspi\right)^2+
\hat\gamma_{se}^{(1)}\aspi\frac{\alpha}{4\pi}+
\hat\gamma_s^{(2)}\left(\frac{\alpha_s}{4\pi}\right)^3+
\hat\gamma_{se}^{(2)}\left(\aspi\right)^2\frac{\alpha}{4\pi}
\end{equation}
with $\hat\gamma_s^{(0)}$, $\hat\gamma_s^{(1)}$  and $\hat\gamma_s^{(2)}$ being
 anomalous 
dimension matrices that are generalizations of the corresponding coefficients 
in (\ref{gammaexp}) to include mixing among operators under QCD 
renormalization. If
$\ord(\alpha)$ effects are included in the
coefficients at scales $\ord(\mw)$, the anomalous dimension matrix
must also include $\ord(\alpha)$ contributions which are represented
by $\hat\gamma_e^{(0)}$, $\hat\gamma_{se}^{(1)}$ and  $\hat\gamma_{se}^{(2)}$ 
at LO, NLO and NNLO, respectively.

The corresponding generalization of the Wilson coefficients in (\ref{init})
takes the form
\be\label{initvector}
\vec C(\mw)=\vec C_0+\frac{\as(\mw)}{4\pi} \vec C_1+
\left(\frac{\alpha_s (M_W)}{4\pi}\right)^2\vec C_2 + \frac{\alpha}{4\pi} \vec C_1^{es} + \frac{\as(\mw)}{4\pi}\frac{\alpha}{4\pi} \vec C_2^{es}\,,
\ee
where now the coefficients are column vectors and  the evolution in (\ref{CC})
generalizes for $\alpha=0$ to the one in (\ref{CV}). For $\alpha\not=0$ 
the corresponding formulae are much more complicated. They can be found in \cite{Buras:1999st}.

\subsubsection{Renormalization Scheme Dependence}
As already stated above, beyond LO various quantities like 
Wilson coefficients
and the anomalous dimensions depend on the renormalization scheme for
operators.
This dependence  arises because the renormalization
prescription involves an arbitrariness in the finite parts to be
subtracted along with the ultraviolet singularities.  
Two different schemes are then related by a finite renormalization.  

I have discussed this issue in detail in Section 6.7 of my Les Houches lectures  \cite{Buras:1998raa}. See also Section 5.2.7 in my recent book
 \cite{Buras:2020xsm}. 
Here I just want to recall one NLO formula to which I 
will refer from time to time. It is a relation between  
anomalous dimension matrices in two different renormalization schemes:
\begin{equation}\label{gpgs}
\hat\gamma^{(0)\prime}=\hat\gamma^{(0)} \qquad
 \hat\gamma^{(1)\prime}=\hat\gamma^{(1)}+[\Delta\hat r,\hat\gamma^{(0)}]+
2\beta_0 \Delta\hat r ,
\end{equation}
where the {\it prime} distinguishes the two schemes and $\Delta\hat r$ is 
a shift at $\ord(\alpha_s)$ in the matrix elements of
 operators calculated 
in these two renormalization schemes:
\begin{equation}\label{rsqc}
\langle\vec Q\rangle'
=(1+\aspi \Delta\hat r)\langle\vec Q\rangle, \qquad 
\vec C'
=(1-\aspi (\Delta\hat r)^T)\vec C.
\end{equation}

\subsection{Inclusive Decays}
So far I have discussed only  {\it exclusive} decays. It turns out that
in the case of {\it inclusive} decays of heavy mesons, like $B$-mesons,
things turn out to be easier. In an inclusive decay one sums over all 
(or over
a special class) of accessible final states and eventually one can show 
 that 
 the resulting branching ratio can be calculated
in the expansion in inverse powers of $\mb$ with the leading term 
described by the spectator model
in which the $B$-meson decay is modelled by the decay of the $b$-quark.
Very schematically one has then for the decay rate
\be\label{hqe}
\Gamma(B\to X)=\Gamma(b\to q) +\ord(\frac{1}{\mb^2})~. 
\ee
This formula is known under the name of the Heavy Quark Expansion (HQE)
\cite{Chay:1990da}. Pedagogical reviews on this topic and heavy quark 
effective theories can be found in 
\cite{Chay:1990da,Neubert:1993mb,Manohar:2000dt,Shifman:1995dn,Falk:2000tx,Lenz:2014jha}.

Since the leading term in this expansion represents the decay of the quark,
it can be calculated in perturbation theory or more correctly in the
renormalization group improved perturbation theory. It should be realized
that also here the basic starting point is the effective Hamiltonian 
 (\ref{b1})
and that the knowledge of the couplings $C_i(\mu)$ is essential for 
the evaluation of
the leading term in (\ref{hqe}). But there is an important difference 
relative to the
exclusive case: the matrix elements of the operators $Q_i$ can be 
``effectively"
evaluated in perturbation theory. 
This means, in particular, that their $\mu$ and renormalization scheme
dependences can be evaluated and the cancellation of these dependences by
those present in $C_i(\mu)$ can be investigated.

Clearly in order to complete the evaluation of $\Gamma(B\to X)$ also the 
remaining terms in
(\ref{hqe}) have to be considered. These terms are of a non-perturbative 
origin, but
fortunately they are suppressed often by  two powers of $m_b$. 
They have been
studied by several authors in the literature with the result that they affect
various branching ratios by less than $10\%$ and often by only a few percent.
Consequently the inclusive decays give generally more precise theoretical
predictions at present than the exclusive decays. On the other hand their
measurements are harder. There are of course some important theoretical
issues related to the validity of HQE in (\ref{hqe}) which appear in the 
literature under the name of quark-hadron duality but  
I will not discuss them here.

The very rough appearance of the second term on the r.h.s of (\ref{hqe}) 
totally underrepresents the efforts which have been made over many years 
to calculate these contributions. But as I was not involved in these efforts 
and they contain some non-perturbative aspects, I will not discuss them here.
With the improved precision of experimental data the uncertainties related 
to these terms become a problem and it may well turn out one day that 
exclusive decays will be under better control provided lattice QCD will 
provide precise values of the relevant form factors.
We will summarize the status of the first term in (\ref{hqe}) in some detail in Sections~\ref{sec:6} and \ref{sec:8}, where we will also briefly comment on the non-leading terms.

\subsection{The Structure and the Status of the NLO and NNLO Corrections}\label{sec:2.6}
\subsubsection{General Comments}
As we will see in the following sections
during the last 35 years the NLO corrections to $C_i(\mu)$ have
been calculated within the SM
for the most important and
interesting decays. Also several NNLO calculations have been performed.
In tables~\ref{TAB1}-\ref{TAB5}  we give
references to all NLO and NNLO calculations within the SM done until now
 that deal with the processes discussed by us.
While these calculations improved considerably the precision of
theoretical predictions in weak decays and can be considered as an
important progress in this field, the pioneering LO calculations
for current-current operators \cite{Altarelli:1974exa,Gaillard:1974nj}, penguin operators 
\cite{Shifman:1976ge,Gilman:1979bc},
$\Delta S=2$ operators \cite{Vysotsky:1979tu,Gilman:1982ap} and rare $K$ decays 
\cite{Dib:1989cc}
should not be forgotten.

\boldmath
\subsubsection{Different Classes of QCD Corrections}
\unboldmath
The structure of QCD corrections to various decays depends on the 
decay considered. In particular the expansion in $\alpha_s$ can vary from 
decay to decay. Moreover even within a given decay the structure of QCD 
corrections to internal charm and top contributions differ from each other.
Let us then classify various cases beginning with the simplest ones and 
systematically increasing the complexity.

{\bf Class 1}

The simplest situation arises when there is only one contributing operator 
with a {\em vanishing anomalous dimension }
and in addition there is no mixing of this operator with operators which 
have non-vanishing anomalous dimensions. Moreover
the diagrams in the full theory from which this operator was born 
have only heavy internal 
particles like $W^\pm$, the top quark and generally 
heavy particle exchanges in the extensions of the SM.

This is the case of the operators $Q_{\nu\bar\nu}(B)$, $Q_{\mu\bar\mu}(B)$ in 
{(\ref{10V})} and $\tilde Q_{\nu\bar\nu}(B)$, $\tilde Q_{\mu\bar\mu}(B)$ in 
(\ref{11VR})  
contributing to rare decays $B\to X_s\nu\bar\nu$, $B\to K^*\nu\bar\nu$,
$B\to K\nu\bar\nu$ and $B_{s,d}\to\mu^+\mu^-$. If the operators 
$Q_{\nu\bar\nu}(K)$ and $Q_{\mu\bar\mu}(K)$ in (\ref{11V}) and
$\tilde Q_{\nu\bar\nu}(K)$ and $\tilde Q_{\mu\bar\mu}(K)$ in (\ref{11VR})
 originate
in the internal 
top quark contributions and other heavy particle contributions to 
 rare decays $K\to\pi\nu\bar\nu$ and $K_{L,S}\to\mu^+\mu^-$ then also these 
contributions belong to this class. The case of internal charm quark 
contributions is classified separately below.

Denoting a given loop function in the absence of QCD corrections by $F_1(x_t)$,
with $x_t=m_t^2/M_W^2$ as an example,
the decay amplitudes in this case 
have the following perturbative expansion in $\alpha_s$ 
\be\label{E1}
A_1=F_1(x_t)+\ord(\alpha_s)+\ord(\alpha_s^2)\,,
\ee
with $\alpha_s$ evaluated at the high scale where the operator has been 
generated. We drop the Lorentz structure for simplicity. 
Therefore $\ord(\alpha_s)$ corrections are generally small and 
no large logarithm related to operator renormalization is present in them 
due to the vanishing of the anomalous dimension of the contributing operator.
However  $x_t$ depends on the scale $\mu_t$ present in $m_t(\mu_t)$ 
with similar comments applying to other coloured heavy particles present beyond 
the SM.
The corresponding logarithm involving this scale and present in the 
$\ord(\alpha_s)$ correction in (\ref{E1}) cancels this scale dependence 
present in the leading term $F_1(x_t)$ so that up to higher order corrections 
$A_1$ is independent of $\mu_t$. On the other hand the size of the $\ord(\alpha_s)$ 
correction in (\ref{E1}) clearly depends on the chosen $\mu_t$. It turns out 
that it is useful to set $\mu_t=m_t(m_t)$. Then 
the result can be summarized by
\be\label{E11}
A_1=F_1(x_t)\eta_{\rm QCD}\,,
\ee
with $\eta_{\rm QCD}$  close to unity and practically independent of the measured
top quark mass. For other choices of $\mu_t$ the factor $\eta_{\rm QCD}$ 
can {differ} significantly from unity but then also the numerical value of 
$F_1(x_t)$ is different so that $A_1$ remains the same up to higher order 
corrections. This removal of order $10\%$ dependence on $\mu_t$ in the 
LO formulae for rare $K$ and $B$ decays was the basic motivation for the calculations in 
\cite{Buchalla:1992zm,Buchalla:1993bv}. The QCD calculations in this class
are described in Section~\ref{sec:5} and the relevant references are collected 
in Table~\ref{TAB3}.

{\bf Class 2}

This class is constructed from Class 1 by giving the operator an anomalous 
dimension but still requiring that it does not mix with other operators 
and all particles in loops generating this operator are heavy. This is 
the case of $\Delta S= 2$ and $\Delta B=2$ operators in the SM
when only top quark contributions in the box diagrams are considered. In this case 
we have
\be\label{E2}
A_2=F_2(x_t)\times 1_{\rm QCD}+\ord(\alpha_s)+\ord(\alpha_s^2),
\ee
where the funny $1_{\rm QCD}$ represents the leading RG factor like the 
one involving $P$ in (\ref{UMNLO}). Now the $\ord(\alpha_s)$ and higher 
order terms involve not only $M_W$ but also low energy scale $\mu$ at the 
end of the RG evolution as seen explicitly in (\ref{UMNLO}).
Moreover the  
$\ord(\alpha_s(M_W))$ correction involves now two logarithms multiplied by 
two different anomalous dimensions, one anomalous dimension of the mass 
operator related to the $\mu_t$ dependence present already in Class 1 and 
the second present in $P$ involving the anomalous dimension of $Q(\Delta F = 2)$ 
in (\ref{O7}). The latter logarithm cancels the $\mu_W$ dependence present in 
the funny factor $1_{\rm QCD}$ in (\ref{E2}) so that $A_2$ does not 
depend on the precise value of the scale at which the Wilson coefficients 
are defined. Again one can summarize the result schematically by 
\be\label{E21}
A_2=F_2(x_t)\eta_{\rm QCD}.
\ee
However, this time $\eta_{\rm QCD}$ can depart significantly from unity 
as summation of large logarithms in the process of RG evolution is 
involved. $\eta_{\rm QCD}$ depends as seen in (\ref{UMNLO}) on the lower 
scale $\mu$ and this dependence cancels the one present
 in the hadronic matrix elements. This {latter} dependence in $\eta_{\rm QCD}$ 
is often factored out so that the known factors $\eta_2\equiv\eta_{tt}$ in 
$\varepsilon_K$ and $\eta_B$ in $B_{d,s}^0-\bar B_{d,s}^0$ mixing are $\mu$-independent 
and this also applies to the $\hat B_i$ factors that up to factors involving 
weak decay constants represent hadronic matrix elements. Explicit expressions 
are given in Section~\ref{sec:4}.

This discussion applies also to the $\Delta F=2$ operators in (\ref{normal}) 
except that now mixing under renormalization between operators $Q_1^{\rm LR}$
and $Q_2^{\rm LR}$ and similarly between $Q_1^{\rm SLL}$ and $Q_2^{\rm SLL}$ 
 takes place. Explicit formulae for this case can be found in \cite{Buras:2001ra}. 
In some extensions of the SM FCNC operators are generated already at tree level
 but also in this case an analogous discussion 
can be made. We will be more explicit about this during our presentation.

The QCD calculations in this class
are described in Section~\ref{sec:4} and the relevant references are collected 
in Table~\ref{TAB2}.

{\bf Class 3}

We next consider QCD corrections to charm contributions to $\kpn$ and 
$\kmm$ decays. The corresponding operators are those of Class 1 but now
a light charm quark mass is present in the loop. Consequently even 
without QCD corrections a large logarithm $\ln m_c/M_W$ is present and 
on the way to low scales  bilocal operators enter the game. They undergo 
a rather complicated renormalization \cite{Buchalla:1993wq}. Only when the charm is integrated out 
we obtain the local operators 
$Q_{\nu\bar\nu}(K)$ and $Q_{\mu\bar\mu}(K)$ in (\ref{11V}). From the point of view of the renormalization group analysis, the expansion in $\alpha_s$ takes in this case the following form
\be\label{E3}
A_3= \ord (\frac{1}{\alpha_s}) + \ord (1) + \ord(\alpha_s).
\ee 
Thus the NLO corrections to the charm part of the amplitudes for 
$\kpn$ and $\kmm$ amount to the $\ord(1)$ term \cite{Buchalla:1993wq}, while the NNLO 
corrections amount to the $\ord(\alpha_s)$. Still to get this term three-loop
diagrams have to be evaluated \cite{Buras:2006gb,Gorbahn:2006bm} for $\kpn$ and $\kmm$, respectively. The LO 
term has been calculated in \cite{Dib:1989cc}.

The QCD calculations in this class
are described in Section~\ref{sec:5} and the relevant references are collected 
in Table~\ref{TAB3}.

{\bf Class 4}

We next consider the set of $\Delta F=1$ current-current operators in 
(\ref{O1}) that both have nonvanishing anomalous dimensions and
mix under renormalization. After the diagonalization 
of this system one gets two operators $Q_\pm$ in (\ref{F4}) which 
evolve without 
mixing. The situation for each operator is then similar to Class 2 except 
that no box 
diagrams involving heavy particles have to be evaluated to get a nonvanishing result. Consequently 
the expansion is as follows
\be\label{E4}
A_4=1_{\rm QCD}+\ord(\alpha_s)+\ord(\alpha_s^2),
\ee
where $\alpha_s$ terms are evaluated both at $M_W$ and the low scale 
$\mu$ according to the evolution in (\ref{UMNLO}). The leading term is 
again the one involving $P$ in (\ref{UMNLO}).

The QCD calculations in this class
are described in Section~\ref{sec:3} and the relevant references are collected 
in Table~\ref{TAB1}.

{\bf Class 5}

We next consider  QCD penguin and electroweak penguin operators 
in (\ref{O2})-(\ref{O5}) contributing to non-leptonic decays. 
These operators mix under QCD and QED
renormalization and evidently the QCD penguin and electroweak penguin 
diagrams in the full theory 
are $\ord(\alpha_s)$ and $\ord(\alpha)$, respectively. The 
Wilson coefficients of the corresponding operators after the top quark and 
$W^\pm$ have been integrated out are also of the same order respectively. 
This mismatch of powers in $\alpha_s$ can be overcome in the process of
renormalization group analysis by multiplying the electroweak operators 
by $1/\alpha_s$ and compensating this rescaling by multiplying  
their Wilson coefficients by $\alpha_s$. As $\alpha$ is from the 
point of view of QCD a fixed number, the RG can now be performed  as in 
class 2
except for the following changes. $Q_1$ and $Q_2$ operators have to be 
included as they mix into $Q_3-Q_{10}$ operators  affecting their QCD
evoloution. Thus we deal with $10\times 10$ anomalous dimension matrices,
but because of the presence of electroweak penguins also terms 
$\ord(\alpha\alpha_s)$ at NLO order have to be considered and 
$\ord(\alpha\alpha_s^2)$ at NNLO in addition to the usual $\ord(\alpha_s^2)$ 
and $(\alpha_s^3)$ terms, respectively. See (\ref{ggew}).

We observe that now the formulae are a bit more involved but what is more 
important are the following facts which apply for instance to the evaluation 
of the ratio $\epe$ in $K_L\to\pi\pi$:
\begin{itemize}
\item
At LO there is no top quark mass dependence nor any heavy particle mass 
dependence from penguin diagrams.
\item
At NLO these mass dependences enter for the first time.
\item
 The NLO matching conditions
for electroweak penguin operators  do not involve QCD corrections to box and
penguin diagrams and  consequently the renormalization scale 
dependence in the top quark mass in these processes
is not negligible at the NLO. In order to reduce this unphysical dependence,
QCD corrections to the relevant box and penguin diagrams have
to be computed. 
In the renormalization group improved perturbation
theory these corrections are a part of NNLO
corrections. In \cite{Buras:1999st}  such corrections
have been computed for $\epe$. Their inclusion 
 allowed to reduce renormalization 
scheme dependence present in the electroweak penguin sector. 
\end{itemize}

The QCD calculations in this class
are described in Section~\ref{sec:3} and the relevant references are collected 
in Table~\ref{TAB1}.

{\bf Class 6}

We next come to dipole penguin operators in (\ref{O6}) restricting 
the discussion to the $B\to X_s\gamma$ decay. These two operators mix 
under renormalization with each other and are also influenced by the 
mixing with  the current-current operators and QCD penguin operators.
Thus we deal here with a $8\times 8$ anomalous dimension matrix. 

I will report on the heroic efforts to calculate the QCD corrections to 
the $B\to X_s\gamma$ rate in Section~\ref{sec:6}.
Let me here only write down symbolically the general structure of the 
amplitudes in this class:
\be\label{E6}
A_6=F_6(x_t)\times 1_{\rm QCD}+ \tilde 1_{\rm QCD}  +\ord(\alpha_s)+\ord(\alpha_s^2),
\ee
with the first two terms representing LO and the $\ord(\alpha_s)$ and 
$\ord(\alpha^2_s)$
 terms representing  NLO and NNLO
corrections, respectively. The $\tilde 1_{\rm QCD}$ term represents the mixing of dipole 
penguin operators with the 
current-current and QCD-penguin operators. In fact this term is responsible 
for the strong enhancement of the $B\to X_s\gamma$ decay rate.

The QCD calculations of the $B\to X_s\gamma$ decay rate
are described in Section~\ref{sec:6} and the relevant references are collected 
in Table~\ref{TAB4}.

{\bf Class 7}

In  rare decays $K_L\to\pi^0\ell^+\ell^-$ and
$B \to X_s \ell^+\ell^-$  semileptonic  operators 
in (\ref{9V}) are involved. 
In addition to these operators also current-current non-leptonic
operators (\ref{O1}) and QCD penguin operators in (\ref{O2}) and (\ref{O3})
have to be taken into account. The electroweak penguins and dipole 
penguins turn out to be irrelevant in $K_L\to\pi^0\ell^+\ell^-$ but the dipole 
penguins have to be included in $B \to X_s \ell^+\ell^-$.

The new feature is that although the operator $Q_{9V}$ has no anomalous 
dimension by itself it mixes with the operators $Q_1-Q_6$ and consequently 
a $7\times 7$ anomalous dimension matrix has to be considered. The resulting 
structure of the coefficient $C_{9V}$ is then
\be\label{E7}
C_{9V}= \ord (\frac{1}{\alpha_s})+ F(x_t) +\tilde 1_{\rm QCD}+ \ord(\alpha_s).
\ee
Consequently the structure is similar to Class 3 but at the NLO level in this 
case heavy quark mass dependence enters.

The operator $Q_{10 A}$ has no anomalous dimension and  similarly to operators
in (\ref{10V}) and (\ref{11V}) does not mix with anybody. Therefore its 
Wilson coefficient has the structure in (\ref{E1}). Evidently this 
operator enters the amplitude for $K_L\to\pi^0\ell^+\ell^-$ at the NLO level.

In the case of $B \to X_s \ell^+\ell^-$ the situation is complicated by the 
presence of dipole operators but the structure of the corresponding Wilson 
coefficients $C_{9V}$ and $C_{10A}$ is the same. The $\ord(\alpha_s)$ 
corrections to the penguin and box diagram relevant for the NNLO evaluation 
of  $B \to X_s l^+l^-$ rate have been calculated in \cite{Bobeth:1999ww}.

The QCD calculations of $K_L\to\pi^0l^+l^-$ and $B \to X_s l^+l^-$
are described in Sections~\ref{sec:7} and \ref{sec:8} respectively and the relevant references are collected 
in Table~\ref{TAB5}.

{\bf Class 8}

Finally I give the structure of QCD corrections for the charm-charm 
($\eta_1\equiv\eta_{cc}$) and charm-top ($\eta_3\equiv\eta_{ct}$) contributions 
to the $\Delta S=2$ Hamiltonian. The structure of these corrections differs from 
$\eta_2\equiv\eta_{tt}$ discussed in Class 2 and also from each other but 
I think it is instructive to put them together in order to see 
the difference. We have
\be\label{eta1}
\eta_1=(\alpha_s)^{P_+}\left(1_{\rm QCD}+\ord(\alpha_s)+\ord(\alpha_s^2) \right)\,,
\ee
\be\label{eta3}
\eta_3=(\alpha_s)^{P_+}\left(\frac{1}{\alpha_s}1_{\rm QCD}+\tilde 1_{\rm QCD}+\ord(\alpha_s)\right)\,.
\ee

The references to QCD calculations of $\eta_1$ and $\eta_3$ are collected 
in Table~\ref{TAB2} and in Section~\ref{sec:4} we will make  several remarks on  these calculations.

\subsubsection{Two-Loop Anomalous Dimensions Beyond the SM}
In the extentions of the SM new operators are present. The
two loop anomalous dimensions for the $\Delta F=2$ and 
$\Delta F=1$ four-quark
dimension-six operators listed in (\ref{normal}) and (\ref{normal1}) 
have been computed in \cite{Ciuchini:1997bw,Buras:2000if}.
In \cite{Buras:2000if} also the remaining anomalous dimensions 
of $\Delta F=1$ non-leptonic penguin operators can be found.
These days they are relevant for  the NLO analysis 
within the WET \cite{Aebischer:2021raf}
and the SMEFT \cite{Aebischer:2022anv}.

\subsubsection{Two-Loop Electroweak Corrections}
In order to reduce scheme and scale dependences related to the
definition of electroweak parameters like $\sin^2\theta_W$,
and $\alpha_{QED}$,  two-loop electroweak contributions to rare decays
have to be computed. For $K^0_L \rightarrow \pi^0\nu\bar{\nu}$, 
$B_{d,s} \rightarrow l^+l^-$ and $B \rightarrow X_{\rm s}\nu\bar{\nu}$ they
can be found in \cite{Buchalla:1997kz,Brod:2010hi,Bobeth:2013tba}, for 
$B_{d,s}^0-\bar B_{d,s}^0$ mixing in 
\cite{Gambino:1998rt}
and for $B\rightarrow X_s \gamma$ in 
\cite{Czarnecki:1998tn,Strumia:1998bj,Kagan:1998ym,Gambino:2000fz,Gambino:2001au}. For $\varepsilon_K$ such corrections have been calculated in \cite{Brod:2019rzc,Brod:2021qvc}.

\subsubsection{NLO QCD Calculations Beyond the SM}
There exist also a number of partial or complete NLO QCD calculations 
within the Two-Higgs-Doublet Model (2HDM) and the MSSM. In the case
of the Two-Higgs-Doublet Model such calculations for $B_{d,s}^0-\bar B_{d,s}^0$
mixing, $B\to X_s\gamma$ and $B\to X_s\ell^+\ell^-$ 
can be found in \cite{Urban:1997gw},
\cite{Bobeth:1999ww,Borzumati:1998tg,Borzumati:2003rr,Ciuchini:1997xe,Ciafaloni:1997un} and \cite{Schilling:2004gk}, respectively. 
In 2HDM also NNLO QCD corrections to $B\to X_s\gamma$ have been 
calculated \cite{Hermann:2012fc}.

The corresponding NLO calculations for $B_{d,s}^0-\bar B_{d,s}^0$ and
$B\to X_s\gamma$ in the MSSM can be found in \cite{Ciuchini:2005kp,Bertuzzo:2010un,Virto:2009wm,Virto:2011yx} and 
\cite{Bobeth:1999ww,Borzumati:1999qt,Feng:2000kg,Ciuchini:1998xy,Greub:2011ji,Greub:2011qt}, respectively.  The paper \cite{Bobeth:1999ww} gives also the results for
$B\to X_s~ {\rm gluon}$. In fact Bobeth, Misiak and Urban \cite{Bobeth:1999ww} 
present rather general formulae for Wilson coefficients relevant for $B\to X_s\gamma$ and $B\to X_s~ {\rm gluon}$ evaluated 
at high scale 
(matching conditions) at the LO and NLO level that can be used for other 
extensions of the SM. 

Finally, I would like to mention calculations of 
 NLO QCD corrections to rare $K$ and $B$ decays in the 
MSSM at low $\tan\beta$ \cite{Bobeth:2001jm} and of  NNLO QCD corrections to $B\to X_s \ell^+\ell^-$ in the MSSM \cite{Bobeth:2004jz}.

\subsubsection{Penguin-Box Expansion}
The rare and CP violating decays of $K$ and $B$ mesons as well as $\varepsilon_K$, 
$\epe$   and $B^0_q-\bar B^0_q$ mixings are governed in the SM
by various penguin and box diagrams with internal top quark and charm quark
exchanges. Some examples are shown in Fig.~\ref{fig:fdia}. 
Evaluating these diagrams one finds
a set of basic universal (process independent) 
$\mt$-dependent functions $F_r(x_t)$ \cite{Inami:1980fz,Buras:1998raa} where $x_t=\mt^2/\mw^2$. 
Explicit expressions for these
functions can be found in \cite{Buras:1998raa,Buras:2020xsm}.

It is useful to express the OPE formula (\ref{amp5}) directly in terms
of the functions $F_r(x_t)$ \cite{Buchalla:1990qz}:
\begin{equation}
A({M\to F}) = P_0(M\to F) + \sum_r P_r(M\to F) \, F_r(x_t),
\label{generalPBE1}
\end{equation}
where the sum runs over all possible functions contributing to a given
amplitude. 
$P_0$  summarizes contributions stemming from internal charm quark. 
In the OPE formula (\ref{amp5}), the functions $F_r(x_t)$ are
hidden in the initial conditions for $ C_i(\mu)$ represented
by $\vec C(\mw)$ in (\ref{CV}).

The coefficients $P_0$ and $P_r$ are process dependent and
include QCD corrections contained in the evolution matrix
$\hat U(\mu,\mw)$. They depend also on hadronic matrix
elements of local operators and the relevant CKM factors.
An efficient and straightforward method for finding the coefficients
$P_r$ is presented in \cite{Buchalla:1990qz}.
As the expansion in (\ref{generalPBE1}) involves basic one-loop 
functions from penguin and box diagrams it was naturally given the 
name of the  {\it Penguin-Box Expansion} (PBE).

Generally, several basic functions contribute to a given decay,
although decays exist which depend only on a single function.
We have the following correspondence between the most interesting FCNC
processes and the basic functions within 
models with constrained Minimal Flavour Violation \cite{Blanke:2006ig}: 
\begin{center}
\begin{tabular}{lcl}
$K^0-\bar K^0$-mixing ($\varepsilon_K$) 
&\qquad\qquad& $S(v)$ \\
$B_{d,s}^0-\bar B_{d,s}^0$-mixing ($\Delta M_{s,d}$) 
&\qquad\qquad& $S(v)$ \\
$K \to \pi \nu \bar\nu$, $B \to X_{d,s} \nu \bar\nu$ 
&\qquad\qquad& $X(v)$ \\
$K_{\rm L}\to \mu \bar\mu$, $B_{d,s} \to \ell^+\ell^-$ &\qquad\qquad& $Y(v)$ \\
$K_{\rm L} \to \pi^0\ell^+ \ell^-$ &\qquad\qquad& $Y(v)$, $Z(v)$, 
$E(v)$ \\
$\varepsilon'$, Nonleptonic $\Delta B=1$, $\Delta S=1$ &\qquad\qquad& $X(v)$,
$Y(v)$, $Z(v)$,
$E(v)$ \\
$B \to X_s \gamma$ &\qquad\qquad& $D'(v)$, $E'(v)$ \\
$B \to X_s~{\rm gluon}$ &\qquad\qquad& $E'(v)$ \\
$B \to X_s \ell^+ \ell^-$ &\qquad\qquad&
$Y(v)$, $Z(v)$, $E(v)$, $D'(v)$, $E'(v)$,
\end{tabular}
\end{center}
where $v$ denotes collectively the arguments of a given function with $v=x_t$ 
in the SM. In these models the 
operator structure of the SM remains intact and NP modifies only the 
basic functions.
It should be mentioned that this correspondence is strictly valid 
at LO. At NLO in processes in which mixing between different operators is 
present new loop functions can contribute to a given process but these 
contributions are generally small.

Originally  PBE was designed to expose the $\mt$-dependence
of FCNC processes \cite{Buchalla:1990qz} which was hidden in the Wilson coefficients. 
In particular in the case of $\epe$, where many of these functions enter, this 
turned out to be very useful.
After the top quark mass has been
measured precisely this role of PBE is less important.
On the other hand,
PBE is very well suited for the study of the extensions of the
SM in which new particles are exchanged in the loops.
If there are no new local operators beyond those present in the SM
the mere change is to modify the functions $F_r(x_t)$ which now
acquire the dependence on the masses of new particles such as
charged Higgs particles and supersymmetric particles. The process
dependent coefficients $P_0$ and $P_r$ remain unchanged. 
The effects of new physics can then be seen transparently. Many 
examples of the applications of PBE can be found in the literature. 
In particular in the last two  decades we have used this method for the 
study of FCNC processes in the MSSM, a model with a universal extra 
dimension, littlest Higgs model (LH), littlest Higgs model with T parity (LHT),
MSSM at low $\tan\beta$, the SM with four generations, $Z^\prime$
model and 331 models. In these papers 
compilations of the functions $F_r$ in a given model can be found. 
A complete list of references can be found in \cite{Buras:2010wr,Buras:2012ts,Buras:2013ooa,Buras:2020xsm}. 

One virtue of this method is a transparent study of the departure from 
MFV. In this framework, as discussed in detail in \cite{Buras:2003jf},
 the basic loop 
functions are universal with respect to the system considered. Indeed 
as seen above the same function $S(v)$ enters $\varepsilon_K$ and 
$B_{d,s}-\bar B_{d,s}$ mixings. Similarly the same function $X(v)$ enters 
$\kpn$ and $B\to X_{d,s}\nu\bar\nu$ decays. Indeed in MFV the flavour 
dependence resides fully in the CKM matrix elements. Moreover, all these
functions are real as the sole complex CP-violating phase resides in the CKM 
matrix and in flavour blind CP phases. In the presence of new sources of flavour and CP violation things are 
different:
\begin{itemize}
\item
The universality in question is broken and many MFV relations between 
various branching ratios are generally violated.
\item
The basic functions become complex quantities leading often to new CP-violating  effects 
not encountered in the CKM framework.
\end{itemize}
All these new effects can be transparently seen in this framework.

If
new effective operators with different Dirac and colour structures
are present, new functions multiplied by new coefficients
$P_r(M\to F)$ contribute to (\ref{generalPBE1}). Still one is free 
to add these contributions to the functions of the SM, but then 
the dependence on the new $P_r(M\to F)$ is included in the basic 
functions. This is not always convenient if one wants to see
explicitly the effects of new right-handed operators or scalar operators 
$Q_i^{LR}$, $Q_i^{SLL}$ and $Q_i^{SRR}$ given previously.
Therefore in this case 
it is better to proceed as follows.

We can start with (\ref{amp5}) but instead of evaluating it at the 
low energy scale we choose for $\mu$ the high energy scale to be called 
$\mu_H$ at which heavy particles are integrated out. Then absorbing 
$G_F/\sqrt{2}$ and   $V^i_{CKM}$ in the Wilson coefficients $C_i(\mu_H)$
the amplitude for $M-\overline{M}$ mixing ($M= K, B_d,B_s$) is 
simply given by
\be\label{amp6}
A(M\to \overline{M})
=\sum_{i,a} C^a_i(\mu_H)\langle \overline{M} |Q^a_i(\mu_H)|M\rangle.
\ee
Here the sum runs over all the operators in (\ref{normal}), that is 
 $i=1,2$ and $a=VLL,VRR,LR, ...$. The 
matrix elements for $B_d-\bar B_d$ mixing are for instance given as 
follows \cite{Buras:2001ra}
\be\label{eq:matrix}
\langle \bar B_d^0|Q_i^a(\mu_H)|B_d^0\rangle = \frac{2}{3}M_{B_d}^2 F_{B_d}^2 P_i^a(B_d),\qquad ({\rm Version~1})
\ee
where the coefficients $P_i^a(B_d)$ 
collect compactly all RG effects from scales below $\mu_H$ as well as
hadronic matrix elements obtained by lattice methods at low energy scales.
 When using the formula like (\ref{eq:matrix}) one should check how 
the external states are normalized. In \cite{Buras:2001ra} we have used 
the normalization leading to (\ref{eq:matrix}) but in my recent papers 
the  following formula 
\be\label{eq:matrix1}
\langle \bar B_d^0|Q_i^a(\mu_H)|B_d^0\rangle = \frac{1}{3}M_{B_d} F_{B_d}^2 P_i^a(B_d), \qquad ({\rm Version~2})
\ee
can be found. The coefficients  $P_i^a(B_d)$  are the same in these expressions 
and the difference in the factor in front of them is canceled when 
physical amplitudes are evaluated.  This issue  is independent of  $P_i^a(B_d)$ but can be rather confusing for non-experts and I wanted to mention it here. More details on it can be found in my Les Houches lectures and in my book.

Analytic formulae for   $P_i^a(B_d)$,  $P_i^a(B_s)$  and  $P_i^a(K)$ 
 are given in \cite{Buras:2001ra} 
while the applications of this method in various beyond SM analyses 
can be found in 
\cite{Buras:2010mh,Buras:2010zm,Buras:2010pz}. As the 
Wilson coefficients $ C_i(\mu_H)$ depend directly on the loop functions 
and fundamental parameters of a given theory, this dependence can be 
exhibited if necessary. The most recent values for the related hadronic 
matrix elements for $K^0-\bar K^0$ mixing have been obtained by three LQCD collaborations: ETM \cite{Carrasco:2015pra},  SWME \cite{Jang:2015sla} and RBC-UKQCD \cite{Boyle:2017skn,Boyle:2017ssm}. An insight in these results has been obtained
with the help of the DQCD approach \cite{Buras:2018lgu}.

The following points should be emphasized:
\begin{itemize}
\item The expressions (\ref{amp6}) and (\ref{eq:matrix}) are valid 
for any model with the model dependence entering only the Wilson coefficients 
$C_i^a(\mu_H)$, 
which generally also depend on the meson system considered. In particular, 
they are valid both within and beyond the MFV framework. In MFV models
CKM factors and Yukawa couplings 
define the flavour dependence of these coefficients, while in 
non-MFV models additional flavour structures are present in $C_i^a(\mu_H)$.
\item The coefficients $P_i^a$ are model independent and include 
the renormalization group evolution from the high scale $\mu_H$ down to low
 energy $\mathcal O(\mu_K,\mu_B)$. As the physics cannot depend on
 the renormalization scale $\mu_H$, the 
$P_i^a$ depend also on $\mu_H$ so that the scale dependence present in 
$P_i^a$ is canceled by the one in $C_i^a$. 
Explicit formulae for $\mu_H$ dependence of $P_i^a$ can be found 
in \cite{Buras:2001ra}. It should be stressed that here we are talking about 
logarithmic dependence on $\mu_H$. The power-like dependence 
(such as $1/M_H^2$, \ldots) is present only in the $C_i^{a}$.
\item The $P_i^a$ depend however on the system considered as the hadronic matrix elements of the operators in (\ref{amp6}) relevant for  instance for 
$K^0-\bar K^0$ mixing differ from the matrix elements of analogous operators 
relevant for $B_{s,d}^0-\bar B_{s,d}^0$ systems. Moreover whereas the 
RG evolution in the latter systems stops at $\mu_B=\mathcal O(M_B)$, 
in the case of $K^0-\bar K^0$ system it is continued down to 
$\mu_K\sim 2$ GeV, where the hadronic matrix elements are 
evaluated by lattice methods.
\end{itemize}

 After this rather heavy material we are ready to return to the main story of 
this paper: the evaluation of higher order QCD and electroweak corrections 
to weak decays. The results of these efforts played already 
an important role in the tests of the SM in the last 35 years. But their role will be even more important in the coming flavour precision era in which 
hopefully we will identify new physics at very short distance scales. 
The results discussed below are crucial in this indirect search for new physics. 
Indeed, one requires a high precision for SM contributions to flavour observables
in order to identify the new phenomena through the differences between the data
and SM predictions.

\section{\boldmath{$\Delta S=1$} and \boldmath{$\Delta B=1$} Non-Leptonic 
and Semi-Leptonic Decays}\label{sec:3}
\setcounter{equation}{0}
\subsection{Effective Hamiltonians}
The effective Hamiltonian for non-leptonic $\Delta S=1$ transitions is given 
in the SM as follows:
\begin{equation}
\Heff(\Delta S=1) = 
\frac{G_{\rm F}}{\sqrt{2}} V_{us}^* V_{ud}^{} \sum_{i=1}^{10}
\left( z_i(\mu) + \tau \; y_i(\mu) \right) Q_i(\mu) 
\label{eq:HeffdF1:1010}
\end{equation}
with $\tau=-V_{ts}^* V_{td}^{}/(V_{us}^* V_{ud}^{})$.
The operators $Q_i$ are the analogues of the ones given in 
(\ref{O1})-(\ref{O5}). In the case of $\Delta B=1$ transitions 
the flavours have to be changed appropriately
and the effective Hamiltonian is usually written as follows
\begin{eqnarray}\label{HB4} 
H_{eff}(\Delta B=1) &=&
\frac{G_F}{\sqrt{2}} \Bigg[ \lambda_u(C_1(\mu_b) 
Q^u_1+C_2(\mu_b)Q^u_2) 
+\lambda_c( C_1(\mu_b) Q^c_1+C_2(\mu_b)Q^c_2)
\nonumber\\  
&& -\lambda_t \sum_{i=3}^{10,8G} C_i(\mu)Q_i \Bigg] ,
\end{eqnarray}~ 
where
\be\label{lb}
\lambda_q=V_{qs}^{*}V_{qb}
\ee
and
\be\label{Qq}
Q_1^q=(\bar q_\alpha b_\beta)_{V-A}(\bar s_\beta q_\alpha)_{V-A}~,
\quad\quad
Q_2^q=(\bar q_\alpha b_\alpha)_{V-A}(\bar s_\beta q_\beta)_{V-A}~.
\ee
In particular $Q_i^c=Q_i$ in (\ref{O1}).
\subsection{Current-Current Operators}
Let me begin our NLO story with the first climb within the MNLC which Peter 
Weisz and myself started in December 1988:
the calculation of two-loop anomalous dimensions of $Q_1$ and $Q_2$ operators.
It involves 
28 two-loop diagrams shown in Fig.~\ref{fig:4}  and the corresponding counter terms. 
We decided to 
perform it in three schemes for $\gamma_5$: anticommuting (NDR), 
't Hoof-Veltman scheme 
and the DRED scheme used by the Italian group in 1981. These schemes are
discussed in detail in Section~4.1.3 in my book \cite{Buras:2020xsm}.

I am sure that 
presently a calculation of that type is done 
fully automatically by means of appropriate computer programs but in the
winter 1988/1989 and in the spring of 1989 
we did it almost entirely by hand. This has the advantage that each step of 
the calculation can be followed and enjoyed in contrast to looking 
constantly at the computer.

Working then entirely by hand the first step is the calculation of two-loop momentum integrals keeping 
$1/\varepsilon^2$ and $1/\varepsilon$ terms {with $\epsilon=(4-D)/2$.} 
In doing this, it is useful to factor out any Dirac and colour structures. 
In this 
manner the results for the two-loop integrals in question can be used for the 
calculations of two-loop anomalous dimensions of other operators and in 
different renormalization schemes in which $\gamma_5$ is treated differently.
This first step is rather tedious but 
straightforward and it is not surprising that working independently
 we obtained the same results 
for all 
28 diagrams already in the first comparison. I have used these results for the 
calculation of two-loop anomalous dimensions of other operators in 1991 and 
1999. I will discuss these calculations later on.
In calculating colour factors I found the paper
\cite{MacFarlane:1968vc} very useful. 

\begin{figure}[hbt]
\vspace{0.10in}
\centerline{
\epsfysize=1.25in
%\rotate[r]{
\epsffile{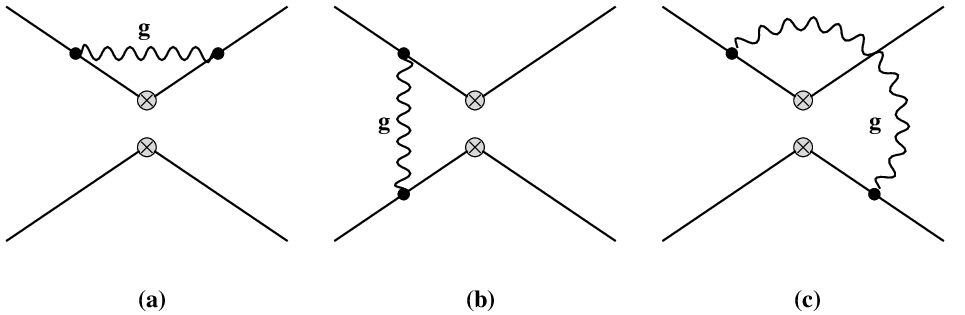}
}%}
\vspace{0.06in}
\caption[]{One loop current-current  diagrams that contribute to one-loop 
anomalous dimensions and enter as subdiagrams the two-loop 
calculations of two-loop anomalous dimensions.
The 4-vertex ``$\otimes\,\,\otimes$'' denotes the
insertion of a 4-fermion operator $Q_i$.
\label{L:15}}
\end{figure}

The last step
of the calculation, the manipulation of Dirac structures turned out to be the 
crucial part of our work. When calculating the subdiagrams in Fig.~\ref{L:15} 
we encountered structures like 
 \be\label{F1}
\Gamma_\nu\gamma_\rho\gamma_\mu\otimes\Gamma^\nu\gamma^\rho\gamma^\mu ,
\qquad \Gamma_\nu=\gamma_\nu(1-\gamma_5)
\ee
that we had to reduce to the operators $Q_1$ and $Q_2$ which have the 
structure $\Gamma\otimes\Gamma$. At one-loop one can do 
this in $D=4$ dimensions as $1/\varepsilon$ is the leading singularity, but 
in a two-loop calculation the $1/\varepsilon$
singularity, from which the anomalous dimensions are extracted, is 
next-to-leading. Consequently $\ord(\varepsilon)$ terms in Dirac structures
multiplied  by the leading $1/\varepsilon^2$  
singularity from the momentum integrals have an impact on two-loop 
anomalous dimensions and have to be taken properly into account.

As the first scheme we considered the one with anticommuting $\gamma_5$ 
in $D\not=4$ 
dimensions giving it the name NDR (naive dimensional regularization). In 
order to reduce the structures like the one in (\ref{F1}) to $Q_1$ and 
$Q_2$ we first 
followed the procedure of Tracas and Vlachos \cite{Tracas:1982gp} who simply wrote
\be\label{F2}
\Gamma_\nu\gamma_\rho\gamma_\mu\otimes\Gamma^\nu\gamma^\rho\gamma^\mu=
A~\Gamma_\nu\otimes\Gamma^\nu
\ee
and found the coefficient $A$ by replacing $\otimes$ by $\gamma_\tau$ and 
contracting the Dirac 
indices on both sides. This procedure, to be called the ``Greek method" in what follows, 
gives $A=4(4-\epsilon)$ and consequently
\be\label{F3}
\Gamma_\nu\gamma_\rho\gamma_\mu\otimes\Gamma^\nu\gamma^\rho\gamma^\mu=
    4(4-\epsilon)\Gamma_\nu\otimes\Gamma^\nu.
\ee

This method is very efficient and can be applied to Dirac structures with 
many $\gamma_\mu$ that appear at two-loop level. It can also be easily  
generalized to other operators. 
For 
instance in the case of $\gamma_\mu(1-\gamma_5)\otimes\gamma^\mu(1+\gamma_5)$
one should replace $\otimes$ by 1.

Applying this method to 28 diagrams in question and using it also for the 
counter diagrams we could readily find the total $1/\varepsilon$ singularity.
Subsequently 
including the two-loop wave function renormalization for the external quarks 
we found the two-loop anomalous dimension matrix in the basis
\be\label{F4}
Q_+=\frac{Q_2+Q_1}{2}, \qquad Q_-=\frac{Q_2-Q_1}{2}~.
\ee

We expected this matrix to be diagonal but we both found it to contain 
non-diagonal terms. There was a hope for an hour. We disagreed on four 
diagrams. After a fortunate draw $2:2$ we were certain that our calculation 
was algebraically correct but unfortunately the unwanted non-diagonal terms 
were still there. 
The only place, we could think, something went wrong was the ``Greek method" 
for the reduction of the complicated Dirac structures to the physical 
operators $Q_1$ and $Q_2$ as given in (\ref{F3}).

Indeed in addition to the physical operators 
$Q_{1,2}$ on the r.h.s of (\ref{F3}) 
one could have other operators with different Dirac structures that vanish in 
$D=4$. They had to vanish in $D=4$ because in $D=4$ the formula (\ref{F3}) 
is correct as can be easily checked by using standard manipulations of 
Dirac matrices. 
Peter called these operators ``effervescent" operators. I did not object to 
this 
name. How could I? After all he is English, not me. Later we were told that 
the proper name is ``evanescent" and we used this name in the following papers 
but in our first paper published in Nucl. Phys. B still "effervescent" 
operators appear \cite{Buras:1989xd}. Amusingly our Rome competitors 
used our wording still several years later \cite{Ciuchini:1993fk}.

Explicit expressions for the evanescent operators can be worked out. 
We have done it in our paper. But in practice it is more convenient 
to define them simply as the difference between the r.h.s. and l.h.s. of 
(\ref{F3}) and 
to insert them like that in the relevant two-loop diagrams. Therefore instead of 
(\ref{F3}) we have 
\be\label{F3A}
\Gamma\gamma_\rho\gamma_\mu\otimes\Gamma\gamma^\rho\gamma^\mu=
    4(4-\epsilon)\Gamma\otimes\Gamma + E^{\rm NDR}\,,
\ee
with the evanescent operator  $E^{\rm NDR}$ defined simply 
by this equation. As discussed in detail in \cite{Dugan:1990df,Herrlich:1994kh} 
 this is not the only 
possible definition of evanescent operators but possibly the most convenient 
one. The unphysical arbitrariness in the definition of evanescent operators 
has also been emphasized by Jamin and Pich \cite{Jamin:1994sv}.

Having indentified the possible origin of our problems we have incorporated 
the evanescent operators into our calculation. In particular we derived, to my 
knowledge for the first time, formulae that allow the extraction of the 
two-loop anomalous dimensions of physical operators in the presence of 
evanescent 
operators. The outcome of these efforts is section 4 of our paper. We have 
invested plenty of time in writing this section but apparently 
several 
of my colleagues had a difficult time in following it. I tried to improve on 
it in my Les Houches lectures \cite{Buras:1998raa}, where a systematic procedure for the inclusion 
of evanescent operators in the calculation of two loop anomalous dimensions of
local operators can be found.

\begin{figure}[ht]
\vspace{0.15in}
\centerline{
\epsfysize=4.6in
\epsffile{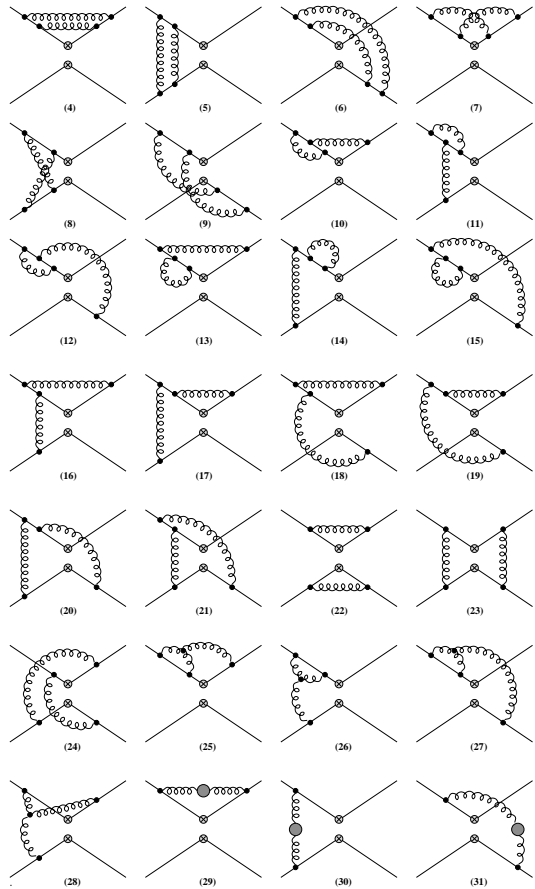}
}
\vspace{0.15in}
\caption{Two--loop current--current diagrams contributing to 
$\hat\gamma_s^{(1)}$.
The curled lines denote gluons. The 4-vertices
``$\otimes\ \otimes$'' denote standard operator insertions.
In addition shaded blobs stand for self-energy insertions.
Possible left-right or up-down reflected diagrams are not shown.
\label{fig:4}}
\end{figure}

Having the full machinery for the evaluation of the contributions 
of evanescent operators at hand, we 
could now find that their presence not only modified the diagonal terms in the 
$2\times 2$ matrix in question but also canceled the off-diagonal terms. 
We were now 
sure that the first NLO summit has been conquered: we knew the $(Q_+,Q_-)$  
or equivalently the $(Q_1,Q_2)$ matrix at the two-loop level in the NDR scheme. 
As at no place in the 
calculation it was necessary to evaluate the dangerous traces  
$Tr(\gamma_\mu\gamma_\rho\gamma_\nu\gamma_\lambda\gamma_5)$, we were 
quite confident that this result was correct.

The calculation in the 't Hooft-Veltman (HV) scheme for $\gamma_5$ was 
technically 
more difficult because of the horrible Dirac algebra for which we had to use a 
computer program written by Peter. As I did the NDR calculation entirely by 
hand, including Dirac algebra, I could test the correctness of this program. 
Otherwise we did not encounter any obstacles and we soon had the two-loop 
anomalous dimension matrix of $(Q_1, Q_2)$ in the HV scheme. By calculating 
the one-loop diagrams in Fig.~2 we found the matrix $\Delta\hat r$ in (\ref{rsqc}), relating the matrix elements of the operators in question
in 
the NDR and HV schemes.  Inserting it in (\ref{gpgs}) we could indeed 
verify that our 
results for the two-loop anomalous dimensions in these two
schemes were compatible with each other.

Strictly speaking this was the end of the story as we had the two-loop
anomalous dimension matrix 
in the HV scheme that did not have any mathematical inconsistencies related 
to $\gamma_5$. We have also demonstrated that at least in this case a 
consistent 
calculation in the simpler  NDR scheme could be made. Still we were curious 
whether the 
calculation of the Italian pioneers in 1981 was compatible with our results. 
Actually it would suffice to calculate the relevant matrix $\Delta \hat r$
 relating HV or NDR 
scheme to the DRED scheme to find out that the calculation of 1981 was 
correct, but for reasons that I do not understand today, we repeated the 
1981 
calculation confirming diagram by diagram the results of the Italian team. 
Most probably we were feeling very strong in such calculations and we
were simply delighted in producing results for these 28 diagrams in a 
different renormalization scheme.

We have submitted our paper \cite{Buras:1989xd} to Nucl.~Phys.~B in June 
1989 and sent our 
preprint to CERN and SLAC libraries. One should recall that in 1989 the Los 
Alamos Archive did not exist yet. Few weeks later our preprint has been 
distributed by the ordinary mail around the world.

At the end of July 1989 I attended the Photon-Lepton conference in Stanford. 
On the first day of the conference the two Guidos of the 1981 team 
congratulated me on our paper. They were truly delighted. They were 
apparently not sure that their paper was correct. Indeed the calculation 
in the DRED scheme is very involved as also some aspects of QCD coupling
renormalization had to be modified.

At the same conference I met for the first time Matthias Jamin, who just got 
his PhD in Heidelberg and was supposed to join my group in Munich two 
months later. I told him about the MNLC and he immediately became the 
third member of the club. 
Already in October of the same year we were climbing together the 
{\it second} NLO 
summit, the 
QCD corrections to $\Delta F=2$ processes that I will describe  briefly in 
the next section. For the time being I will continue with the $\Delta F=1$ 
effective Hamiltonian for non-leptonic decays
including now the QCD and electroweak penguin operators.

\begin{table}[thb]
\caption{NLO and NNLO Calculations for Non-leptonic and Semi-Leptonic $\Delta F=1$ Transitions. More references on semi-leptonic $B$ decays can be found in Sections~\ref{Gambino1} and \ref{Gambino2}.}
\label{TAB1}
\begin{center}
\begin{tabular}{|l|l|l|}
\hline
\bf \phantom{XXXX} Decay &  {\bf NLO} & {\bf NNLO}  \\
\hline
\hline
Current-Current $(Q_1,Q_2)$      &\cite{Altarelli:1980te,Buras:1989xd}& \cite{Gorbahn:2004my}\\
QCD penguins $(Q_3,Q_4,Q_5,Q_6)$  &\cite{Buras:1991jm,Buras:1992tc,Buras:1993dy,Ciuchini:1992tj,Ciuchini:1993vr,Chetyrkin:1997gb},\cite{Fleischer:1992gp}& \cite{Gorbahn:2004my,Cerda-Sevilla:2016yzo}\\
electroweak penguins $(Q_7,Q_8,Q_9,Q_{10})$  &\cite{Buras:1992zv,Buras:1993dy,Ciuchini:1992tj,Ciuchini:1993vr} & \cite{Buras:1999st}\\
inclusive non-leptonic decays       & \cite{Altarelli:1980te,Altarelli:1980fi,Buchalla:1992gc,Bagan:1994zd,Bagan:1995yf,Krinner:2013cja}; \cite{Jamin:1994sv}  & \\
Two-Body B-Decays in QCDF   & 
\cite{Beneke:1999br}--\cite{Bartsch:2008ps}&
 \cite{Beneke:2005vv}-\cite{Beneke:2009ek}\\
Current-Current (BSM) & \cite{Ciuchini:1997bw,Buras:2000if}    & \\
Penguins (BSM)  &\cite{Buras:2000if}  &\\
Semi-Leptonic $B$ Decays ($\vcb$, $\vub$)&
\cite{Ali:1979is,Altarelli:1982kh,Jezabek:1988iv,Jezabek:1988ja,Czarnecki:1989bz,Czarnecki:1994pu,Nir:1989rm,Bagan:1994qw,Lenz:1997aa,Lenz:1998qp,Falk:1995me,Trott:2004xc,Aquila:2005hq,DeFazio:1999ptt,Gambino:2006wk}& 
\cite{Aquila:2005hq,Luke:1994yc,Ball:1995wa,Gremm:1996gg,Pak:2008qt,Melnikov:2008qs,Biswas:2009rb,Czarnecki:1998kt,Czarnecki:1997hc,Czarnecki:1997cf,Czarnecki:1996gu,vanRitbergen:1999gs,Bigi:1993ex,Bosch:2004th,Neubert:1993ch,Andersen:2005mj,Beneke:2008ei,Greub:2009sv}\\
\hline
\end{tabular}
\end{center}
\end{table}

In Table~\ref{TAB1} we collect the references to the papers which calculated 
NLO and NNLO corrections to $\Delta F=1$ processes except for rare and radiative decays discussed in Sections 6--8. 
Two-Body B Decays in QCD Factorization (QCDF)  are discussed in Section 9. 
I thank Gerhard Buchalla and Martin Beneke for helping me in  collecting the references to 
NLO and NNLO calculations in QCDF given in this table.

\subsection{QCD Penguin Operators}

In the fall of 1990, after two successful expeditions, it was time to return 
to the QCD penguin operators that were the main topic of the seminal supper
with 
Guido Martinelli in the Ringberg castle two and a half years before. There 
were no signs coming from Rome that the Italian team was making any 
progress on penguins but I started worrying that they were far ahead of us. 
For this reason I decided to increase our team. Markus Lautenbacher, my 
former diploma student and since April 1990 my PhD student, became the 
fourth member of the MNLC and its first PhD student. Markus did not have 
any experience with two-loop calculations but his high computer skills and an 
impressive discipline in doing research convinced me that he would be a great 
help in our project. Our first goal was the calculation of the $6\times 6$ 
two-loop 
anomalous dimension matrix $\hat\gamma_s^{(1)}$ describing the mixing under 
renormalization 
of the operators $Q_1,Q_2,..Q_6$ in the NDR and HV schemes. 
The calculation of $\hat\gamma_s^{(1)}$ involves the insertions of all these
operators into vertex diagrams considered already by Peter and myself 
in our first paper and into two-loop penguin 
diagrams in Fig.~\ref{fig:5} to be 
considered for the first time. The latter diagrams do not 
have any impact on the sector $(Q_1,Q_2)$ so that the corresponding 
$2\times 2$ submatrix of $\hat\gamma_s^{(1)}$ 
calculated by Peter and myself remained untouched. 

In the first month I 
worked closely with Markus helping him in making first steps on this new 
ground, whereas Peter and Matthias worked independently by themselves. 
Later Matthias 
and Markus worked closely together and constructed  an efficient program for 
Dirac algebra manipulations in $D\not=4$  in the NDR and HV schemes \cite{Jamin:1991dp}.  This 
program written in Mathematica became an important part of our project in 
particular in the case of the HV scheme and even I used it despite of my 
previous comments on computer manipulations. In this scheme
 the calculations of two-loop 
penguin diagrams by hand were 
prohibitive as even computer manipulations required in 1990 a good PC. The 
evaluation of two-loop momentum integrals in a few penguin diagrams that we 
did mostly by hand turned out to be rather involved and a method by Peter to 
find the coefficients of the divergences in these particular diagrams 
numerically was very helpful. The corresponding calculation of the vertex 
diagrams was simple as we had already all integrals from \cite{Buras:1989xd}. Only the Dirac 
structures were different.

There were two new features with respect to the calculation of the $(Q_1,Q_2)$ 
system of 1989. First we had to face the dangerous traces 
$Tr(\gamma_\mu\gamma_\rho\gamma_\nu\gamma_\lambda\gamma_5)$. In the HV 
scheme they can be straighforwardly evaluated but their evaluation in the NDR scheme 
could lead to wrong results. For a few weeks we thought that we had to 
abandon the calculation in the NDR scheme but at the end we solved the 
problem in two ways. My solution was to work, dependently on the diagram 
considered, with a second operator basis $\{\tilde Q_i\}$ with 
$\tilde Q_i$ being Fierz transformed operators of $Q_i$.
With the help of these operators it was possible to avoid the appearance of 
the dangerous traces. However, simply replacing $ Q_i$ by 
$\tilde Q_i$ in order to avoid dangerous
traces with $\gamma_5$ and inserting it into a two-loop penguin diagram would
eventually 
give the wrong result for $\hat\gamma_s^{(1)}$. This is the second new feature of
the presence of 
penguin diagrams: the insertion of 
$(\tilde Q_1,\tilde Q_2,\tilde Q_3,\tilde Q_4)$  into a two-loop diagram 
gives different result from the insertion of 
$(Q_1, Q_2, Q_3, Q_4)$  and consequently 
the two-loop mixing between $(Q_1,Q_2)$ and $(Q_3,Q_4)$ differs from the one 
between $(\tilde Q_1,\tilde Q_2)$ and $(\tilde Q_3,\tilde Q_4)$. Fortunately,
similary to the renormalization scheme 
dependence of $\hat\gamma_s^{(1)}$, the difference in 
question could be found by 
performing a 
one-loop calculation that did not involve dangerous traces with $\gamma_5$. 
Incorporating this difference properly into the calculation that involved 
both the 
original operators and their Fierz transforms allowed then to obtain 
$\hat\gamma_s^{(1)}$ for the 
original basis in the NDR scheme without any problems with $\gamma_5$. 

\begin{figure}[ht]
\vspace{0.15in}
\centerline{
\epsfysize=4.6in
\epsffile{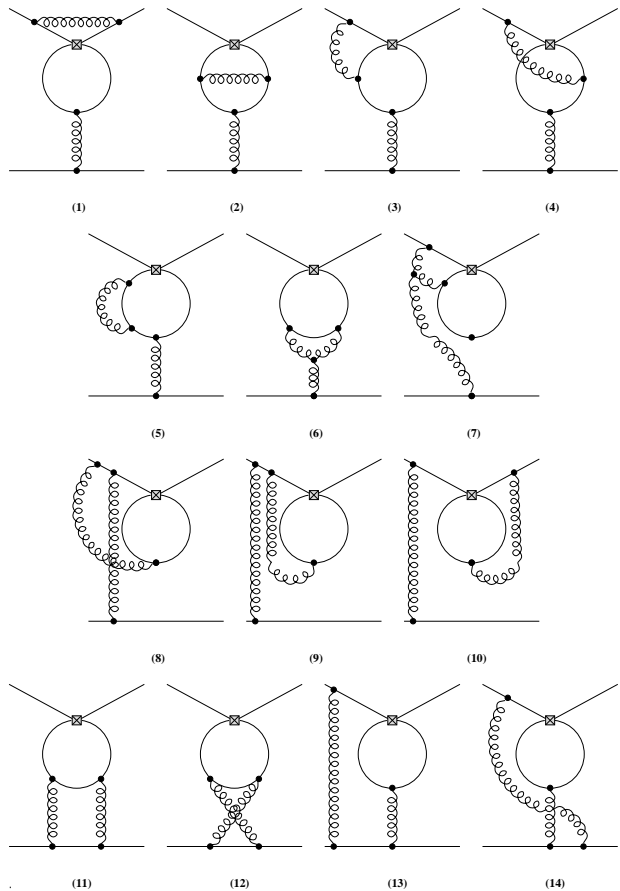}
}
\vspace{0.15in}
\caption{Two--loop penguin diagrams contributing to $\hat\gamma_s^{(1)}$.
The curled lines denote gluons. Square-vertices stand for 
penguin insertions.
Possible left-right reflected diagrams are not shown.
\label{fig:5}}
\end{figure}

This procedure is described in detail in \cite{Buras:1992tc}. In the case of the 
evaluation of the two-loop anomalous dimensions of electroweak penguin 
operators it had to be generalized to include more bases because of the more 
complicated flavour structure of these operators. This is described in 
\cite{Buras:1992zv}.
This procedure was followed by Matthias and Markus. Peter succeeded to 
avoid the dangerous traces in a different manner but I do not remember how. 
This is however immaterial as our independent results for $\hat\gamma_s^{(1)}$ 
in the NDR 
scheme agreed with each other. 

I must admit that at that time I was very 
satisfied with my procedure of working simultaneously in $D\not=4$ with 
the original 
basis and the Fierz transformed basis and making finite renormalizations 
through one-loop calculations at the end as explained above. However, in 
1994 a more elegant and a more systematic procedure with the same 
outcome has been proposed by Matthias Jamin and Toni Pich \cite{Jamin:1994sv} 
 in the course of 
their NLO analysis of inclusive $\Delta F=1$ transitions. Basically one can 
make four 
dimensional Fierz transformations in a $D\not=4$  calculation provided the 
evanescent 
operators that vanish in $D=4$ under the Fierz transformation are also 
included in 
the analysis. They are simply given by the difference of a given operator and 
its Fierz transformed operator. This method has been rediscovered by Mikolaj 
Misiak and J\"org Urban in the process of another NLO climb that I did with 
them in 1999. I will return to it in Section~\ref{sec:3.5}.

Still my cooking recipe that involved calculating the differences between the 
one-loop insertion of an operator and of its Fierz transformed operator is 
very 
useful for finding out whether the evanescent operators of that type are 
relevant for the calculation of two-loop anomalous dimensions or not. If this 
difference vanishes, the evanescent operators in question 
do not contribute.
This is the case of all operators $Q_1,..Q_{10}$ inserted into the 
current-current diagrams and of $Q_5,..Q_8$ inserted into the penguin diagrams.

We thus had in the Spring of 1991 the full $6\times 6$ anomalous dimension matrix $\hat\gamma_s$ at 
$\ord(\alpha_s^2)$ in the NDR 
scheme.\footnote{It's generalization to a $10\times 10$ matrix is discussed in the 
next subsection.} The calculation in the HV scheme was time-consuming because of the
difficult Dirac algebra but with the computer program developed by Markus 
and Matthias we could calculate $\hat\gamma_s^{(1)}$ in the HV scheme 
both directly by 
calculating the traces with $\gamma_5$ and by using my procedure discussed 
above, obtaining the same 
result. Finally calculating the relevant one loop shift 
$\Delta \hat r$ we verified that our results for $\hat\gamma_s^{(1)}$ in the 
NDR and HV schemes were consistent with each other.

Next we calculated the initial conditions for the Wilson coefficient 
functions at $\mu=\ord(M_W)$ 
 both in the NDR and the HV scheme and verified that the scheme 
dependence of these coefficients cancelled the one of the evolution matrices 
at $\mu=\ord(M_W)$ as explained in section 2.

Thus the third NLO summit has been reached. As usually reaching a summit 
a photo is taken, I thought it was appropriate to show the resulting 
 two-loop anomalous dimension matrix $\hat \gamma^{(1)}_s$ in the
NDR scheme {with $f$ being the number of quark flavours:}
\begin{equation}
\left(
\begin{array}{cccccc}
-{{21}\over 2} - {{2\,f}\over 9} & {7\over 2} + {{2\,f}\over 3} & {{79}\over\
  9} & -{7\over 3} & -{{65}\over 9} & -{{7}\over{3}} \\ \mvs
{7\over 2} + {{2\,f}\over 3} & -{{21}\over 2} - {{2\,f}\over 9} &\
  -{{202}\over {243}} & {{1354}\over {81}} & -{{1192}\over {243}} &
{904 \over 81} \\ \mvs
0 & 0 & -{{5911}\over {486}} + {{71\,f}\over 9} & {{5983}\over {162}} +\
  {f\over 3} & -{{2384}\over {243}} - {{71\,f}\over 9} &
{1808 \over 81} - {f \over 3} \\ \mvs
0 & 0 & {{379}\over {18}} + {{56\,f}\over {243}} & -{{91}\over 6} +\
  {{808\,f}\over {81}} & -{{130}\over 9} - {{502\,f}\over {243}} &
-{14 \over 3} + {{646\,f} \over 81} \\ \mvs
0 & 0 & {{-61\,f}\over 9} & {{-11\,f}\over 3} & {{71}\over 3} + {{61\,f}\over\
  9} & -99 + {{11\,f} \over 3} \\ \mvs
0 & 0 & {{-682\,f}\over {243}} & {{106\,f}\over {81}} & -{{225}\over 2} +\
  {{1676\,f}\over {243}} & -{1343 \over 6} + {{1348\,f} \over 81}
\end{array}
\right).
\label{eq:gs1ndrN3Kpp}
\end{equation}
This matrix looks truly horrible and I will in what follows refrain from 
showing other photos of this type. In fact after the inclusion of 
electroweak penguins and going to three loops the results, although very 
impressive, cannot be easily digested. On the other hand 
the $2\times2$ submatrix in the 
upper left corner, our first summit conquered  by Peter Weisz and 
myself in June of 
1989, looks beautiful and simple.

Our paper has been submitted to Nucl. Phys. B in May 1991 \cite{Buras:1991jm}. 
In addition to the 
$6\times 6$ matrices  and the Wilson coefficients of $(Q_1,Q_2,...Q_6)$ 
in the NDR and HV schemes
for both $\Delta S=1$ and $\Delta B=1$ decays 
contained general expressions for the evolution matrices $\hat
U(\mu_1,\mu_2)$  including NLO 
corrections. In order to derive these expressions we have used the general
all order formulae 
of my 1980 review on asymptotic freedom in deep inelastic scattering 
\cite{Buras:1979yt}. Finally 
we have demonstrated the scheme independence of the resulting decay 
amplitudes. Thus at last, three years after the Ringberg workshop, the Wilson 
coefficients of current-current and QCD penguin operators were known at 
NLO in the NDR and HV schemes.

I have presented these results in a parallel session at the joined 
Photon-Lepton and European Physical Society Meeting that in 1991 took place in 
Geneva, Switzerland \cite{Buras:1991vc}. Rather disappointigly only few of my colleagues 
appreciated these 
results. In particular Eduardo de Rafael thought it was an overkill in view 
of the uncertainties in the hadronic matrix elements of the operators in 
question. Eduardo got interested in our work only ten years later, when 
he wanted to know more about the role of evanescent operators and Fierz 
relations in our calculations that he and his collaborators wanted to combine 
with their calculations of hadronic matrix elements within the large N approach.

Also to my great surprise and true disappointment, there was essentially no reaction from Guido 
Martinelli. He only informed me that his PhD students are working on this 
project as well and that in order to complete the project one needs the 
hadronic matrix elements of QCD penguin operators. In fact in 1991 I knew 
these matrix elements in the context of the Dual QCD approach developed in collaboration 
with Bardeen and G\'erard several years earlier, but Guido meant here 
the ones obtained by lattice QCD methods. Unfortunately Guido's dream to 
achieve the latter result did not materialize until 2020, but the uncertainties
in the 2020 lattice result are so large that we still do not know whether
 the QCD-penguin 
hadronic parameter $B_6$  agrees with the one obtained in the Dual QCD approach earlier. But I am rather confident that this will turn out to be the case, similar to  $\hat B_K$ relevant for the CP-violating 
parameter $\varepsilon_K$ in $K_L\to\pi\pi$ and the electroweak penguin parameter $B_8$ entering $\epe$.  I will write more about it at the end of this review.

As far as QCD penguins  are concerned the first NLO analysis of penguin induced 
$B$-decays and the related CP-asymmetries using our two-loop results has 
been performed by my student, Robert Fleischer, in the Summer of 1992 \cite{Fleischer:1992gp}.
Robert combined his one-loop calculations of matrix elements 
with the two-loop anomalous dimensions discussed above and demonstrated the 
scheme independence of the final result. It was his Diploma thesis. During 
his Phd studies Robert fell in love with electroweak penguins and investigated 
their role in non-leptonic decays. Therefore he did not have time to 
participate in the subsequent papers on NLO QCD corrections to weak decays 
except 
for his second paper
in \cite{Fleischer:1992gp} which is  a proof of his new interests in 
1994.

Further progress in the evaluation of the $6 \times 6$ anomalous dimension
  matrix for the operators $Q_1$, \ldots, $Q_6$ will be described in
  Sections~\ref{subsec:admbsg} and \ref{sec:bsgNNLO}.

\subsection{Electroweak Penguin Operators}
Despite this rather moderate interest in our work in 1991 I was convinced 
that we should continue our project. The next step was to extend our 
calculation of $\hat\gamma^{(1)}_s$ in (\ref{ggew}) to electroweak 
penguin operators $Q_7$, $Q_8$, $Q_9$ and $Q_{10}$ and to calculate the ten 
dimensional two-loop anomalous 
dimension matrix    $\hat\gamma^{(1)}_{se}$  that is 
necessary for the inclusion 
of the electroweak penguin operators at the NLO level with the goal to 
calculate the CP-violating ratio $\epe$. Moreover, we wanted to
write up the 
details of all these calculations. We have not done this in 
\cite{Buras:1991jm}. 

Unfortunately, 
Peter told me that he would only be involved in the 
calculation of  $\hat\gamma^{(1)}_s$ as he was 
again very much involved in the collaboration with Martin L\"uscher. Thus 
only 
Matthias, Markus and me were involved in the  $\hat\gamma^{(1)}_{se}$ project
that amounted in particular to the calculation of the two-loop diagrams 
in Fig.~\ref{fig:4ew}. 
Having all the 
machinery at hand we performed both calculations during the fall of 1991 and 
the winter 1991/1992 so that in March 1992 we had $\hat\gamma_s^{(1)}$ and 
$\hat\gamma_{es}^{(1)}$ including all the ten
operators in the NDR and HV schemes. Moreover we calculated $\ord(\alpha)$ 
corrections to the Wilson coefficients at $\mu=M_W$, an ingredient of the NLO 
analysis, that is necessary to remove the renormalization scheme 
dependence from the decay amplitudes.

Unfortunately, there was a problem with our result for $\hat\gamma_{es}^{(1)}$. 
While the $[\hat\gamma_{s}^{(1)}]_{\rm NDR}$ and $[\hat\gamma_{s}^{(1)}]_{\rm HV}$ 
were compatible with each other, $[\hat\gamma_{es}^{(1)}]_{\rm NDR}$ and 
$[\hat\gamma_{es}^{(1)}]_{\rm HV}$ were not. That is  $[\hat\gamma_{es}^{(1)}]_{\rm HV}$ 
obtained by 
the direct two-loop calculation differed by a small amount from 
$[\hat\gamma_{es}^{(1)}]_{\rm HV}$ found 
from $[\hat\gamma_{es}^{(1)}]_{\rm NDR}$ by means of a formula analogous to 
(\ref{gpgs}) that is given in \cite{Buras:1992zv}.

\begin{figure}[ht]
\vspace{0.15in}
\centerline{
\epsfysize=4.6in
\epsffile{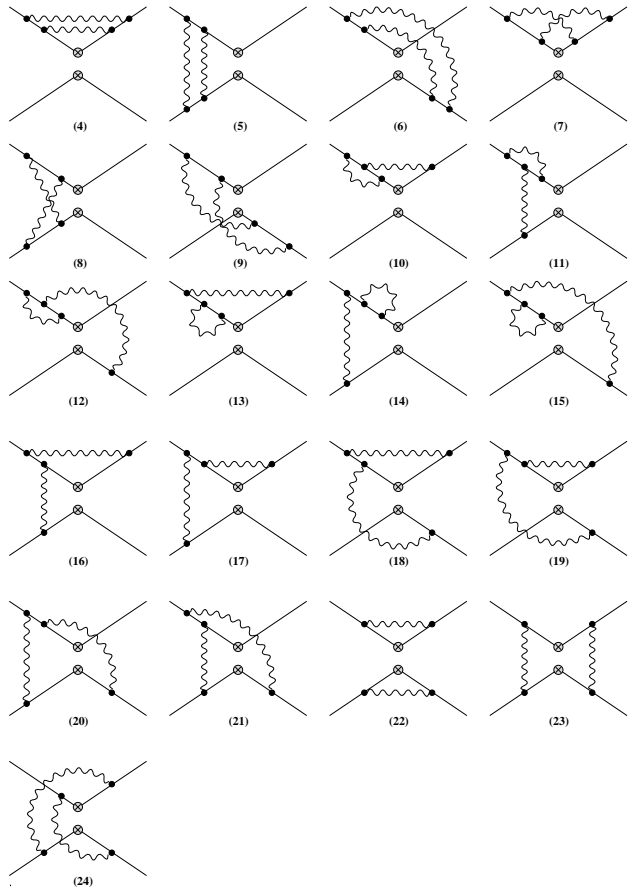}
}
\vspace{0.15in}
\caption{Two--loop current--current diagrams contributing to $\hat\gamma_{se}^{(1)}$.
The wavy lines denote gluons or photons. The 4-vertices
``$\otimes\ \otimes$'' denote standard operator insertions.
Possible left-right or up-down reflected diagrams are not shown.
\label{fig:4ew}}
\end{figure}

We spent some 
time in order to clarify this 
discrepancy but after a few weeks we made a pause in our search for the 
error. I think we were simply exhausted. Moreover everyone was involved 
simultaneously in other projects: Peter with Martin L\"uscher, Matthias and 
Markus with writing up their paper on "TRACER", the program, written in 
Mathematica, for Dirac algebra in $D\not=4$ for NDR and HV schemes 
\cite{Jamin:1991dp} and in my 
case  in addition to being the head of the theory 
institute at our university and finishing a review article with Michaela 
Harlander, I started a new NLO climb:
in the spring of 1992 our  NLO club got the {\it fifth} member, a very 
important one, namely my new PhD student 
Gerhard Buchalla with whom I planned to attack at the 
NLO level all rare semi-leptonic $K$ and $B$ decays dominated by $Z^0$-penguins. 
More about this in Section~\ref{sec:5}.

Fortunately before making a pause in our climb we decided to write up the two 
papers, one including Peter on $\hat\gamma_s^{(1)}$ that did not have any 
problems and the 
second one without him on $\hat\gamma_{es}^{(1)}$ that had the problem 
mentioned above. Thus 
already in April 1992 our papers were essentially finished but before we could 
show them to the public we had still to solve the remaining problem in the 
second paper.

From the beginning I was fully confident that our results in the NDR scheme 
were correct. The calculations were simpler than in the HV scheme and I was 
able to make several tests that all worked. As the calculations of 
$\Delta\hat r$ matrices 
relating the NDR and HV schemes is a one loop affair I was also confident 
that our results in the HV scheme obtained from the NDR scheme by means 
of a relation similar to (\ref{gpgs}) were also correct. However, the game of making a given NLO 
calculation in various schemes and checking the compatibility of the results 
started by Peter and myself three years before somehow fooled us and we did 
not send the papers for publication although we had all results already in 
April 1992. In the language of {mountain} climbing it is afterall irrelevant 
whether the 
first climb of a summit  was done using the NDR ``climbing style'' or the
 HV one.

Fortunately, the Rome group did not present their results at the 1992 summer  
conferences and consequently we were still in the game. Moreover, Matthias 
became  CERN fellow and could inform us in the first days of November 1992
that Guido Martinelli will give a seminar on $\epe$ beyond leading logarithms
four 
weeks later. It was time to be active again. Feeling like colonel 
Hunt before the final attack to conquer the Mount Everest summit
I convinced my collaborators to
send out the paper on $\hat\gamma_s^{(1)}$ in the existing form and to present 
the details of the 
calculation of $\hat\gamma_{es}^{(1)}$ in the second paper only in the NDR scheme
making the shift 
$\Delta \hat r$ to obtain it in the HV scheme. Our two papers 
\cite{Buras:1992tc,Buras:1992zv}
appeared in 
the second half of November 1992, 
roughly two weeks before Guido's CERN talk and three weeks before the 
Rome group sent out their letter to the Los Alamos archive 
\cite{Ciuchini:1992tj}.

To our delight the Rome team consisting of Marco Ciuchini, Enrico Franco, 
Guido Martinelli and Laura Reina agreed with our results on the anomalous 
dimension matrices in both NDR and HV schemes but to our surprise they did 
not present any details of their calculations of these matrices. Instead they 
presented their analysis of $\epe$ including NLO QCD and QED corrections. Thus, 
although the Munich team has published as the first group all ingredients of a 
NLO analysis of $\Delta F=1$ processes: two-loop anomalous dimensions and the 
Wilson coefficients at $\mu=M_W$, the Rome group was the first to present a 
NLO 
analysis of $\epe$ that included both QCD and electroweak penguin contributions.

\begin{figure}[ht]
\vspace{0.15in}
\centerline{
\epsfysize=4in
\epsffile{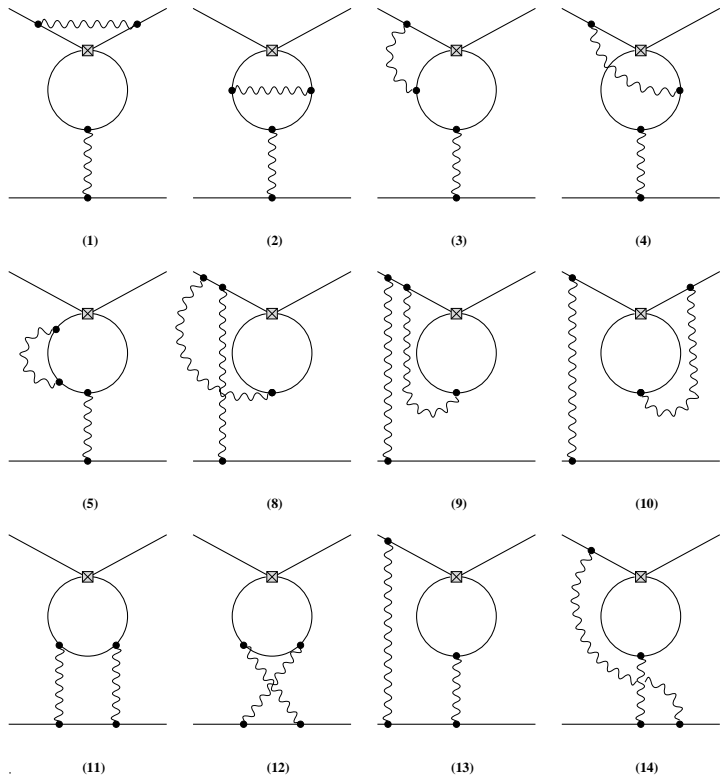}
}
\vspace{0.15in}
\caption{Two--loop penguin diagrams contributing to $\hat\gamma_{se}^{(1)}$.
The wavy lines denote gluons or photons with each diagram containing 
one gluon and one photon. Square-vertices stand for two types 
of penguin insertions.
Possible left-right reflected diagrams are not shown.
\
\label{fig:5ew}}
\end{figure}

Our NLO analysis of $\epe$ appeared in March 1993 in a very long paper (112 
pages) that fortunately was accepted in this form in Nucl.~Phys.~B.
\cite{Buras:1993dy}.
 We have 
presented there very explicit formulae for the Wilson coefficients of all 
operators and studied also the $\mu$ dependence of the hadronic $B_i$ 
parameters. 
Moreover, we have proposed a method for extracting some of these 
parameters from the CP-conserving amplitudes. In this manner we could 
incorporate the $\Delta I=1/2$ rule into the analysis of $\epe$. The results of
this analysis have 
been used by us in the 1990s to gradually develop an approximate but rather 
accurate formula for $\epe$ that depends on three parameters $\bsi$, $\bei$ and
$\Omega_{\rm eff}$ with 
the first two representing the relevant matrix element of $Q_6$ and $Q_8$ 
operators 
and the last one summarizing isospin breaking corrections \cite{Buras:1993bi,Buras:1996dq,Buras:2003zz,Buras:2014sba}. The last update
of this formula can be found in \cite{Aebischer:2020jto} that includes most
recent evaluation of isospin breaking corrections \cite{Buras:2020pjp}.

As the formulae presented by the Rome group were not as explicit as ours, an 
analytic comparison between their final results and ours was not possible but 
numerically the agreement was within a few percent.

In our long paper \cite{Buras:1993dy}
  on $\epe$ we have also presented for the first time 
NLO Wilson coefficients of
the operators relevant for non-leptonic $\Delta B=1$ decays. Amusingly the 
interest of 
particle physics community in $\epe$, in the middle of the 1990s, was rather moderate 
and our $\epe$ paper of 1993 was cited in this period mainly for our 
$B$ physics results. This changed in 1999 when NA48 and KTeV began to present 
their final results for $\epe$.

One month after our final paper of this period the Rome group presented in great details their calculation of two-loop 
anomalous dimensions concentrating on the HV scheme \cite{Ciuchini:1993vr}. This helped us to 
identify the error in our direct calculation of 
$[\hat\gamma_{es}^{(1)}]_{\rm HV}$. Their 
result for this matrix 
indeed agreed with ours obtained in our paper indirectly through 
$[\hat\gamma_{es}^{(1)}]_{\rm NDR}$.

The most important result of this Munich-Rome competition, that lasted for 
several years, is the agreement on the two-loop anomalous dimensions of the 
operators $Q_i\,(i=1,...10)$ that have been used by the flavour 
community since then. While they were calculated with the aim to 
find $\epe$ at the NLO 
level, they play also an essential role in all $\Delta F=1$ transitions, not 
only the non-leptonic ones. Thus they also enter the NLO calculations of 
$B\to X_s\gamma$, $B\to X_se^+e^-$ and even $K\to\pi\nu\bar\nu$ 
 decays. This is why in planning the grand expedition in 1988 it was 
essential to calculate these anomalous dimensions first.

In March 1993 I gave a seminar on our $\epe$ analysis in the theory group 
at CERN. I was  
approached there by two young physicists whom I did not meet before. It was 
Marco Ciuchini and Enrico Franco who came from Rome on a night train for a 
day to CERN to listen to my talk. I was really impressed that our competition 
went so far. However, it turned out that this friendly competition did not 
end in 1993.

\subsection{More Operators}\label{sec:3.5}

In 1997 an Italian group consisting of six climbers \cite{Ciuchini:1997bw}, including three from the 
expedition of early 1990s, calculated two-loop anomalous dimensions of a set 
of operators relevant at NLO for $\Delta \Gamma_{s,d}$ in 
$B^0_{s,d}-\bar B^0_{s,d}$ mixings and in particular for $\Delta F=1$ and
$\Delta F=2$ 
non-leptonic decays and transitions in the extensions of the SM. Good examples are  the MSSM at 
large $\tan\beta$, multi-Higgs models and generally models that include 
in addition to left-handed currents also right-handed currents and scalar 
currents. These operators are given in the operator basis of 
\cite{Buras:2000if} in (\ref{normal}) and (\ref{normal1}).

I was a bit surprised that we {had} not looked at these operators, except for 
the first one, in Munich before. Otherwise we would calculate their anomalous 
dimensions already in 1993. The whole machinery developed by us at the 
beginning of the 1990s could be used here. Knowing this, there was 
essentially no point in repeating this new Italian calculation. 
The results given 
in \cite{Ciuchini:1997bw} were bound to be correct. However, one result presented in this paper 
and even pointed out by the authors in the abstract made me interested 
in looking at this 
analysis closer.

The Italian group calculated the two-loop anomalous dimensions of the 
operators in (\ref{normal}) and (\ref{normal1}) in the so-called RI scheme that is  apparently useful for lattice 
calculations and in the NDR scheme. The calculation in the RI scheme was 
entirely new. The calculation for the operators $Q^{LR}_{1,2}$ in the NDR 
scheme was
really not new as the 
anomalous dimensions of these operators can be directly obtained from our 
earlier calculations of QCD penguin operators. What was new in the NDR scheme were the two-loop 
anomalous 
dimensions of the operators $Q_{1,2}^{SLL}$. Here the Italian group found a surprising 
result: the {analogues} of the $Q_+$ and $Q_-$ operators in (\ref{F4}) mixed in this 
sector 
under 
renormalization. Their two-loop anomalous dimension matrix was 
non-diagonal and the results for the NDR presented in the appendix of this 
paper 
looked very complicated. If this was indeed true, the NDR scheme, as defined 
by Peter and myself in \cite{Buras:1989xd}, would not be an elegant scheme.

Also Mikolaj Misiak, who joined MNLC in 1995, was interested in this 
result. Together with J\"org Urban, a PostDoc in my group, 
we decided to look at the Italian paper 
closer. I should have noticed it right away, but it was Mikolaj who reminded 
me of the ``Greek story" of missing evanescent operators in the case of 
$Q_1$ 
and $Q_2$ and the resulting mixing of $Q_+$ and $Q_-$ found by Peter and 
myself in 
1989. This time the Fierz vanishing evanescent operators were involved. I 
have discussed them already in section 3.1. We soon suspected that the Italian 
masters were performing Fierz transformations in a $D\not=4$ calculation  
without 
including the evanescent operators in question. We have all three 
independently 
performed the calculation of the anomalous dimensions of this sector now 
including the Fierz vanishing evanescent operators obtaining the result in
the NDR scheme
that was much simpler than the one of the Italian group, in particular 
the non-diagonal entries disappeared as they did in 1989 in the case of 
the current-current operators $(Q_+,Q_-)$.

Mikolaj and J\"org confirmed the RI calculation of \cite{Ciuchini:1997bw}
 in an arbitrary 
covariant 
gauge and found the matrices $\Delta\hat r$ relating the RI scheme and 
the NDR scheme. 
In this manner we could also find the compatibility of the RI result in 
\cite{Ciuchini:1997bw}
and 
our NDR result but the one loop matrices connecting these two schemes 
were clearly different from those given in \cite{Ciuchini:1997bw}.

 Unfortunately, our Italian colleagues did not agree with our interpretation 
of the strange form of their result, 
 but they admitted that the renormalization 
scheme they used was not the standard NDR scheme of \cite{Buras:1989xd}
 and that our result 
was more elegant and phenomenologically more useful. In summary: the first 
NLO climb in the RI scheme related to the operators 
(\ref{normal}) and (\ref{normal1})
 should be credited to 
the Italian group while in the NDR scheme to us.

The operators in (\ref{normal}) and (\ref{normal1}) do not constitute the full set of six-dimensional four 
quark operators contributing to $\Delta F=1$ processes. In addition to 
QCD penguins and electroweak penguins of the SM there are other penguin 
operators.
In our paper \cite{Buras:2000if} we have therefore generalized our  analysis of 
two-loop anomalous dimensions to the full set of $\Delta F=1$ four-quark
 operators. These 
results are much less known but should be useful in the extensions of the SM 
one day. Indeed they played recently a very important role in the calculations
of NLO QCD corrections to the WCs of non-leptonic operators both
in the WET and the SMEFT as described briefly in Section~\ref{sec:SMEFT}.

\subsection{QCD Corrections to Semi-leptonic B Decays}
This is probably a good place to summarize the QCD corrections to semi-leptonic 
$B$ decays that are necessary for an accurate determination of the CKM elements  $\vcb$ and $\vub$. As I did not take part in these calculations I consulted 
a very prominent member of the MNLC club, Paolo Gambino, who contributed in 
an important manner to these calculations. As we will soon see another member 
of our club, Andrzej Czarnecki, who will enter the scene later on, also made 
important contributions to this field.  In order to appreciate
the progress made in this field in the last decade,
it is useful to look at older nice summaries of this topic as far as 
$b\to c l\nu$ is concerned in \cite{Aquila:2005hq,Gambino:2011fz}. More
recent papers are given below.

In what follows I  will list papers where the purely perturbative corrections to semileptonic $B$ decays have been computed including also those to power corrections. 
\subsubsection{NLO Corrections to  Inclusive $B\to X_c \ell \nu$ Decays}
The pioneering first steps in this field can be found in \cite{Ali:1979is,Altarelli:1982kh}.
 The first complete analytic calculations, not only of the rate but  also of a few differential distributions, have been performed much later by Andrzej Czarnecki, 
 Marek Jezabek and Hans K\"uhn \cite{Jezabek:1988iv,Jezabek:1988ja,Czarnecki:1989bz,Czarnecki:1994pu}. In this context also the corrections 
to the rate in a compact form found by Yossi Nir should be mentioned \cite{Nir:1989rm}.

Next corrections to moments of hadronic spectra which used the results 
of previous calculations can be found in \cite{Falk:1995me},
while full triple differential distribution at $\ord(\alpha_s)$ that are
necessary for realistic experiments have been obtained in \cite{Trott:2004xc} and in 
particular in \cite{Aquila:2005hq}.

\subsubsection{NNLO Corrections to  Inclusive $B\to X_c l\nu$ Decays}\label{Gambino1} First BLM-NNLO corrections to the rate have been obtained in \cite{Luke:1994yc} and 
in particular in \cite{Ball:1995wa}. The BLM-$\ord(\alpha_s^2 \beta_0)$ corrections
to the lepton spectrum can be found in \cite{Gremm:1996gg} and to
 triple differential distributions in \cite{Aquila:2005hq}.

Next non-BLM two-loop corrections (analytic, zero cut) have been obtained 
in \cite{Pak:2008qt} and in numerical form but with realistic cuts in 
\cite{Melnikov:2008qs,Biswas:2009rb}. Finally complete two-loop corrections at specific kinematic 
points (zero recoil, which is important for the determination of 
$\vcb$ from $B\to D^{(*)} l \nu$) can be found in \cite{Czarnecki:1998kt,Czarnecki:1997hc,Czarnecki:1997cf,Czarnecki:1996gu}.

In the last decade and in particular recently the following advances have been
made in this field\footnote{I thank Paolo Gambino for  constructing the following two paragraphs.}.

Several results are now available at  NNNLO or $\ord(\alpha_s^3)$.  After the pioneering work in \cite{Archambault:2004zs} for the zero-recoil limit with $m_b=m_c$,
we now have the total rate \cite{Fael:2020tow} and some of the moments without cuts \cite{Fael:2022frj}, in an expansion around $m_b=m_c$, which converges
well  
at the physical point. {More recently, three-loop corrections to the muon and heavy quark decay rates have been computed in \cite{Czakon:2021ybq} confirming  part of the results of \cite{Fael:2020tow}.}
The calculation of the rate, together with the corresponding relations between on-shell and kinetic scheme parameters \cite{Fael:2020iea}, has allowed for an important  reduction of the uncertainty in the determination of $|V_{cb}|$ from inclusive semileptonic $B$ decays, now at the 1.2\% level \cite{Bordone:2021oof}.

At this level of precision, perturbative corrections to the Wilson coefficients of power suppressed operators become important. The complete 
$\ord(\alpha_s)$ corrections to the  $1/m_b^2$ corrections have been computed in \cite{Alberti:2012dn,Alberti:2013kxa}  
(total rate in \cite{Mannel:2015jka} as well), and also the dominant $\ord(\alpha_s/m_b^3)$ 
contributions to the total rate and $q^2$ moments are known \cite{Mannel:2019qel}. 
In the case of inclusive radiative decays the $\ord(\alpha_s/m_b^2)$ have been computed in \cite{Ewerth:2009yr}.

\subsubsection{NLO and NNLO Corrections to Inclusive 
$B\to X_u l \nu_l$ Decays}\label{Gambino2}
NNLO complete calculation of the width has been performed in \cite{vanRitbergen:1999gs}.
NLO  and BLM-NLO full triple differential distributions have been calculated 
in \cite{DeFazio:1999ptt} and \cite{Gambino:2006wk}, respectively. Leading 
shape functions and resummation in $B\to X_u l \nu_l$ has been done in 
\cite{Bigi:1993ex,Bosch:2004th,Neubert:1993ch,Andersen:2005mj} and non-leading shape functions in \cite{Leibovich:2002ys}.

Despite partial results {\cite{Czarnecki:2001cz,Asatrian:2008uk,Beneke:2008ei,Bell:2008ws,Bonciani:2008wf},}
the complete NNLO corrections to the partonic triple differential distribution are not yet known analytically, 
but numerical results for the moments were given in \cite{Brucherseifer:2013cu}.
The $\ord(\alpha_s/m_b^2)$ corrections have also been recently computed in \cite{Capdevila:2021vkf}.

\boldmath
\section{\boldmath{$\Delta S=2$}, \boldmath{$\Delta B=2$} and \boldmath{$\Delta B=0$} Transitions}\label{sec:4}
\setcounter{equation}{0}
\boldmath
\subsection{Effective Hamiltonians for $\Delta F=2$ Transitions}
\unboldmath
Let us begin this section by recalling the effective Hamiltonians for 
$\Delta S=2$ and $\Delta B=2$ transitions in the SM. 
We have first \cite{Buras:1983ap,Buras:1984pq,Buras:1990fn}
\begin{eqnarray}\label{hds2}
{\cal H}^{\Delta S=2}_{\rm eff}&=&\frac{G^2_{\rm F}}{16\pi^2}M^2_W
 \left[\lambda^2_c\eta_1 S_0(x_c)+\lambda^2_t \eta_2 S_0(x_t)+
 2\lambda_c\lambda_t \eta_3 S_0(x_c, x_t)\right] \times
\nonumber\\
& & \times \left[\as^{(3)}(\mu)\right]^{-2/9}\left[
  1 + \frac{\as^{(3)}(\mu)}{4\pi} J_3\right]  Q(\Delta S=2) + h. c.
\end{eqnarray}
where
$\lambda_i = V_{is}^* V_{id}^{}$. Here
$\mu<\mu_c=\ord(m_c)$.
In (\ref{hds2}),
the relevant operator
\begin{equation}\label{qsdsd}
Q(\Delta S=2)=(\bar sd)_{V-A}(\bar sd)_{V-A}\,,
\end{equation}
is multiplied by the corresponding coefficient function.
This function is decomposed into a
charm-, a top- and a mixed charm-top contribution with $S_0(x_i)$  and 
$S_0(x_i,x_j)$ being one-loop 
box functions in the SM.

Short-distance QCD effects are described through the correction
factors $\eta_1$, $\eta_2$, $\eta_3$ and the explicitly
$\alpha_s$-dependent terms in the last line of (\ref{hds2}). This 
factor allows to introduce 
the renormalization group 
invariant parameter $\hat B_K$ by 
\begin{equation}
\hat B_K = B_K(\mu) \left[ \alpha_s^{(3)}(\mu) \right]^{-2/9} \,
\left[ 1 + \frac{\alpha_s^{(3)}(\mu)}{4\pi} J_3 \right],
\label{eq:BKrenorm}
\end{equation}
\begin{equation}
\langle \bar K^0| (\bar s d)_{V-A} (\bar s d)_{V-A} |K^0\rangle
\equiv \frac{8}{3} B_K(\mu) F_K^2 m_K^2.
\label{eq:KbarK}
\end{equation}
Note that here the normalization  of external states is like in (\ref{eq:matrix}). The change from $2/3$ to $8/3$ is simply related to the fact that in
 (\ref{eq:matrix}) projections $P_L$ are used while here $V-A$ which leads to an additional factor of four.

The corresponding Hamiltonian for $B^0_{d,s}-\bar B^0_{d,s}$ 
mixing has a similar structure but it is simpler as only the top contribution 
matters. We have for  $B_q^0-\bar B_q^0$
mixing 
\begin{eqnarray}\label{hdb2}
{\cal H}^{\Delta B=2}_{\rm eff}&=&\frac{G^2_{\rm F}}{16\pi^2}M^2_W
 \left(V^\ast_{tb}V_{tq}\right)^2 \eta_{B}
 S_0(x_t)\times
\nonumber\\
& &\times \left[\alpha^{(5)}_s(\mu_b)\right]^{-6/23}\left[
  1 + \frac{\alpha^{(5)}_s(\mu_b)}{4\pi} J_5\right]  Q^q(\Delta B=2) + h. c.
\end{eqnarray}
Here $\mu_b=\ord(m_b)$,
\begin{equation}\label{qbdbd}
Q^q(\Delta B=2)=(\bar bq)_{V-A}(\bar bq)_{V-A}, \qquad q=d,s~.
\end{equation}

The renormalization group invariant parameters $\hat B_q$ are defined 
by
\begin{equation}
\hat B_{B_q} = B_{B_q}(\mu) \left[ \as^{(5)}(\mu) \right]^{-6/23} \,
\left[ 1 + \frac{\as^{(5)}(\mu)}{4\pi} J_5 \right]\,,
\label{eq:BBrenorm}
\end{equation}
\begin{equation}
\langle \bar B^0_q| (\bar b q)_{V-A} (\bar b q)_{V-A} |B^0_q\rangle
\equiv \frac{8}{3} B_{B_q}(\mu) F_{B_q}^2 m_{B_q}^2\,,
\label{eq:BbarB}
\end{equation}
where
$F_{B_q}$ is the $B_q$-meson decay constant.

Numerical values for the Bag parameter $B_{B_q}$ and the decay constant $F_{B_q}$ have to be obtained by non-perturbative methods. For the decay constants lattice determinations show by far the lowest uncertainties, an average of these is provided by the FLAG collaboration
\cite{FlavourLatticeAveragingGroupFLAG:2021npn}. In the case of the Bag parameters HQET sum rule calculations \cite{Kirk:2017juj,King:2019lal} obtain a similar precision as the most recent lattice evaluations \cite{FermilabLattice:2016ipl,Dowdall:2019bea}. Averages of both methods have been presented in \cite{DiLuzio:2019jyq}.

We are now ready to discuss the history of the NLO QCD calculations of 
\begin{equation}\label{etass}
\eta_1\equiv\eta_{cc}, \qquad \eta_2\equiv\eta_{tt},\qquad \eta_3\equiv\eta_{ct}, 
\qquad \eta_B.
\end{equation}

\subsection{The Top Quark Contributions}
In the fall of 1989 Matthias Jamin joined my group becoming the third member 
of the MNLC. Instead of continuing our NLO calculations for $\Delta F=1$
transitions we decided to calculate the NLO QCD corrections to top quark 
contributions to the 
effective Hamiltonians for $B_{d,s}^0-\bar B_{d,s}^0$ mixings and $K^0-\bar K^0$ mixing. 
We were not the first to do 
this climb but the first two attempts were unsuccessful. The calculations were 
plainly wrong with the results for the Wilson coefficients exhibiting infrared 
regulator dependence that was a consequence of an incorrect matching of the
full and effective theories. Moreover these calculations did not include
the two-loop anomalous dimension of the 
operator $Q(\Delta F=2)$. As I have a high respect for the 
leaders of these 
two expeditions, that otherwise had significant contributions to our field, I 
prefer  not to refer to these papers.

\begin{figure}[hbt]
\vspace{0.10in}
\centerline{
\epsfysize=1.5in
%\rotate[r]{
\epsffile{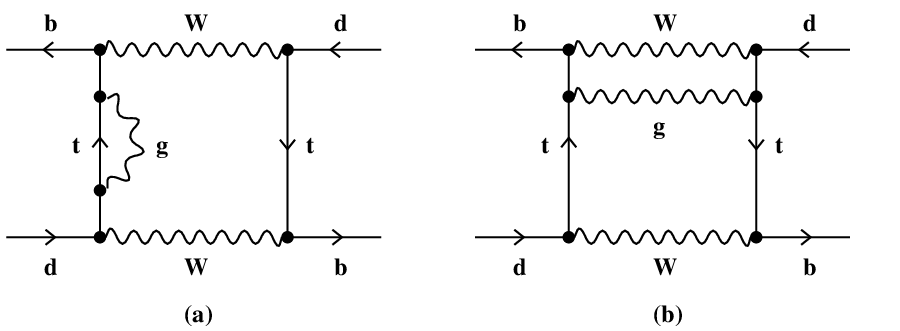}
}%}
\vspace{0.08in}
\caption[]{Examples of two-loop diagrams contributing to 
$B_d^0-\bar B_d^0$ mixing.
\label{L:8a}}
\end{figure}
Our new NLO project \cite{Buras:1990fn} involved two parts. One very easy, the other one 
rather difficult. The first one was the calculation of the two-loop anomalous 
dimension of the relevant operator $Q(\Delta F=2)$. It took no time as 
the anomalous 
dimension of $Q(\Delta F=2)$ is simply equal to $\gamma_+$, the anomalous 
dimension of $Q_+$ in (\ref{F4}) calculated at the 
two-loop level by Peter 
and myself half a year before. The second part was much more time consuming. 
It involved the 
calculation of $\ord(\alpha_s)$ QCD corrections to the box diagrams involving internal $W^\pm $
bosons, the Goldstone bosons $\phi^\pm$ and the top quark exchanges. 
This is a two-loop calculation in a full theory with massive heavy 
particles. Two examples  of contributing diagrams 
are shown in Fig.~\ref{L:8a}. The remaining diagrams can be found in 
Fig. 2 of \cite{Buras:1990fn}. There are 8 classes of diagrams in total 
shown in that figure from which the remaining diagrams can easily be 
obtained.

In order to extract the relevant Wilson coefficient of 
the operator $Q(\Delta F=2)$ from this calculation the proper 
matching to the effective theory has to be made. This requires the 
calculation of
 the 
$\ord(\alpha_s)$ corrections to the matrix 
element of this operator  between the external quark states. 
The two-loop calculation 
of the $\ord(\alpha_s)$ corrections to the box diagrams involves infrared 
divergences. We 
decided to set the external momenta to zero and regulate the infrared 
divergences by the masses of the external quarks. Working off-shell 
introduces necessarily gauge dependence (gluon propagator) in the final 
result. The final result of the box diagram calculation was therefore gauge 
and 
infrared regulator dependent. These dependences certainly did not belong to 
the Wilson coefficient of the operator $Q(\Delta F=2)$ but to its matrix 
element and were removed in 
the process of 
matching of the full theory to the effective theory with the latter exhibiting
 precisely the same 
gauge and infrared regulator dependences.

The main results of our paper were the values of the QCD factors $\eta_B$ 
and $\eta_2$ 
for $B_{d,s}^0-\bar B_{d,s}^0$ and $K^0-\bar K^0$  Hamiltonians, respectively:
\be\label{etas}
\eta_B=0.55\pm 0.01, \qquad \eta_2=0.577\pm 0.007~.
\ee
Our result for $\eta_B$ has been confirmed several years later in
\cite{Urban:1997gw}.

These QCD factors  have been used frequently in the 
literature in the last 30 years. While using these 
particular values one should remember that they should be used in 
conjunction with the renormalization group invariant  parameters $\hat B_K$ and
$\hat B_{s,d}$ in (\ref{eq:BKrenorm}) and (\ref{eq:BBrenorm}), respectively.
 These parameters have been introduced 
at the LO  already in 1983 by Wojtek Slominski, Herbert Steger,       
and myself \cite{Buras:1983ap}
and have been generalized to NLO in the present 
calculation. I do not remember who is the father of the $B_K$ parameter without 
the ``hat", but I think it is John Donoghue.

This takes care of the $\mu$ dependence at the lower end of the RG evolution.
$\eta_i$ resulting from the calculation of $C(\mu)$ and $\hat B_i$
representing the matrix element $\langle Q(\mu)\rangle$ are defined in such a manner  
that they 
separately are $\mu$-independent. However, $\eta_i$ depend on the scale 
$\mu_W=\ord(M_W)$ at 
which  the matching between the full and the effective theory is made. This 
dependence is canceled by the $\mu_W$ dependence of the initial conditions 
$C(\mu_W)$ and 
this cancellation is only meaningfull at  the NLO. More importantly the $\eta_i$
 are 
defined such that they multiply the leading order box functions and consequently 
they include the $\ord(\alpha_s)$ corrections to the box diagrams that Peter,
Matthias and I calculated. Consequently they depend also on the scale 
$\mu_t$ at which the 
running top quark mass $m_t(\mu_t)$ used in the calculation is defined. 
This $\mu_t$  
dependence of $\eta_i$ is important as it cancels up to higher order 
corrections the $\mu_t$ dependence of the Inami-Lim function 
$S_0(x_t(\mu_t))$ that remained 
uncompensated at LO. The values in (\ref{etas}) correspond to $\mu_t=m_t$. 
This turns out to 
be a convenient choice as with $\mu_t=m_t$, the QCD factors are practically 
independent of the actual measured value of $m_t(m_t)$. The 
leading logarithm
multiplying a large anomalous dimension of the mass operator vanishes at this 
scale. I mention this issue again because it is important.

To my knowledge the issue of the $\mu_t$ uncertainty in $\Delta F=2$
transitions in the LO 
and its reduction through the inclusion of NLO QCD corrections, was for the 
first time addressed in \cite{Buras:1990fn}. I have described it here 
because it enters all the calculations presented below and our paper 
completed in the Spring of 1990 can be considered as the prototype of 
analogous  two-loop calculations of the Wilson coefficients for any FCNC 
process that is 
sensitive to the top quark mass or any other heavy particle with colour, 
for instance squarks in the MSSM.

I should emphasize that our calculation of NLO corrections to box diagrams 
with top 
quark exchanges performed in 1989-1990 had much simpler structure than 
the LO calculation of renormalization group effects done by Fred Gilman and 
Mark Wise already in 1982 \cite{Gilman:1982ap}. The reason is that in 1983 the typical values of 
$m_t$ 
considered in the literature were substantially lower than $M_W$ and Fred and 
Mark in their 1983 LO calculation had to integrate out first $W^\pm$ and 
subsequently the top quark at much lower scales. In the range 
$m_t\le \mu\le M_W$ they had to 
deal with  QCD corrections to bilocal operators originating in the 
contraction of the $W^\pm$ propagator in a box diagram to a point but leaving 
the top 
quark propagator as it is. The renormalization and the calculation of 
QCD corrections to these 
bilocal structures are rather involved even in the LO. Fortunately in 1989 it 
was already known that $M_W \le m_t \le 200\gev$. 
Consequently $W^\pm$ and the top quark could be integrated out simultaneously 
generating a local operator from the beginning and the QCD renormalization of the bilocal 
structures at NLO, a non-trivial  task, could be avoided in our calculation. 

Finally, recently two-loop electroweak corrections to the top-quark contribution to $\varepsilon_K$ have been calculated in  \cite{Brod:2021qvc}.
 They turned
out to be below $1\%$ but this increased the confidence that most important
corrections have been taken into account.

\begin{table}[thb]
\caption{NLO and NNLO Calculations for $\Delta F=2$ and $\Delta F=0$ Transitions}
\label{TAB2}
\begin{center}
\begin{tabular}{|l|l|l|l|}
\hline
\bf \phantom{XXXXX} Decay &  {\bf NLO QCD} & {\bf NNLO QCD} & {\bf EW} \\
\hline
\hline
$\eta_1$                   & \cite{Herrlich:1993yv}&  \cite{Brod:2011ty}&\\
$\eta_2,~\eta_B$           & \cite{Buras:1990fn} & & {\cite{Gambino:1998rt,Brod:2021qvc}}\\
$\eta_3$                   & \cite{Herrlich:1995hh,Herrlich:1996vf}& 
\cite{Brod:2010mj}& \cite{Brod:2022har}\\
$\eta_{tt}$         & \cite{Brod:2019rzc}      & & \cite{Brod:2021qvc}  \\
$\eta_{ut}$         &   \cite{Brod:2019rzc}     & \cite{Brod:2019rzc} & \cite{Brod:2022har} \\
ADMs BSM & \cite{Ciuchini:1997bw,Buras:2000if}  & & \\
\hline
$\Delta\Gamma_{B_s}$      & 
\cite{Beneke:1998sy,Ciuchini:2003ww,Beneke:2003az,Lenz:2006hd,Asatrian:2017qaz,Asatrian:2020zxa,Gerlach:2021xtb,Gerlach:2022wgb}&\cite{Gerlach:2022hoj} & \\
$\Delta\Gamma_{B_d}$      & \cite{Ciuchini:2003ww,Beneke:2003az}&& \\
Lifetime Ratios &  \cite{Keum:1998fd,Ciuchini:2001vx,Beneke:2002rj,Franco:2002fc,Kirk:2017juj,King:2021jsq} &&\\
$\Delta F=2$ Tree-Level & \cite{Buras:2012fs} & &  \\
\hline
\end{tabular}
\end{center}
\end{table}

\subsection{Charm and Top-Charm Contributions}

In the case of box diagrams with two charm quark exchanges there is no way 
out. One has to face the bilocal structures at NLO because the simultaneous 
integration of the charm quark and $W^\pm$ would lead to $\ln M_W/m_c$ terms 
and consequently to the 
breakdown of perturbation theory. This rather difficult project was assigned 
in 1992 to my PhD student Stefan Herrlich. Stefan was a very good student and 
possibly he would succeed this climb by himself but fortunately for him and for
the project, he was joined early 1993 by Ulrich Nierste. Ulrich got 
his diploma in W\"urzburg working with the loop masters like Manfred B\"ohm 
and Ansgar Denner and consequently he was fit for this difficult climb almost 
immediately after his arrival in Munich. In fact Ulrich was soon leading this 
important climb and developed to one of the most prominent members of the 
MNLC.

The calculations of $\eta_1=\eta_{cc}$ and $\eta_3=\eta_{ct}$ QCD factors at NLO  were 
completed in 1993 \cite{Herrlich:1993yv} and 1995 \cite{Herrlich:1995hh}, respectively and the  full analysis of $\Delta S=2$ Hamiltonian at the NLO 
level could be performed soon after \cite{Herrlich:1996vf}. These were truly heroic climbs that were 
not repeated by anybody until 2010, when two younger members of the MNLC 
performed  the NNLO calculation 
of $\eta_3$: one of my many physics sons, Martin Gorbahn, a big star these days in multiloop calculations and my physics grandson, 
a PhD student of Uli Nierste, Joachim Brod \cite{Brod:2010mj}. Finally, one 
year later, Joachim and Martin calculated $\eta_1$ at NNLO \cite{Brod:2011ty}. 
This was the hardest of the calculations of $\eta_i$ but the result turned out
to be rather disappointing as the NNLO calculation did not reduce the uncertainties in the charm part. A solution to this problem had to wait for several years. I will report on it later on.

Even Guido Martinelli was impressed by the Herrlich-Nierste calculations and 
he told me this at least three times at different occasions. The values for 
$\eta_1$ and $\eta_3$ enter the analysis of the CP-violating parameter
$\varepsilon_K$  and are relevant ingredients of any analysis of the 
unitarity triangle, in particular after the lattice value for $\hat B_K$ 
   and the estimate of the long distance effects in $\varepsilon_K$ improved \cite{Buras:2008nn,Buras:2010pza}.
 I was through all these years convinced, knowing Stefan and Ulrich, that the NLO values for $\eta_1$ and $\eta_3$ found by them were correct. Indeed 
Brod and Gorbahn in their  NNLO calculations of $\eta_1$ and $\eta_3$ 
had to repeat Herrlich-Nierste NLO calculations, confirming their results 
and my expectations. In any case it is interesting to observe that 
 the $\eta_i$ factors at NLO and NNLO remained in the possession of our 
physics family. 

In summary the NNLO values of $\eta_1$ and $\eta_3$ read  in 2011 as follows \cite{Brod:2010mj,Brod:2011ty}
\be
\eta_1=1.87\pm0.76, \qquad \eta_3=0.496\pm0.47\,,
\ee
and $\eta_B$ and $\eta_2$ are given in (\ref{etas}). The large error in $\eta_1$
was  clearly disturbing.

All phenomenology papers 
on $\Delta S=2$ processes were using these numbers until in 2019  Brod, Gorbahn and Stamou
\cite{Brod:2019rzc} have presented a more accurate formula for $\varepsilon_K$. It uses
the unitarity relation $\lambda_c=-\lambda_u-\lambda_t$ instead of
$\lambda_u=-\lambda_c-\lambda_t$ as done in the previous literature. This allows
to remove significant theoretical uncertainties from charm contribution to $\varepsilon_K$. Simply various uncertainties present in different contributions in the previous formulation cancel each other in the case of $\varepsilon_K$ in the new formulation.

The effective Hamiltonian of (\ref{hds2}) is now replaced by \cite{Brod:2019rzc}
\begin{eqnarray}\label{hds2new}
{\cal H}^{\Delta S=2}_{\rm eff}&=&\frac{G^2_{\rm F}}{16\pi^2}M^2_W
 \left[\lambda^2_u\eta_{uu} S_{uu}(x_c)+\lambda^2_t \eta_{tt} S_{tt}(x_t,x_c)+
 2\lambda_u\lambda_t \eta_{ut} S_{ut}(x_c, x_t)\right] \times
\nonumber\\
& & \times \left[\as^{(3)}(\mu)\right]^{-2/9}\left[
  1 + \frac{\as^{(3)}(\mu)}{4\pi} J_3\right]  Q(\Delta S=2) + h. c.
\end{eqnarray}
where
$\lambda_i = V_{is}^* V_{id}^{}$. Here
\be
S_{uu}(x_c)=S_0(x_c),\quad  S_{tt}(x_t,x_c)=S_0(x_t)+S_0(x_c)-2 S_0(x_c,x_t),\quad  S_{ut}(x_c, x_t)=S_0(x_c)-S_0(x_c,x_t),
\ee where
$S_0(x_i)$ and $S_0(x_i,x_j)$ are the standard Inami-Lim functions \cite{Inami:1980fz,Buras:1983ap} that also enter the expression in (\ref{hds2})
used until 2019.

The QCD factors in (\ref{hds2new})  read \cite{Brod:2019rzc} 
\be
\eta_{uu}=\eta_1=1.87\pm0.76,\qquad \eta_{tt}=0.55(2),\qquad \eta_{ut}=0.402(5).
\ee
$\eta_{uu}$ and $\eta_{ut}$ are calculated at NNLO while $\eta_{tt}$ is known
only at the NLO level.

The first term in (\ref{hds2new}) is real and does not contribute to $\varepsilon_K$ but it affects $\Delta M_K$ so that the large uncertainty there remains. On the other {hand} the error in $\eta_{ut}$ is reduced by an order of magnitude relative to the one
in $\eta_3$ so that there is 
an impressive reduction of theoretical uncertainties in $\varepsilon_K$  relative to the  previous formulation.
Most of the literature these days uses this new strategy  which promotes
$\varepsilon_K$ to a quantity with small theoretical errors.

The new SM expression for  $\varepsilon_K$ reads
\cite{Brod:2019rzc}
\begin{equation}\label{BGS}
    |\epsilon_K|
=  \kappa_\epsilon C_\varepsilon \hat{B}_K
|V_{cb}|^2 \lambda^2 \bar \eta 
 \times \Big[|V_{cb}|^2(1-\bar\rho)
\eta_{tt} S_{tt}(x_t,x_c) - \eta_{ut} S_{ut}(x_c, x_t) \Big]\,.
\end{equation}
It replaces the usual phenomenological expression given in \cite{Buras:2008nn}. 

Finally, as already mentioned, two-loop electroweak corrections to charm-top contribution have been calculated in \cite{Brod:2022har}. They turned
out to be below $1\%$ but this increased the confidence that most important
corrections have been taken into account. 

\boldmath
\subsection{$\Delta\Gamma_s$, $\Delta\Gamma_d$, and CP violation in Mixing at NLO and Beyond}\label{sec:4.4}
\unboldmath
The $B^0_q$--$\bar B^0_q$ mixing system (with $q=d$ or $s$)
involves two hermitian $2\times2$ matrices, the mass matrix $M_{12}^q$
and the decay matrix $\Gamma_{12}^q$. The former quantity determines $\Delta M_q$
and is calculated from ${\cal H}^{\Delta B=2}$ in
(\ref{hdb2}). $\Gamma_{12}^q$ instead involves two insertions of ${\cal H}^{\Delta B=1}$
and determines the width difference $\Delta\Gamma_q$
between the two  $B^0_q$ eigenstates and the CP asymmetry in
flavour-specific decays, $a_{\rm fs}^q$, characterising CP violation in
mixing. It is usually measured in semileptonic decays. 

In {that} winter 1997/1998 I have been asked by Gerhard Buchalla, whether I 
would be interested in joining him, Martin Beneke, Christoph Greub, Alexander 
Lenz and Uli Nierste in the calculation of NLO QCD corrections to the life-time 
 difference or equivalently $\Delta\Gamma_s$ in the $B_s^0-\bar B_s^0$ system. 
At first I found it an interesting idea. Afterall $\Delta\Gamma_s$ is much 
larger than $\Delta\Gamma_d$ and working with my physics sons Alex and Uli, my 
physics stepson Martin and the Swiss master Christoph 
for the first time would be 
a real fun. In 1984 I studied $\Delta\Gamma_d$ at LO with Slominski and 
Steger \cite{Buras:1984pq} and in addition after nine years of NLO climbing I was 
well prepared for this new expedition. Yet, this winter I was busy with 
writing up my Les Houches lectures and other projects and I did not join 
Gerhard et al.

The paper \cite{Beneke:1998sy} as well as improvements of the operator basis \cite{Lenz:2006hd}
and numerous phenomenological analyses of Lenz and
Nierste (see e.g.\ \cite{Lenz:2011ti,Esen:2019jjy}) resulted from this project.
Equally important, this topic was the subject of the PhD thesis of Alexander 
Lenz.

Interestingly, also the younger generation of Rome masters got involved 
in these efforts \cite{Ciuchini:2003ww}, in particular Cecilia 
Tarantino, who became in 2006 my close one-loop collaborator within the 
Littlest Higgs Model with T-parity. I hope one day somebody will report 
on this competition, although as most of the authors are at least twenty 
years younger than me, it will still take some time. The results of these 
papers played already an important role in the analyses of the 
Tevatron data and presently are also very important for the LHCb.

In  \cite{Beneke:1998sy} the $(Q_{1,2},Q_{1,2})$ contributions to $\Gamma_{12}^q$ with two insertions of the current-current
operators  were presented at NLO, i.e.\ two-loop level,
together with the one-loop NLO contributions for the  $(Q_{1,2},Q_{8G})$
insertion, both only for the CKM-favoured term with two 
charm quarks describing $b\to c\bar c q$ decays. 
It was sufficient to include the terms with  penguin
operators $Q_{3,\ldots 6}$ at LO, because their coefficients are much
smaller than $C_{1,2}$. For the prediction of   $a_{\rm fs}^q$ one also
needs the $(u,c)$, $(c,u)$, and $(u,u)$ diagrams and the analogous NLO
calculation has been obtained at the same time by the two groups \cite{Ciuchini:2003ww,Beneke:2003az}. In \cite{Ciuchini:2003ww}, also the calculation of  \cite{Beneke:1998sy} has been confirmed.

Next steps addressed the calculation of NNLO corrections to the
$(Q_{1,2},Q_{1,2})$ and $(Q_{1,2},Q_{8G})$ terms in $\Gamma_{12}^q$ as
well as NLO corrections to the contributions with one or two
$Q_{3,\ldots 6}$ insertions. The corresponding contributions in the
large-$N_f$ limit were presented in \cite{Asatrian:2017qaz,Asatrian:2020zxa} and
the full three-loop NNLO result for $(Q_{1,2},Q_{1,2})$ was recently
completed in \cite{Gerlach:2022hoj}.
All two-loop contributions involving one or two
insertions of  $Q_{3,\ldots 6}$ and $Q_{8G}$ were calculated in
\cite{Gerlach:2021xtb,Gerlach:2022wgb}, which not only completed the NLO calculation in the penguin
sector but also  the NNLO calculation of all contributions with
$Q_{8G}$. The two-loop results for the $(Q_{8G},Q_{8G})$ insertion even goes
beyond NNLO. All results in \cite{Asatrian:2017qaz,Gerlach:2021xtb,Gerlach:2022wgb,Gerlach:2022hoj} are obtained in an expansion in $m_c/m_b$ up to second order. While the NNLO result for $\Gamma_{12}^q$ removes an important 
source of the theoretical uncertainty, the latter is still not competitive
with the current experimental error of $\Delta \Gamma_s$. The remedy is
expected from a calculation of NLO corrections to the $1/m_b$-suppressed term
in $\Gamma_{12}^s$.

\boldmath
\subsection{NLO QCD Corrections to Tree-Level $\Delta F=2$ Processes}\label{sec:4.5}
\unboldmath
NLO QCD corrections to box diagrams are rather involved and it appears 
a bit premature to calculate them for the extensions of the SM. Therefore, 
between 2006 and 2011, when I exclusively studied extensions of the SM, 
I decoupled from calculations of NLO QCD corrections. However, in the fall 
of 2011, I noticed that for $\Delta F=2$ processes mediated at tree-level 
by a colourless neutral gauge boson and a colourless neutral scalar exchanges the matching conditions at 
NLO require only one-loop calculations and can be done model independently 
without too much effort. 
Somehow QCD experts including myself, did not notice it before. 

In April 2011, Jennifer Girrbach, PhD student of Ulrich Nierste and consequently 
my granddaughter in physics, joined my group. Jennifer had no experience in 
QCD calculations but I thought it would be fun to perform this climb with 
her and to teach her this field. Moreover, she was one of the stars of 
Ulrich's group and I was sure that we would reach this summit together. 
Indeed Jennifer learned the QCD technology  in a short time and performed 
all necessary calculations independently of me. We published our results 
on $\Delta F=2$ transitions already in January 2012 \cite{Buras:2012fs}. Since 
then we could use these results for concrete models with $Z'$ 
tree-level exchanges. We could subsequently extend this calculation to 
non-leptonic $\Delta F=1$ processes mediated by colourless 
neutral gauge boson and colourless neutral scalar exchanges \cite{Buras:2012gm}.

By now QCD corrections to the WCs resulting from the matching of other 
NP models entering FCNCs already at tree-level to the effective low energy theory
have been computed. Here we mention the work by Aebischer, Crivellin and Greub
in which such corrections have been calculated for scalar and vector leptoquarks
\cite{Aebischer:2018acj}. Note that this time the exchanged particle carries
colour. 

\subsection{Lifetime Ratios at NLO}
$b$-flavoured hadron lifetimes can be expanded in powers of $1/m_b$. In
each order new perturbatively calculable coefficients appear; they
multiply effective operators whose dimensions increase with increasing
powers of $1/m_b$. Similarly to the case  of $\Gamma_{12}^q$ the
calculation involves two insertions of ${\cal H}^{\Delta B=1}$, but
contrary to the cases discussed in this section  one finds $\Delta B=0$ 
operators on the effective side of the matching calculation. The
dominant contributions to the lifetime splittings in the SU(3)$_{\rm F}$
(anti-)triplets $(B^+,B_d^0,B_s^0)$ and $(\Xi^-,\Xi^0,\Lambda_b^0 )$ involve
operators with flavour structure $\bar b b\, \bar q q$, where $q=u,d$ or
$s$, whose
coefficients depend on $q$.  In the hadronic matrix element the light
quark $q$ is contracted with the valence quark of the decaying hadron,
leading to different total decay rates for different hadrons. The
desired contributions are spectator effects. 

The dominant spectator contributions to $B_d^0$  and $B_s^0$ decays
involve $Q_{1,2}$ and are $b \bar d \to c \bar u$ and  $b \bar s \to c
\bar c$, respectively. Since the coefficients are similar in size, the
lifetimes of $B_d^0$  and $B_s^0$ are equal up to ${\cal O}(1\%)$.
Penguins contribute to $\tau(B_s^0)$, but not to   $\tau(B_d^0)$, and turn out to
be equally important numerically for  $\tau(B_s^0)/\tau(B_d^0)$. The
penguin contribution include one-loop
$(Q_2, Q_{8G})$ and two-loop double-penguin
$(Q_2, Q_2)$  diagrams, which were  the first calculated NLO contributions
to a heavy lifetime ratio \cite{Keum:1998fd}. 

The NLO calculation of the lifetime ratio $\tau(B^+)/\tau(B_d^0)$ was
performed in \cite{Ciuchini:2001vx} for the case $m_c=0$ and with the full charm mass
dependence in \cite{Beneke:2002rj,Franco:2002fc}. The same coefficients also
enter  $\tau(\Xi_b^-)/\tau(\Xi_b^0)$ \cite{Beneke:2002rj}.  In \cite{Franco:2002fc}
also $\tau(B_s^0)/\tau(B_d^0)$ is presented at NLO and a partial NLO
result for $\tau(\Lambda_b)/\tau(B_d^0)$ is given. The $\Delta B=0$
matrix elements  are difficult to compute on the lattice, so that QCD
sum rules are the method of choice, with Alex Lenz being the leading
player  \cite{Kirk:2017juj}.

Recently the Wilson coefficient of the Darwin operator arising at order $1/m_b^3$ was determined for the first time for non-leptonic decays \cite{Lenz:2020oce,Mannel:2020fts}. This contribution turned out to be large and it dominates the theory prediction of $ \tau(B_s) / \tau(B_d)$, see \cite{Lenz:2022rbq}.

\section{Rare K and B Decays}\label{sec:5}
\setcounter{equation}{0}
\boldmath
\subsection{Effective Hamiltonians for $\kpn$ and $\klpn$}
\unboldmath
These decays have been with us already for more than forty years and 
the world of particle physics is waiting for their precise measurements. 
The first detailed review was published in 2008 \cite{Buras:2004uu}.
 More recent developments are presented below and in my book \cite{Buras:2020xsm}.

The effective Hamiltonian for $\kpn$  can
be written as
\begin{equation}\label{hkpn} 
{\cal H}_{\rm eff}={G_{\rm F} \over{\sqrt 2}}{\alpha\over 2\pi 
\sin^2\theta_{\rm W}}
 \sum_{l=e,\mu,\tau}\left( V^{\ast}_{cs}V_{cd} X^l(x_c)+
V^{\ast}_{ts}V_{td} X(x_t)\right)
 (\bar sd)_{V-A}(\bar\nu_l\nu_l)_{V-A} \, .
\end{equation}
The index $l$=$e$, $\mu$, $\tau$ denotes the lepton flavour.
The dependence on the charged lepton mass resulting from the box-graph
is negligible for the top contribution. In the charm sector this is the
case only for the electron and the muon but not for the $\tau$-lepton.

The function $X(x_t)$ relevant for the top part is given by
\begin{equation}\label{xx9} 
X(x_t)=X_0(x_t)+\aspi X_1(x_t) +\left(\aspi\right)^2 X_2(x_t)\,,
\end{equation}
with the leading contribution $X_0(x)$ resulting from $Z$ penguin diagrams and 
box-diagrams 
and  $X_{1,2}(x_t)$ denoting QCD corrections to these diagrams that will be 
described below.

In the case of charm contributions it is  useful to define the parameter 
\be \label{eq:defPcX}
P_c(X) = \frac{1}{\lambda^4} \left ( \frac{2}{3} X^e (x_c) + \frac{1}{3} X^\tau
(x_c) \right ) \, , 
\eeq 
with $\lambda = | V_{us} |$ being the Wolfenstein parameter 
($\lambda\approx 0.225$).

Keeping terms to first order in $\as$, the perturbative
expansion of $P_c(X)$ has the following general structure 
\be\label{eq:PcXPT} 
P_c(X) = \frac{4 \pi}{\as (\mu_c)} P_c^{(0)}(X) + P_c^{(1)}(X) +
\frac{\as (\mu_c)}{4 \pi} P_c^{(2)}(X) \, . 
\ee

In the case of the decay $\klpn$ only the top function $X(x_t)$ matters. 
Similarly in the case of $B\to X_s\nu\bar\nu$ only this function matters.
\subsection{Effective Hamiltonians for $K_L\to\mu^+\mu^-$ and 
$B_{s,d}\to\mu^+\mu^-$}
In the case of $K_L\to\mu^+\mu^-$ only the subleading short distance (SD) part 
can be computed. 
The analysis of this part proceeds in essentially the same
manner as for $\kpn$. The only difference is introduced through the
reversed lepton line in the box contribution. In particular there is
no lepton mass dependence, since only massless neutrinos appear as
virtual leptons in the box diagram.

The effective Hamiltonian  can be written as
follows:
\begin{equation}\label{hklm}{\cal H}_{eff}(K_L\to\mu^+\mu^-)=
 -\frac{G_{\rm F}}{\sqrt 2} 
\frac{\alpha}{2\pi \sin^2 \theta_{\rm W}}
 \left( V^{\ast}_{cs}V_{cd} Y(x_c)+
V^{\ast}_{ts}V_{td} Y(x_t)\right)
 (\bar sd)_{V-A}(\bar\mu\mu)_{V-A} + h.c. 
\end{equation}
The function $Y(x)$ is given by
\begin{equation}\label{yyx}
Y(x_t) = Y_0(x_t) + \aspi Y_1(x_t)+ \left(\aspi\right)^2 Y_2(x_t)\,,
\end{equation}
and
\be
P_c(Y)=\frac{Y(x_c)}{\lambda^4}
\ee
has an expansion similar to $P_c(X)$ in (\ref{eq:PcXPT}).

Only the function $Y(x_t)$ is relevant for $B_{s,d}\to\mu^+\mu^-$ for 
which the effective Hamiltonian reads
\begin{equation}\label{hyll}
{\cal H}_{\rm eff}(B_{s}\to\mu^+\mu^-) = -\frac{G_{\rm F}}{\sqrt 2} 
\frac{\alpha}{2\pi \sin^2 \theta_{\rm W}} V^\ast_{tb} V_{ts}
Y(x_t) (\bar bs)_{V-A} (\mu^+\mu^-)_{V-A} + h.c.   
\end{equation}
with $s$ replaced by $d$ in the
case of $B_d\to \mu^+\mu^-$.

In Fig.~\ref{L:17}   we give  only the 
diagrams contributing to $Y_0(x_t)$ to emphasize that only the box diagrams 
have to be calculated relative to the $\nu\bar\nu$ case as the direction of 
the internal lepton line differs in this case (compare with Fig.~\ref{L:7}).

\begin{figure}[hbt]
\vspace{0.10in}
\centerline{
\epsfysize=1.5in
%\rotate[r]{
\epsffile{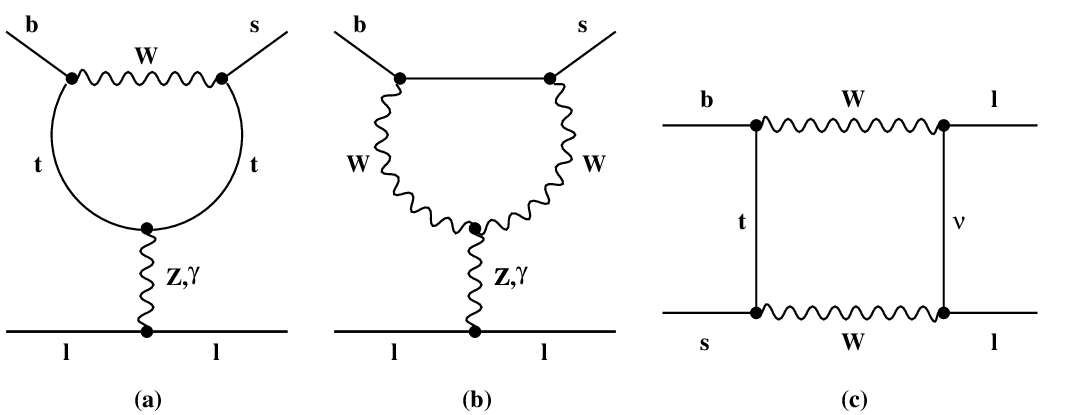}
}%}
\vspace{0.08in}
\caption[]{One loop diagrams contributing to rare decays
with charged leptons in the final state.
\label{L:17}}
\end{figure}

We are now ready to discuss the calculations of NLO and NNLO QCD corrections to these decays in which I took  part. Subsequently we will discuss briefly 
NLO electroweak corrections calculated in the last decade by my former PhD students.

\subsection{NLO and NNLO  QCD Calculations}
The story of NLO QCD corrections to rare $K$ and $B$ decays begins in the fall of 
1990 when I started the calculation of $\ord(\alpha_s)$ corrections to the 
flavour changing $Z^0$-penguin one-loop diagrams that dominate semi-leptonic 
rare decays like $K\to\pi\nu\bar\nu$, $B_{s,d}\to\mu^+\mu^-$ and $B\to
X_{s,d}\nu\bar\nu$. The calculation involves 30 two-loop diagrams with three 
 examples shown in Fig.~\ref{L:7}. Most of 
them are free from infrared divergencies so that the external masses and 
momenta can be neglected. The infrared divergent diagrams can be regulated by
non-vanishing external masses but a more elegant method is the dimensional 
regularization.

I have started this climb by myself but after roughly twenty two-loop diagrams 
I stopped. I found this solo climb doable but I was already involved in the 
calculation of two-loop anomalous dimensions of penguin operators described
previously  and 
moreover it is always safer to have a partner in the climbs of that sort.
Fortunately, in contrast to the ordinary climbing, in situations like that 
there is no need to return to the base camp. With good notes the climb can be 
continued whenever one decides to do it.

\begin{figure}[hbt]
\vspace{0.10in}
\centerline{
\epsfysize=3in
%\rotate[r]{
\epsffile{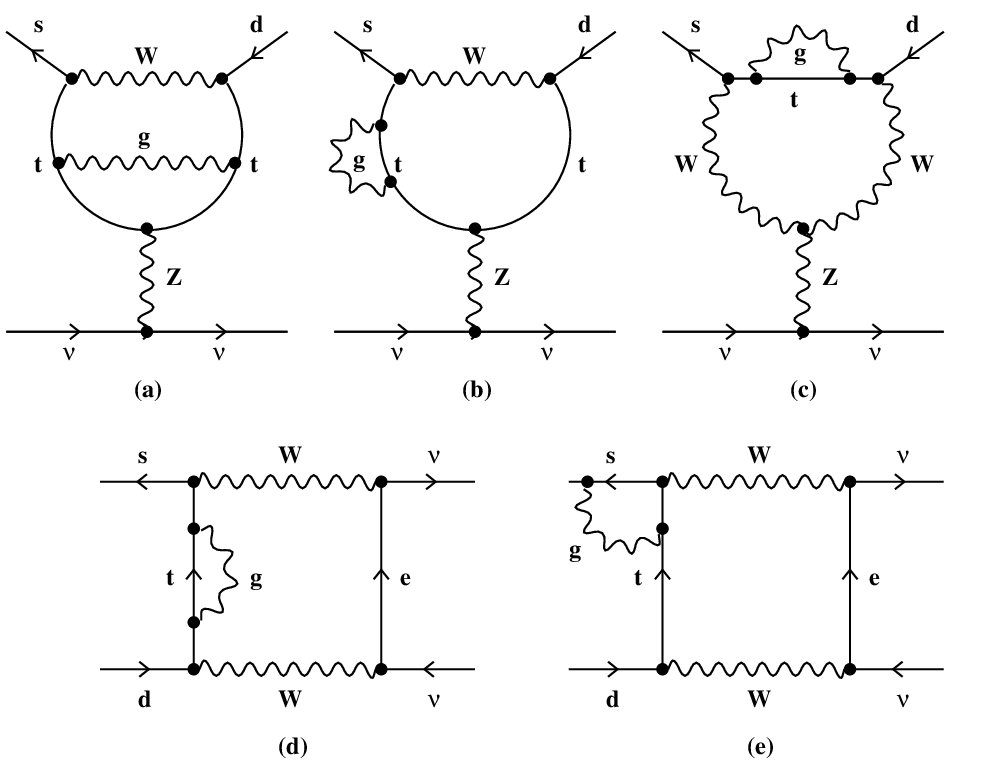}
}%}
\vspace{0.08in}
\caption[]{Examples of two-loop diagrams contributing to $X_1(x_t)$.
\label{L:7}}
\end{figure}
 
In July 1991 Gerhard Buchalla returned from the military service, fit to 
begin his PhD studies and to
join me in the $Z^0$-penguin climb. However, as a warmup I suggested to 
him a one-loop calculation: $\ord(\alpha_s)$ corrections to non-leptonic 
$c$-quark decay at $\mu=\ord(m_c)$ 
in the NDR and HV schemes that in 1981 was done by Altarelli et al. in the 
DRED scheme. Gerhard found the compatibility of his results with those of 
 \cite{Altarelli:1980te,Altarelli:1980fi}, 
published them \cite{Buchalla:1992gc} \footnote{Such calculations have been 
refined in the context of $b\to ccs$ \cite{Bagan:1994zd,Bagan:1995yf} with the participation of my 
assistant, Patricia Ball.}, and started from the base camp to climb the $Z^0$-penguin NLO 
QCD summit in the early summer of 1992. This clearly motivated me to continue my 
climb of 1990. Gerhard was one of my best PhD students ever and even if we were climbing separately, I knew that after reaching 
the summit, I would meet him there to compare my results with his.

As a CERN fellow from 1975 to 1977 I had the opportunity to talk to one of the 
old masters of higher order calculations of $g-2$, A. Peterman. He told me that 
in order to be sure that a result of a lengthy multi-loop calculation is 
correct, the climbing partners, should have as little contact with each other
as possible, comparing their results only at the end of the climb.

While in most calculations, I have done in the 1990s, I followed Peterman's 
advice as much as possible, the calculation of $\ord(\alpha_s)$ corrections to
$Z^0$-penguin diagrams with Gerhard could be considered as a perfect example 
of such an approach. Our calculations were totally independent. I have no 
idea how he got the final result and the same applies to him with respect to 
my calculation. 
In the fall of 1992 we compared our results diagram by diagram reaching full 
agreement on all 30 diagrams except for one term in one diagram, that turned 
out to be a misprint in my notes. We were rather confident that our result was 
correct.

As a byproduct we could extract from our $\ord(\alpha_s)$ corrections to 
the $Z\bar b s$ vertex the $\ord(\alpha_s)$ corrections 
to the flavour 
conserving $Z^0b\bar b$ penguin diagram in the large $m_t$ limit, being in 
fact the first 
group that confirmed the results of \cite{Fleischer:1992fq}
 done in the context of electroweak 
precision studies three months earlier. In the first years after the 
appearance of our paper \cite{Buchalla:1992zm}, it got most citations precisely for this additional 
calculation. After the discovery of $\kpn$ in 1997 things of course changed.

In the following paper \cite{Buchalla:1993bv}, after calculating $\ord(\alpha_s)$ QCD corrections to
$\Delta F=1$ box diagrams, we could finally present $\ord(\alpha_s)$ 
corrections to all rare $K$ and $B$ decays 
dominated by internal top quark exchanges: 
$K_L\to\pi^0\nu\bar\nu$, $B_{s,d}\to\mu^+\mu^-$ and $B\to X_{s,d}\nu\bar\nu$.
In the case of $K^+\to\pi^+\nu\bar\nu$ and $K_L\to\mu^+\mu^-$ we had still to
calculate NLO QCD corrections to the internal charm 
contributions. This was the subject of our third and final paper of this 
period \cite{Buchalla:1993wq} which required the inclusion 
of QCD corrections to the
bilocal structures as the ones given in Fig.~\ref{L:11}.

\begin{figure}[hbt]
\vspace{0.10in}
\centerline{
\epsfysize=2in
%\rotate[r]{
\epsffile{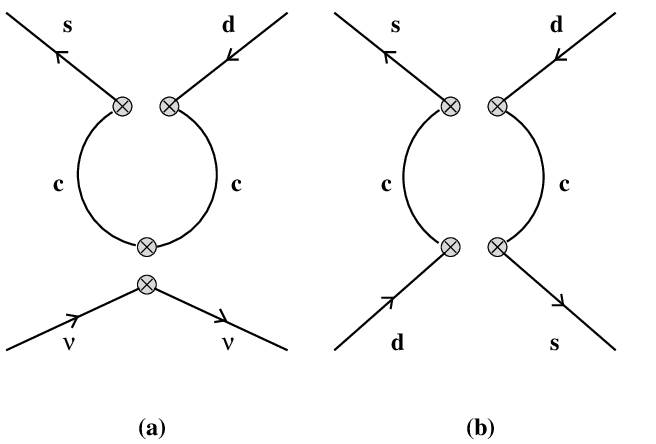}
}%}
\vspace{0.08in}
\caption[]{Bilocal Structures. 
\label{L:11}}
\end{figure}

Our calculations of two first papers reduced the $\ord(15\%)$ 
uncertainty in the branching ratios for
$K_L\to\pi^0\nu\bar\nu$, $B_{s,d}\to\mu^+\mu^-$ and $B\to X_{s,d}\nu\bar\nu$
 due to the choice of 
$\mu_t$ in $m_t(\mu_t)$ 
present in the LO calculations in \cite{Dib:1989cc} down to $\pm 1\%$. 
Amusingly the most important phenomenological contribution of the second 
paper is the realization that  most, if not all, papers in the literature missed an overall factor of 2 in the branching ratio for $B_s\to\mu^+\mu^-$. When 
I mentioned during an experimental discussion at Beauty 1995 in Oxford that 
the simulations for the LHCb for this decay should use a branching ratio by 
a factor of two higher than they used in their presentations, there was a real joy among my experimental colleagues and my comment was declared to be the high-light of the day. 
Afterall a branching ratio of $3.6\times 10^{-9}$ is easier to measure than $1.8\times 10^{-9}$. In fact the most recent experimental result for $B_s\to\mu^+\mu^-$ branching ratio from the LHCb, CMS and ATLAS reads {$(3.45\pm 0.29)\cdot 10^{-9}$ \cite{LHCb:2021awg,CMS:2020rox,ATLAS:2020acx,HFLAV:2022pwe}.} This has to be compared
with the most recent SM prediction $(3.78^{+ 0.15}_{-0.10})\cdot 10^{-9}$
\cite{Buras:2022wpw} so that { the room left for  NP became
  much smaller than one year ago. This is mainly due to
  the most recent result from CMS,
 $(3.83\pm 0.42)\cdot 10^{-9}$, taken into account in the HFLAV average quoted above. It  agrees
very well with the latter SM result, but the large experimental  error does not allow for definite conclusions and NP could still be at work here.}

Concerning QCD calculations the most important achievement of 
these three papers is the the calculation of $P_c^{(1)}(X)$ in (\ref{eq:PcXPT}).
This allowed the reduction of the uncertainty due to 
the choice of $\mu_c$ in $m_c(\mu_c)$ in the $P_c(X)$ from $\pm 26\%$ in 
the LO down to $\pm 10\%$. 
Ten years later this calculation has been extended to the 
NNLO level, the last term in (\ref{eq:PcXPT}), by Martin Gorbahn, still another
 multi-loop 
star among my PhD students, Ulrich Nierste, Ulrich Haisch and myself
 \cite{Buras:2006gb}, reducing the uncertainty in question down to $\pm 2\%$.
This should be a very relevant improvement when data on $\kpn$ will become 
accurate. As in  most extensions of the SM, the charm contribution to 
$\kpn$ 
remains essentially unaffected by new physics, these results are also relevant 
for most extensions of the SM. 

In the case of $K_L\to \mu^+\mu^-$ our NLO calculation reduced the $\mu_c$ 
uncertainty in $P_c(Y)$ from $\pm 44\%$ present in the LO down to $\pm 22\%$. 
Ten years later this calculation has been extended to the 
NNLO level by Martin Gorbahn and Ulrich Haisch \cite{Gorbahn:2006bm} 
reducing this uncertainty 
down to $\pm 7\%$. As the charm contribution is much less relevant in this 
decay than in $\kpn$ and $K_L\to \mu^+\mu^-$ is subject to non-perturbative 
uncertainties, this left over uncertainty is practically negligible. Finally, 
our calculations of NLO QCD corrections to $Y(x_t)$, the term involving $Y_1$ 
brought the 
$\mu_t$ uncertainty down to $1\%$ both in $K_L\to \mu^+\mu^-$ and $B_{s,d}\to\mu^+\mu^-$. 
The calculation of the NNLO QCD correction $X_2$ is presently in progress, while $Y_2$ has 
been calculated already in 2013 eliminating practically QCD uncertainties 
in $Y(x_t)$ \cite{Hermann:2013kca}.

In September 1998, almost five years after the completion of our $Z^0$-trilogy,
that also included QCD corrections to the relevant $\Delta F=1$ 
box diagrams, I 
spent one month in the CERN theory group developing a general 
parametrization for $B\to\pi K$ decays in collaboration with Robert Fleischer
and rather 
decoupled from $K\to\pi\nu\bar\nu$ decays. Also Gerhard Buchalla was there. 
One day we 
received an e-mail from Mikolaj Misiak who in collaboration with J\"org Urban 
calculated $\ord(\alpha_s)$ corrections to the top contributions to rare 
decays \cite{Misiak:1999yg}, the subject of 
our first two papers. Their result for $Z^0$-penguin contributions obtained 
using the external masses of 
quarks as infrared regulator agreed with our result but the one for box
diagrams disagreed with our calculation. The difference was 
phenomenologically irrelevant $(1\%)$, but there was a difference. I was 
already worried that I had to go again through my lengthy two-loop 
calculations of 1992, but fortunately we got a second e-mail. Mikolaj 
and J\"org did also the calculation of box diagrams regulating infrared 
divergences dimensionally, as we did five years before, confirming our results
in the full 
theory. This was a true relief. However, they found that our very simple 
calculation in the effective theory did not include an evanescent operator 
related to the dimensional infrared regulator that had to be added in the 
case of the box part. After including it, the small 
difference between our results disappeared.

The issue of including the evanescent operators in this case has nothing to do with the 
cases discussed in section 3 and is rather sophisticated. It is discussed in 
details in \cite{Misiak:1999yg,Buchalla:1998ba}. This example shows that the method of using dimensional 
regularization to regulate infrared divergences and not distinguishing this 
regulator from the dimensional regulator of ultraviolet divergences, although 
very elegant and correct, does not allow a good test of the final result. In 
this 
respect regulating the infrared divergences by external quark masses, as was 
done in our calculations of $\eta_2$ and $\eta_B$ and also by Mikolaj and 
J\"org in their 
calculation of rare decays, is more difficult but safer.

Mikolaj and J\"org checked only our calculation of top quark contributions 
identifying the small difference mentioned above, but somehow they were not 
interested in looking again at the internal charm contributions. This is what 
Gerhard and I did in \cite{Buchalla:1998ba}
 adding the small contribution of the evanescent 
operator to our old result. Our 1999 paper can be considered as a 
compendium of all expressions for the Wilson coefficients relevant for rare 
decays $K\to\pi\nu\bar\nu$, $B_{s,d}\to\mu^+\mu^-$ and $B\to
X_{s,d}\nu\bar\nu$ in the SM at the NLO level. The hard work has been done in our first 
three papers that are also summarized and discussed in our review \cite{Buchalla:1995vs} 
but the 
final fully correct expressions at the NLO level are given in our 1999 paper.

The NNLO QCD calculations for the charm component in $\kpn$ \cite{Buras:2006gb} and in
$K_L\to\mu^+\mu^-$ \cite{Gorbahn:2006bm} were of course much more involved and 
included several two-loop and in particular three loop diagrams in the bilocal 
operator sector. The result was the  $P_c^{(2)}(X)$ in the case of $\kpn$ 
and the
 $P_c^{(2)}(Y)$ in the case of $K_L\to\mu^+\mu^-$.
I will not describe them here as they have been described in great detail 
in two papers on $\kpn$ I was involved in and in the paper by 
Gorbahn and Haisch \cite{Gorbahn:2006bm} on $K_L\to\mu^+\mu^-$.
These calculations  further reduced the perturbative 
 uncertainties in these decays to the level of $\pm 2\%$ so that the most 
important uncertainties in the corresponding branching ratios reside 
in the CKM element $\vcb$  that enters the branching ratios for $\kpn$
roughly like $\vcb^{2.8}$ and $\klpn$ even like  $\vcb^{4}$.

Recently it
has been demonstrated in  \cite{Buras:2021nns,Buras:2022wpw} that this
strong $\vcb$ dependence can be eliminated with the help of $\varepsilon_K$
and the mixing induced CP-asymmetry $S_{\psi K_S}$ so that finally very
precise SM predictions thanks to all these QCD calculations can be found
\be
\mathcal{B}(\kpn)=(8.60\pm 0.42)\cdot 10^{-11},\qquad \mathcal{B}(\klpn)=(2.94\pm 0.15)\cdot 10^{-11}\,.
\ee
In Table~\ref{TAB3} we collect references to QCD calculations 
for rare $K$ and $B$ decays. A given entry means that full NLO or NNLO corrections 
to the decay in question have been calculated in the quoted paper.

\begin{table}[thb]
\caption{Rare K and B decays}
\label{TAB3}
\begin{center}
\begin{tabular}{|l|l|l|}
\hline
\bf \phantom{XXXXX} Decay &  {\bf NLO} & {\bf NNLO}  \\
\hline
\hline
$K^0_L \rightarrow \pi^0\nu\bar{\nu}$, 
$B \rightarrow X_{\rm s}\nu\bar{\nu}$  &\cite{Buchalla:1992zm,Buchalla:1993bv,Misiak:1999yg,Buchalla:1998ba}& \\
$K^+   \rightarrow \pi^+\nu\bar{\nu}$  & \cite{Buchalla:1993wq,Buchalla:1998ba} & \cite{Buras:2006gb}\\
$K_{\rm L} \rightarrow \pi^0\ell^+\ell^-$  & \cite{Buras:1994qa} & \\
$B_{s,d} \rightarrow l^+l^-$  &\cite{Buchalla:1992zm,Buchalla:1993bv,Misiak:1999yg,Buchalla:1998ba}& \cite{Hermann:2013kca}\\
 $K_{\rm L} \rightarrow \mu^+\mu^-$ & \cite{Buchalla:1993wq,Buchalla:1998ba} &
\cite{Gorbahn:2006bm}\\
$K^+\to\pi^+\mu\bar\mu$               & \cite{Buchalla:1994ix} & \\
EW to Charm in $K^+   \rightarrow \pi^+\nu\bar{\nu}$ & \cite{Brod:2008ss}  &    \\
EW to Top in $K\to\pi\nu\bar\nu$ & \cite{Buchalla:1997kz,Brod:2010hi}  &  \\
EW to Top in $B_{s,d} \rightarrow l^+l^-$ & 
\cite{Buchalla:1997kz,Bobeth:2013tba}  &  \\
\hline
\end{tabular}
\end{center}
\end{table}

\subsection{Two-loop Electroweak Contributions}
The rare decays like $K\to\pi\nu\bar\nu$ are theoretically very clean as all 
non-perturbative 
effects investigated by a number of authors \cite{Buchalla:1998ux,Isidori:2005xm,Mescia:2007kn} have been found to be very small
and definitely below any experimental sensitivity in the coming 
ten years. For this reason it is of interest to investigate also two-loop
electroweak contributions to rare decays. 

At first sight these contributions appear to 
be negligible. This however is not fully true for the following reason. The effective 
Hamiltonian for $K\to\pi\nu\bar\nu$ decays involves electroweak parameters 
like $G_F$, $\alpha$ and in particular $\sin^2\theta_W$ that all depend on the
renormalization scheme used in the usual electroweak precision studies. 
This dependence 
can only be reduced by including higher order electroweak corrections to the 
leading one-loop diagrams in rare decays. This means the calculation of 
two-loop electroweak diagrams.
Consider for instance $\sin^2\theta_W$. 
The Particle Data Group 
gives two values for this parameter: $(\sin^2\theta_W)_{\overline{MS}}=0.231$
and {$(\sin^2\theta_W)_{\rm on-shell}=0.224$}, and of course there are other 
possibilities. As $\mathcal{B}(\kpn)$ is inversely proportional to $\sin^4\theta_W$ the
two choices give two values for $\mathcal{B}(\kpn)$  
that differ by $6\%$. This is clearly irrelevant today but I hope that the 
experimental data will improve in this decade to the extent that this 
difference will matter.

A calculation of two-loop electroweak effects is clearly a very difficult 
affair but 
fortunately Gerhard and me could remove the ambiguity in question approximately 
without doing this 
calculation at all  \cite{Buchalla:1997kz}.
We noticed that the calculations of similar effects in the 
context of electroweak precision studies contained sufficient information to 
find two-loop 
electroweak contributions to $K\to\pi\nu\bar\nu$ in the large $m_t$ limit 
without performing any 
loop calculations. Adding these contributions to our previous result reduced 
the ambiguity in question to $(1-2)\%$. Moreover, with the choice of 
$(\sin^2\theta_W)_{\overline{MS}}$ made by 
us not fully accidentally in 1994, the two-loop electroweak corrections to 
$K\to\pi\nu\bar\nu$ 
can be safely neglected and similarly for other decays considered in this 
section.

A much harder calculation has been done by Paolo Gambino, Axel Kwiatkowski 
and Nicolas Pott \cite{Gambino:1998rt}. They calculated full two-loop electroweak contributions to 
$B^0-\bar B^0$ mixing, finding also in this case a very small effect in the 
$\overline{MS}$ scheme.

Coming back to my paper with Gerhard Buchalla on two-loop electroweak 
corrections to rare $K$ and $B$ decays \cite{Buchalla:1997kz}, 
we warned the readers that our 
large $m_t$ limit calculation of these corrections could miss the 
true value of these corrections by a factor of two. In 2010 the 
younger generation of the Munich club, Joachim Brod, Martin Gorbahn and 
the youngest member of our club Emanuel Stamou, performed full two loop
electroweak calculation to $K\to\pi\nu\bar\nu$ decays, basically reaching 
the conclusion of our large $m_t$ limit calculation but reducing further
the theoretical uncertainty \cite{Brod:2010hi}. 
The electroweak contributions to the charm part in 
$K^+\rightarrow \pi^+\nu\bar\nu$ have been calculated by the {duo}
Brod and Gorbahn \cite{Brod:2008ss} already in 2008.

In the fall of 2012, Jennifer and I started to investigate the uncertainties
in $B_{s,d}\to\mu^+\mu^-$ decays with the goal to highlight the importance of a complete
evaluation of higher-order electroweak corrections that in 2012 were known for this
decays only in the large-$m_t$ limit \cite{Buchalla:1997kz} leaving sizable dependence on the definition of electroweak parameters. Using insights from a complete calculation of such corrections for $K\to\pi\nu\bar\nu$ decays, mentioned above, we found a scheme in which NLO electroweak corrections were likely to
be very small. Similary to a 1993 $B\to X_s\gamma$ story told in the next
section we were not sure that this was sufficient for a publication.
But we learned from Diego Guadagnoli and Gino Isidori that they worked on
different uncertainties related to soft radiation and the related issue of the correspondence between the initial and the final state detected by experiments, and those used in the theoretical prediction. We decided to join forces which
resulted in \cite{Buras:2012ru} which was very timely in 
view of the first measurement of the rate for $B_s\to\mu^+\mu^-$ by LHCb and CMS in 2013 that was announced at the EPS conference in the Summer of 2013.

As the experimental result turned out to be in the ballpark of SM expectations
this motivated my three PhD students Bobeth, Gorbahn and Stamou to complete
the two-loop {electroweak} corrections to this decay for an arbitrary top-quark
mass \cite{Bobeth:2013tba} 
bringing the remaining uncertainties from such contributions below $1\%$. 
A nice summary of the status of uncertainties in $B_s\to\mu^+\mu^-$  as of 2014
when also NNLO QCD corrections have been computed \cite{Bobeth:2014tza,Steinhauser:2014hwa} can be found in these papers. Since then the
main progress on this decay was the inclusion of QED corrections
\cite{Beneke:2017vpq,Beneke:2019slt} and the elimination of the $\vts^2$
dependence with the help of $\Delta M_s$ \cite{Buras:2003td,Bobeth:2021cxm}
so that presently the SM branching ratio for this decay, as already stated above, reads
\cite{Buras:2022wpw}
\be
 \overline{\mathcal{B}}(B_s\to\mu^+\mu^-)=(3.78^{+ 0.15}_{-0.10})\cdot 10^{-9}\,.
\ee

In this context an important contribution to the evaluation of the
decay rate of $B_s\to\mu^+\mu^-$ has been made by Robert Fleischer (another member of the MNLC)
and his collaborators by pointing out, already in the spring of 2012, the relevance of $\Delta \Gamma_s$
in the process of the comparison of the theory with experiment 
\cite{deBruyn:2012wk,Fleischer:2012fy}. Our joined
publication that included also the insights from  \cite{Buras:2012ru}
appeared in \cite{Buras:2013uqa}, few months  before the first experimental result was
announced.

Finally, I would like to note that the NLO and the NNLO summits discussed 
in this section similarly to the $\Delta F=2$ summits within the SM 
have been fully dominated by the members of the  MNLC. However, in 
Sections 6 and 8 we will discuss summits which have also been conquered by 
other groups in particular those led by Christoph Greub and 
Mikolaj Misiak.

\boldmath
\section{The \boldmath{$B\to X_s\gamma$} Decay: The K2 of Weak Decays}\label{sec:6}
\unboldmath
\setcounter{equation}{0}
\subsection{Preliminaries}
The calculations of the NLO and NNLO QCD corrections 
to $B\to X_s\gamma$ decay are probably 
the best known 
to the physics community among all QCD calculations in the field of weak 
decays. One of the reasons is 
the fact that the $b\to s\gamma$ transition was the first penguin-mediated 
transition in $B$ physics to be discovered in 1993 in the exclusive decay 
channel $B\to K^*\gamma$ by the 
CLEO experiment \cite{CLEO:1993nic}. The inclusive branching ratio was measured in 1994 by the 
same group \cite{CLEO:1994veu}. The other reason is the particular structure of the QCD 
corrections to this decay that requires a two-loop calculation in order to 
obtain the anomalous dimension matrix in the LO 
approximation. Because of this, it took six years after the first QCD 
calculations
in ordinary perturbation theory to obtain the correct result for the QCD 
corrections to $B\to X_s\gamma$ in the renormalization group improved 
perturbation theory at 
LO. It involved 5 groups and 16 physicists. It is then not surprising that
the corresponding NLO calculations took nine years. They were dominated by 
the group around Christoph Greub and by the members of the MNLC although a few 
other physicists also contributed to this enterprise, as I will report below. I will 
concentrate here on the inclusive decays, as they are theoretically cleaner than 
the exclusive ones but the effective Hamiltonian is, of course, common to 
inclusive and exclusive rates. A nice review of $B\to X_s\gamma$ including 
exclusive decays 
$B\to K^*\gamma$ and $B\to\rho\gamma$ can be found in the reviews 
\cite{Antonelli:2009ws} and \cite{Hurth:2010tk}.
Some comments on the exclusive radiative decays will be made at the end of 
this Section and in Section 9.

The  effective Hamiltonian for $b\to s \gamma$ is given
 at the scale 
$\mu_b={\cal O}(m_b)$
as follows\footnote{Additional operators are needed to describe NLO electroweak corrections.}
\begin{equation} \label{Heff_at_mu}
{\cal H}_{\rm eff}(b\to s\gamma) = - \frac{G_{\rm F}}{\sqrt{2}} V_{ts}^* V_{tb}
\left[ \sum_{i=1}^6 C_i(\mu_b) Q_i + C_{7\gamma}(\mu_b) Q_{7\gamma}
+C_{8G}(\mu_b) Q_{8G} \right]\,,
\end{equation}
where in view of $\mid V_{us}^*V_{ub} / V_{ts}^* V_{tb}\mid < 0.02$,
we have neglected the term proportional to $V_{us}^* V_{ub}$.
Here $Q_1....Q_6$ are the usual four-fermion operators whose
explicit form is given in (\ref{O1})--(\ref{O3}). 
The remaining two operators,
characteristic for this decay, are the {\it dipole--penguins} defined 
in (\ref{O6}).

\subsection{LO Efforts}\label{LOE}

In 1987, two groups \cite{Bertolini:1986th,Deshpande:1987nr} calculated $\ord(\alpha_s)$ QCD corrections to the 
$B\to X_s\gamma$ rate finding a 
huge enhancement of this rate relative to the partonic result without QCD 
corrections. In 1987, when $m_t\le M_W$ was still considered, this enhancement was 
almost by an order of magnitude. However, with the increased value of $m_t$ in the 
1990s, also 
the partonic rate increased, and the dominant additive QCD corrections, 
although still very important, amount in 2023 roughly to a factor of 2.5.

The additive QCD corrections in question originate in the mixing of the 
operator $Q_2$ with the dipole photon penguin operator $Q_{7\gamma}$ that 
is directly responsible for the decay $b\to s\gamma$. The calculation of the 
relevant anomalous 
dimensions at LO is a two-loop affair. Consequently, it took some time 
before the correct result was obtained. In 1988, the Harvard 
\cite{Grinstein:1987vj,Grinstein:1990tj}  and the Toronto
groups \cite{Grigjanis:1988iq,Grigjanis:1989re} calculated the renormalization group improved QCD 
corrections at LO to $B\to X_s\gamma$ using the NDR and the DRED 
schemes, respectively. The results disagreed with each other. This was clearly
a surprise, as 
the LO result for the Wilson coefficients 
cannot depend on the renormalization scheme. 

In 1990, 
  the Pisa group \cite{Cella:1990sh} confirmed the NDR result of Grinstein et al.\cite{Grinstein:1987vj},
  and extended it to include the mixing of $Q_2$ with $Q_{8G}$. 
The fourth calculation was done by 
Mikolaj Misiak \cite{Misiak:1991dj,Misiak:1992bc}, who in a solo climb
evaluated LO QCD corrections to $B\to X_s\gamma$ decay and NLO corrections to
$B\to X_sl^+l^-$ decay. I will discuss the last calculation in the next
section. In these papers Mikolaj found the fourth LO result for the decay in
question, and explained why the previous NDR calculations were incomplete.
  The fifth calculation was done by Adel and Yao \cite{Adel:1992hh} who ignored
  the evanescent operator contributions~\cite{Misiak:1992bc}, and for this sole
  reason failed to reach the correct final result. Other papers of this period contributing to this
discussion and formulating the effective Hamiltonian for $b\to s\gamma$
transitions are \cite{Cho:1991cj,Grigjanis:1992ch}.

Mikolaj solved all the qualitative issues but being alone he did not succeed to reach 
this LO summit. This was done in the summer of 1993 by the Rome group led by 
Guido Martinelli, with the participation of a rising Italian star, 
Luca Silvestrini, 
whose PhD thesis amounted precisely to solving this problem \cite{Ciuchini:1993ks,Ciuchini:1993fk}. This result was soon confirmed by the Pisa group 
around Curci \cite{Cella:1994np,Cella:1994px}, 
 and subsequently by
Mikolaj in an erratum to his second paper \cite{Misiak:1992bc}.

Each of the LO calculations preceeding the final one \cite{Ciuchini:1993ks} had its own
  missed subtleties. One of the tricky points was
that while in the HV scheme 
the one-loop matrix elements of the QCD penguin operators 
$M_i=\langle s\gamma | Q_i| b \rangle$ with $i=3-6$ vanish including finite 
parts, this does not happen in the NDR scheme in which the finite pieces of 
$M_5$ and $M_6$ are different from zero. Combining this one--loop information
with 
the genuine two--loop calculations of the non-diagonal elements of the 
anomalous dimension matrix involving $Q_i$ with $i=1-6$ and $Q_{7\gamma}$ and 
$Q_{8g}$ allowed to show that a
scheme independent leading order result for the Wilson coefficients of the 
operators $Q_{7\gamma}$ and $Q_{8g}$ can be obtained. This important 
solution prompted Mikolaj Misiak to extend this analysis to the DRED scheme 
 \cite{Misiak:1993es}. Next {Mikolaj}, Stefan Pokorski, Manfred M\"unz and myself \cite{Buras:1993xp} introduced effective 
anomalous dimension matrices that were automatically renormalization scheme 
independent at the leading order even in the $B\to X_s\gamma$ decay.

Thus, by the summer of 1993,  correct results for the Wilson Coefficients 
relevant for $B\to X_s\gamma$ and $B\to X_s g$ decays were
known at LO. However, in this year, an important observation was made 
by 
Ahmed Ali, Christoph Greub  and Thomas Mannel \cite{Ali:1993cj}: the LO rate for $B\to X_s\gamma$
 exhibited a very large renormalization scale dependence. Changing the scale 
$\mu_b$ in the Wilson coefficient from $m_b/2$ to $2 m_b$ changed the rate of
$B\to X_s\gamma$ by roughly $60\%$,
making a detailed comparison of theory with experiment impossible. In 1993 
this was not yet a problem, as the inclusive rate was unknown experimentally 
at that time. However, the discovery of the decay $B\to K^*\gamma$ by the CLEO 
collaboration in the 
summer of 1993 was a signal that the inclusive rate would be known soon as 
well. In any case, this problem applies also to $B\to K^*\gamma$. 

The large $\mu_b$ dependence found at LO in this decay is actually 
not surprising. After all, the QCD effects in this decay are very large, 
which can be traced back, at least in part, to the large anomalous dimensions 
of the involved operators.

In the summer of 1993, motivated by the work of Ali, Greub and Mannel, I started to 
look at the steps necessary to do a complete NLO analysis of the $B\to X_s\gamma$
 decay with 
the aim to reduce the strong $\mu_b$ dependence found by them
 at LO. Manfred 
M\"unz, my very good PhD student, joined me in this enterprise but we were 
not sure that just making an outline of this particular 
NLO calculation would be sufficient 
for a publication. Fortunately Mikolaj Misiak, who joined my group in 1993, 
and Stefan Pokorski, who was a visitor at the MPI for Physics at that time, had a 
complementary problem. They were investigating parametric uncertainties 
$(\alpha_s,m_b,m_c)$ in the $B\to X_s\gamma$ decay but, similarly to us, were 
not sure whether such an analysis would be sufficient for a publication. 
Once we discovered our ``problems", it 
was clear that joining our efforts could result in a useful paper. 
This turned out 
to be an excellent decision. Our paper \cite{Buras:1993xp} appeared in November 1993, just 
seven months before the CLEO's discovery of $B\to X_s\gamma$ rate, and 
became a
standard reference in subsequent papers where the actual climbs of the 
NLO $B\to X_s\gamma$ summit have been done.

Before entering the NLO story of this decay, let me mention that in 2012 the
tree-level contribution of $b\to u\bar u s\gamma$ to $B\to X_s\gamma$ 
were calculated by Kami\'nski, Misiak and Poradzi\'nski 
\cite{Kaminski:2012eb}. However, for the photon energy cutoff of $E_0=1.6\gev$ 
or higher this contribution amounts only to $0.4\%$ at the level of the branching ratio.

\subsection{NLO Efforts}
\subsubsection{The Basic Structure}

The complete NLO calculation for the $B\to X_s\gamma$ decay consists of three 
rather difficult steps:

{\bf Step 1}

The calculation of $\ord(\alpha_s)$ two-loop corrections to Wilson coefficients of 
$Q_{7\gamma}$ and $Q_{8g}$ operators at $\mu=\ord(M_W)$. The coefficients 
of $Q_i~(i=1,..6)$ at this order are known from the $\Delta F=1$ non-leptonic 
Hamiltonian discussed in Section 3.

{\bf Step 2}

The calculation of the relevant $8\times 8$ anomalous dimension matrix at 
$\ord(\alpha_s^2)$. The $6\times 6$ submatrix involving the operators
 $Q_i~(i=1,..6)$ is known from the $\Delta F=1$ non-leptonic 
Hamiltonian discussed in Section 3.

{\bf Step 3}

Calculation of the matrix elements $\langle s\gamma | Q_i| b \rangle$ with 
$i=1,..8$ in perturbation theory in $\alpha_s$.

In what follows, I will discuss these three steps one by one in the order as 
given above \cite{Buras:2002er}.

\boldmath
\subsubsection{Wilson Coefficients of $Q_{7\gamma}$ and $Q_{8g}$ at $\mu=\ord(M_W)$}
\unboldmath
The calculation of these Wilson coefficients is much harder than the calculations of Wilson coefficients 
discussed in Section 5 due to the presence of external photons and gluons: 
the external momenta cannot be set to zero. The two-loop diagrams are then 
calculated with non-vanishing external momenta of quarks, the photon and the 
gluon, the result is expanded to second order in external momenta and 
masses, and is matched to the result of the corresponding 
effective theory calculation. The first calculation of $C_{7\gamma}(M_W)$ 
and $C_{8g}(M_W)$ at $\ord(\alpha_s)$ was
performed by Adel and Yao already in 1993 \cite{Adel:1993ah}. Unfortunately, it was done in the 
Zimmermann's renormalization scheme that was rather unfamiliar to
phenomenologists. Consequently, in 1993, this result was not noticed by 
many.

Right at the beginning of 1997, Axel Kwiatkowski (one of my assistants), my 
PhD student Nicolas Pott and myself decided to calculate these coefficients 
in the NDR scheme, that by then became the standard scheme for all NLO 
calculations. However, in March 1997, a paper by Christoph Greub and 
Tobias Hurth  \cite{Greub:1997hf} appeared on the Los Alamos server, in which
 they calculated $\ord(\alpha_s)$ corrections to
$C_{7\gamma}(M_W)$ and $C_{8g}(M_W)$ in the NDR scheme. Moreover, they
demonstrated that their result 
was compatible with the one of Adel and Yao. As by that time we had invested 
already two and a half months in this calculation, this was truly an 
unpleasant 
surprise. Fortunately, it turned out that we still could contribute something 
to 
this calculation. Greub and Hurth regulated the infrared divergences in the 
full 
and effective theory by using dimensional regularization but, to our surprise,
performed the matching in four dimensions. As a result of this, infrared 
divergences did not cancel in the process of matching, and they had to argue 
that they would be removed by including bremsstrahlung corrections. However, 
they 
did not demonstrate this but only showed that if this left-over divergence was 
removed by hand, the resulting finite result in the NDR scheme was 
compatible with the Adel and Yao result. I am mentioning this here to 
emphasize that the inclusion of gluon bremsstrahlung effects is really 
unnecessary to obtain the correct result because on general grounds the 
calculation of Wilson coefficients can be done by choosing any 
external states. For this reason, we continued our calculation and 
demonstrated in detail that performing the full calculation in $4-2\epsilon$
 dimensions, 
including the matching, directly led to the final result of Greub and Hurth,
without any hand-waving arguments for the disappearance of infrared 
divergences. These cancellations are very clearly seen in the expressions 
presented in \cite{Buras:1997bk,Buras:1997xf}. In any case, by the summer of 1997, three groups found the 
Wilson coefficients $C_{7\gamma}(M_W)$ and $C_{8g}(M_W)$. Consequently the
first step of this K2 climb 
was completed. A few months later, an Italian group consisting of Marco Ciuchini, 
Giuseppe Degrassi, Paolo 
Gambino and Gian Giudice \cite{Ciuchini:1997xe} confirmed these results, working in the off-shell operator basis, in contrast to the on-shell basis used by us and
Greub and Hurth.

Before continuing, I would like to emphasize that, in spite of my critical 
remarks on the Greub-Hurth calculation in question, both authors played very 
important roles in the study of QCD corrections to $B\to X_s\gamma$ and 
$B\to X_sl^+l^-$ decays. In particular, Christoph Greub is one of the 
great masters of these decays, and his group made important contributions here 
both in the SM and beyond it. In this context, his numerous analyses of these 
decays within the 2HDM and the MSSM with Francesca Borzumati should be mentioned \cite{Borzumati:1998tg,Borzumati:1999qt,Borzumati:2003rr}.

\subsubsection{Anomalous Dimension Matrix \label{subsec:admbsg}}

The anomalous dimension matrix relevant for the $B\to X_s\gamma$ decay at the
NLO level consists of 
the $6\times 6$ two-loop mixing matrix of four-fermion operators 
$(Q_1,...Q_6)$ discussed 
already in Section 3, the two-loop $2\times 2$ matrix describing the 
evolution and mixing under renormalization of dipole operators
$Q_{7\gamma}$  and $Q_{8g}$ and, finally, the 
three-loop $6\times 2$ matrix describing the mixing between 
$(Q_1,...Q_6)$ and $(Q_{7\gamma},Q_{8g})$.

In 1994, the two-loop $2\times 2$ and three-loop $6\times 2$ matrices were 
still unknown. Mikolaj Misiak and my PhD student Manfred M\"unz decided to perform the first 
of these calculations in early 1994. While certainly not an easy task,
it was achieved by developing a new technique of regularizing IR
  divergences with a common spurious mass parameter. Such a regularization for
  gluon lines had been previously thought to be prohibited because it breaks the
  QCD gauge invariance. However, the breaking turns out to be harmless for the
  RGE parameter calculations in mass-independent regularization schemes, so long as
  subdivergences are treated in a careful manner
  \cite{Chetyrkin:1997fm}. Understanding this fact was a milestone for subsequent
  calculations of beta functions and anomalous dimensions at the three- and
  four-loop levels. The paper of Mikolaj and Manfred appeared on the Los 
  Alamos server in September 1994 \cite{Misiak:1994zw}. Two months later,
    Mikolaj presented the results at the Ringberg meeting, triggering
    objections concerning the IR regularization from David Broadhurst and
    Kostja Chetyrkin -- the great masters in difficult multi-loop
    integrals. However, Kostja realized very quickly that the
    method is correct, and offered his help in continuation of the NLO project
    at the three-loop level.

 The three-loop calculation of the $6\times 2$ matrix in question can certainly be 
  regarded as the most spectacular achievement in the field of higher order 
  QCD calculations in weak decays in the 1990s. Kostja provided soon a
  very efficient recursive formula for three-loop integrals that allowed to
  begin this project. With 800 three-loop Feynman diagrams this was still a
  very complicated project led by Mikolaj with a great help from
  Manfred. Additional difficulty was the treatment of $\gamma_5$ at the
  three-loop level. The calculation in the t'Hooft-Veltman scheme would be simply 
  too much time-consuming, and Mikolaj decided to use the NDR scheme. 
  Unfortunately, my technique (see Section 3) to avoid the dangerous traces 
  with $\gamma_5$, so successful at the two-loop level, fails at three loops. 
  Kostja insisted that it must be possible to define the effective theory
    in such a manner that no traces with $\gamma_5$ occur, as in the
    corresponding SM diagrams. Following this advice,
  Mikolaj replaced the standard basis of four-fermion 
  operators $(Q_1,...Q_6)$ by another (rather complicated looking) set of 
  operators, that allowed him, Manfred and Kostja to complete this project without any
  $\gamma_5$ problems. The result appeared first in \cite{Chetyrkin:1996vx} and the details of 
  the calculation were published in \cite{Chetyrkin:1997fm,Chetyrkin:1997gb}.

At this point, I should mention that the calculation in question was the first 
three-loop calculation in the field of weak decays. {Its} complexity required 
the use of powerful PCs, and was fully done by using computer programs 
for algebraic manipulations like \cite{Jamin:1991dp}. Such programs were subsequently 
further developed by my PhD students, Ulrich Haisch, Christoph Bobeth, 
Martin Gorbahn, and Thorsten Ewerth, and by my assistant J\"org Urban, so 
that similar techniques could be used since then  for the 
calculation 
of two-loop contributions in supersymmetric theories. More about it later.

The three-loop calculation described above together with the initial conditions 
discussed in 6.3.2 and the two-loop matrix elements of the relevant current-
current operators discussed below allowed in 1996 for the first time an almost 
complete (see below) NLO analysis of the $B\to X_s\gamma$ decay \cite{Chetyrkin:1996vx}. 
Since then all analyses of this decay used the results of
\cite{Chetyrkin:1996vx}. It was then fortunate that this result 
was confirmed by Paolo Gambino, and two of my PhD students
Martin Gorbahn and Ulrich Haisch in 2003 \cite{Gambino:2003zm}, who
subsequently extended these very tedious calculations to other operators as
discussed in section 7 below. They confirmed also the 1994 results of
  Mikolaj and Manfred for two-loop mixing of the dipole operators.

\begin{table}[thb]
\caption{$B\to X_s\gamma$ at NLO and NLLO. $\hat\gamma(\rm{Mixing})$ stands 
 for the mixing between 4-quark operators and dipole penguins. 
For more references to $B\to K^*(\rho)\gamma$ see text.}
\label{TAB4}
\begin{center}
\begin{tabular}{|l|l|l|}
\hline
\bf \phantom{XXXX} Decay &  {\bf NLO} & {\bf NNLO}  \\
\hline
\hline
$C_7(M_W)$, $C_8(M_W)$ & \cite{Adel:1993ah,Greub:1997hf,Buras:1997bk,Buras:1997xf,Ciuchini:1997xe}
& \cite{Bobeth:1999mk,Bobeth:1999ww,Misiak:2004ew} \\
$C_1(M_W)-C_6(M_W)$ &   & \cite{Misiak:2004ew} \\
$\hat\gamma(Q_{7\gamma},Q_{8G})$  &\cite{Misiak:1994zw,Gambino:2003zm}& \cite{Gorbahn:2005sa}\\
$\hat\gamma(\rm{Mixing})$     &
  \cite{Chetyrkin:1996vx,Chetyrkin:1997fm,Gambino:2003zm} & \cite{Czakon:2006ss}
\\
Matrix Elements & 
\cite{Ali:1995bi,Pott:1995if,Greub:1996jd,Greub:1996tg,Buras:2001mq,Buras:2002tp,Kaminski:2012eb,Huber:2014nna,Huber:2019ryr}  & \cite{Asatrian:2005pm,Misiak:2006ab,Boughezal:2007ny,Melnikov:2005bx,Blokland:2005uk,Asatrian:2006ph,Asatrian:2006sm,Asatrian:2006rq,Asatrian:2010rq,Ewerth:2008nv,Ferroglia:2010xe,Misiak:2010tk,Misiak:2010sk,Ligeti:1999ea,Bieri:2003ue,Schutzmeier:2009zz,Czakon:2015exa,Misiak:2015xwa,Misiak:2018gvl,Misiak:2020vlo}\\
$B\to K^*(\rho)\gamma$ & \cite{Beneke:2001at,Bosch:2001gv,Ali:2001ez} & \cite{Beneke:2004dp,Ali:2007sj} \\
\hline
\end{tabular}
\end{center}
\end{table}

\subsubsection{Matrix Elements}
The final step of the NLO program for the $B\to X_s\gamma$ decay is the 
calculation of the 
matrix elements $\langle s\gamma|Q_i|B\rangle$. 
The bremsstrahlung contributions to the matrix 
elements of $Q_{7\gamma}$, $Q_{8g}$ and $Q_2$
 were calculated by Ali and Greub in 1995 \cite{Ali:1995bi}. This 
calculation was confirmed six months later by my diploma student Nicolas Pott, who 
extended the Ali and Greub calculation to penguin operators \cite{Pott:1995if}. 
In these papers, also 
one-loop virtual contributions of the matrix elements of 
$Q_{7\gamma}$  and $Q_{8g}$  were calculated.

Much harder is the calculation of the two-loop matrix elements of the four-quark operators, that are 
relevant for the NLO analysis of $B\to X_s\gamma$. In particular, as already 
stressed in our 1993 paper \cite{Buras:1993xp}, the $\mu$-dependence of two-loop matrix elements 
should significantly cancel the strong $\mu_b$-dependence of the LO  branching 
ratio, pointed out by Ali, Greub and Mannel in 1993 \cite{Ali:1993cj}.

The difficulty in this calculation is that an expansion in external momenta 
cannot be made and one has to face a two-loop calculation with full 
kinematics involved. On the other hand in the case of the matrix element of 
the current-current operator $Q_2$, an expansion in powers of 
$m_c^2/m_b^2$ can be made. 
The first calculation of this type, using the technique of Gegenbauer 
polynomials, was done by Christoph Greub, Tobias Hurth and Daniel 
Wyler already in 1996 \cite{Greub:1996jd,Greub:1996tg}.
 As anticipated in 1993, this contribution decreased 
the strong $\mu_b$-dependence of the  rate from $\pm30\%$ found by Ali, Greub 
and Mannel 
down to approximately $\pm 5\%$.

In the summer of 1997, while lecturing at the Les Houches summer school on 
the weak effective Hamiltonian and higher order QCD corrections, I started 
thinking about repeating the 1996 calculation of Greub, Hurth and Wyler. As 
many of the members of the MNLC were involved already in other
projects, I started looking for additional collaborators. Fortunately among 
the many very good students of this summer school, there was one who was 
already a two-loop climber at that time: Andrzej Czarnecki. Already in 1997 
Andrzej had on his account two very important two-loop calculations \cite{Czarnecki:1998kt,Czarnecki:1997hc,Czarnecki:1997cf,Czarnecki:1996gu}  relevant 
for the extraction of the CKM element $V_{cb}$ and knew another technique that 
could in principle be used for the calculation of the matrix elements in 
question: the technique of heavy 
mass expansions. After a short discussion, we agreed to join our forces as 
soon as we complete the projects we were both involved in at that time. 
Accidentaly, our discussion was documented in the form of a photo taken 
by Rajan Gupta. It can be found in the proceedings of the school just before 
my contribution.

However, in the next two years we were both very busy with other projects. 
Among other things, Andrzej with his well known $g-2$ calculations, and I with 
writing up my Les Houches lectures \cite{Buras:1998raa} which thanks to great generosity of Rajan 
Gupta amounted finally to 250 pages. These lectures were very important for
a number of chapters in my recent book \cite{Buras:2020xsm}.

In 2000, Andrzej Czarnecki and me met again and decided to 
increase our team. We were joined by the master Mikolaj Misiak and by 
experienced, already at that time, young NLO climber J\"org Urban. 
In the first climb \cite{Buras:2001mq}
led by Andrzej Czarnecki, we confirmed the result of Greub, Hurth and Wyler 
by using the method of heavy mass expansions, generalizing their result to a 
few additional terms in the expansion in $m_c^2/m_b^2$ 
 that was found to converge rapidly. 
In the second climb \cite{Buras:2002tp}, led by J\"org and Mikolaj, we succeeded 
to express all the two-loop matrix elements in terms of four compact two- and
  three-fold Feynman-parameter integrals. It 
 allowed us to complete the NLO  $B\to X_s\gamma$ 
project by 
calculating its last ingredient: the two-loop matrix elements of QCD penguin 
operators. This last part involved few diagrams with internal $b$-quark, 
implying 
the replacement of $m_c^2/m_b^2$ by $1$, and making the expansions used in 
\cite{Greub:1996jd,Greub:1996tg} and \cite{Buras:2001mq} useless for the calculation of the matrix elements in question.

This last paper \cite{Buras:2002tp} was completed in the spring of 2002, just a few months 
before the celebration of Stefan Pokorski's 60th birthday at the Architects 
House in Kazimierz in Poland. Thus, eight and a half years after the outline of 
the NLO $B\to X_s\gamma$ program \cite{Buras:1993xp}, Mikolaj and me could summarize the 
result of these 
efforts in the volume of Acta Physica Polonica dedicated to Stefan's 
60th birthday \cite{Buras:2002er}. One should take into account that at that time the NLO corrections were called “complete” even though they did not yet include NLO gluonic corrections to the small tree-level LO contributions that we have mentioned at the end of Section~\ref{LOE}. Such virtual effects were computed only in 2014  \cite{Huber:2014nna}, while the corresponding bremsstrahlung
corrections are still under study  \cite{Huber:2019ryr}.

\boldmath
\subsection{$B\to X_s\gamma$ at NNLO \label{sec:bsgNNLO}}
\unboldmath
The story of higher-order calculations of the $B\to X_s\gamma$ rate is not 
finished yet. 
While the NLO
calculations decreased the $\mu_b$-dependence present in the LO expressions 
significantly, 
a new uncertainty was pointed out by Paolo Gambino and Mikolaj Misiak in 2001
\cite{Gambino:2001ew}.
It turns out that the $B\to X_s\gamma$ rate suffers at the NLO from a significant, 
$\pm6\%$, 
uncertainty due to the choice 
of the charm quark mass in the two-loop matrix elements of the four quark 
operators,  in particular in $\langle s\gamma| Q_2| B\rangle $.

In the first calculations \cite{Greub:1996jd,Greub:1996tg,Buras:2001mq,Buras:2002tp}, the pole charm quark mass was used. 
But, as 
stressed by Mikolaj and Paolo, there is no particular reason why the pole mass
should 
be used instead of the $\overline {MS}$ mass. In  fact, they argued that the latter 
choice is 
more appropriate in the case at hand, since charm appears only as an internal 
particle.

While the arguments of Mikolaj and Paolo are very plausible, it is clear that
finally 
the $B\to X_s\gamma$ rate cannot depend on the choice of the charm quark
mass scheme,  even if this 
dependence is significant at the NLO level.

Thus, in order to remove or reduce this uncertainty, a NNLO calculation is 
necessary, a 
truly formidable task. It requires various calculations in three steps: 

{\bf Step 1:}

$C_{7\gamma}(M_W)$ and $C_{8g}(M_W)$ at the three-loop level, and those 
of $C_i(M_W)$ $(i=1-6)$ at two-loop level.

{\bf Step 2:}

Three-loop $6\times 6$ and $2\times 2$ 
anomalous dimension matrices of the operators $(Q_1,..Q_6)$ and 
$(Q_{7\gamma},Q_{8g})$ as well as the mixing between these two sets of 
operators at the four-loop level!

{\bf Step 3:}

The matrix elements 
$\langle s\gamma| Q_i| B\rangle $ $(i=1,..6)$ at the three loop level and 
of the corresponding matrix elements of the dipole  operators 
$(Q_{7\gamma},Q_{8g})$ at the two-loop level.

In the spring of 2003, another workshop in the Ringberg castle took place. 
This 
workshop, organized by Andre Hoang, Gerhard Buchalla, Thomas Mannel and 
myself, gathered experts in heavy flavour physics and had a much broader 
spectrum of topics than the seminal 1988 workshop at
 which the NLO story has been 
initiated. 
At this workshop, Uli Haisch presented the first steps of the three-loop 
calculations of 
the anomalous dimensions of $Q_i~(i=1,...6)$  relevant for 
$B\to X_s\gamma$  at the NNLO level, and of three-loop 
mixing between $Q_i~(i=1,...6)$ and the semi-leptonic operators relevant for 
the decay $B\to X_sl^+l^-$ at NNLO. We 
will discuss this last topic in Section 8. Moreover, Mikolaj Misiak 
outlined  
three-loop calculations of the relevant matrix elements and of the Wilson 
coefficients $C_{7\gamma}(M_W)$  and $C_{8g}(M_W)$ \cite{Misiak:2003xb}. 

In the following years after this 2003 workshop, considerable progress in 
the NNLO program of $B\to X_s\gamma$ has been made, and the 
$B\to X_s\gamma$ rate at NNLO could already be estimated three years later.
It was an effort of more than
17 theorists \cite{Misiak:2006zs} and required a number of calculations over the period of
six years by several groups. As I was not involved in this impressive 
project I will leave the description to the participants of this K2-like
climb. I found in particular the summaries of Mikolaj Misiak
\cite{Misiak:2010dz}, who led
these efforts, and of Uli Haisch \cite{Haisch:2008ar} very informative and clear.
I would like to thank Mikolaj for improving 
my insight in these calculations. 

Let me then only pay the tribute to the successful climbers of the K2 of 
weak decays by listing hopefully most important calculations which led the 
team of Mikolaj Misiak to this important victory. Here the summaries of 
Uli Haisch and Mikolaj Misiak quoted above were very helpful. Explicitly:
\begin{itemize}
\item
$C_{7\gamma}(M_W)$ and $C_{8g}(M_W)$ at the three-loop level  were 
calculated by Misiak and Steinhauser \cite{Misiak:2004ew} and those 
of $C_i(M_W)$ $(i=1-6)$ at two-loop level by Bobeth, Misiak and Urban
\cite{Bobeth:1999mk}.
\item
The three-loop $2\times 2$ anomalous dimension matrix of dipole operators was
calculated by Gorbahn, Haisch and Misiak \cite{Gorbahn:2005sa} and 
the three loop $6\times 6$ anomalous dimension matrix of $(Q_1,..Q_6)$ operators 
by Gorbahn, Haisch and Misiak \cite{Gorbahn:2004my}. Finally 
the four loop mixing of $(Q_1,Q_2)$ operators with the dipole operators 
was found in 2006 by Czakon, Haisch and Misiak \cite{Czakon:2006ss}.
The latter one is the most impressive part of this grand NNLO project.
It 
involved a computation of more than 20000 four-loop diagrams, and
required a mere computing time of several months on around 100 CPUs.
\item
A very difficult 
part of the NNLO calculation turned out to be related to the last step 
of the program, the calculation of the matrix elements to the desired 
order. The fantastic progress in this calculation, made already by 2010
has been summarized very systematically 
by Mikolaj Misiak \cite{Misiak:2010dz}. 
\end{itemize}

As several groups took part in the latter step it is essential to refer 
to all the existing calculations. To this end I found it most convenient 
to follow the summary of Misiak in  \cite{Misiak:2010dz}. 
The discussion of non-perturbative contributions is also included there.

Once the Wilson coefficients have been calculated to the desired order, 
the partonic decay rate is evaluated
according to the formula

\be
\Gamma(b \to X_s\gamma)_{{}_{E_\gamma > E_0}} = N \sum_{i,j=1}^8
C_i(\mu_b)C_j(\mu_b) G_{ij}(E_0,\mu_b),
\ee
where $N = \left|V_{ts}^\star V_{tb}\right|^2 (G_F^2 m_b^5 \alpha_{\rm
  em})/(32\pi^4)$.  At the Leading Order (LO), we have $G_{ij} =
\delta_{i7}\delta_{j7}$, (up to tree-level contributions that have been mentioned at the end of Section~\ref{LOE}), while the ${\cal O}(\alpha_s)$ NLO contributions have been summarized above and in \cite{Buras:2002er}.

At the NNLO, it is sufficient to restrict the attention to $i,j\in\{1,2,7,8\}$
because the Wilson coefficients of QCD penguin operators 
$C_{3,4,5,6}(\mu_b)$ are very small. 
Treating the two similar operators $Q_1$ and $Q_2$ as
a single one (represented by $Q_2$), we list the papers where 
 six independent cases of the NNLO contributions to $G_{ij}$ were
calculated.

First $G_{77}$, $G_{78}$ and $G_{27}$ involve the
photonic dipole operator $Q_{7\gamma}$. While $G_{77}$ was found already
several years ago~\cite{Melnikov:2005bx,Blokland:2005uk,Asatrian:2006ph,Asatrian:2006sm,Asatrian:2006rq}, the complete
calculation of $G_{78}$ was finalized much later
\cite{Asatrian:2010rq,Ewerth:2008nv}. Evaluation of $G_{27}$ is still in
progress (see below).

 The remaining three cases ($G_{22}$, $G_{28}$ and $G_{88}$) receive
contributions from different classes of diagrams.
Diagrams involving  two-body
  final states are IR-convergent and  are just products of the known
NLO amplitudes.  Three- and four-body final state contributions remain
unknown at the NNLO beyond the BLM approximation~\cite{Brodsky:1982gc}. The
BLM calculation for them was completed in
\cite{Ferroglia:2010xe,Misiak:2010tk} providing new results for
$G_{28}$ and $G_{88}$, and confirming the old ones~\cite{Ligeti:1999ea} for
$G_{22}$. The overall NLO + (BLM-NNLO) contribution to the decay rate from
three- and four-body final states in $G_{22}$, $G_{28}$ and $G_{88}$ remains
below 4\% due to the phase-space suppression by the relatively high photon
energy cut $E_0$.  Thus, the unknown non-BLM effects here can hardly cause
uncertainties that could be comparable to higher-order
${\cal O}(\alpha_s^3)$ uncertainties in the dominant terms ($G_{77}$ and
  $G_{27}$).  Misiak  concludes that the considered $G_{ij}$ are known
  sufficiently well. 
  
Thus, the only contribution that is numerically relevant but yet
unknown at the NNLO is $G_{27}$. So far, it has been evaluated for arbitrary
$m_c$ in the BLM approximation~\cite{Bieri:2003ue,Ligeti:1999ea} supplemented
by quark mass effects in loops on the gluon lines~\cite{Boughezal:2007ny}.
Non-BLM terms were calculated only in the two limits: $m_c \gg m_b/2$
\cite{Misiak:2006ab,Misiak:2010sk} and $m_c=0$ \cite{Schutzmeier:2009zz,Czakon:2015exa}. Next, an interpolation between these two limits was performed 
\cite{Czakon:2015exa}. Such a procedure introduces a
non-negligible additional uncertainty to the calculation, which Misiak
estimates to be at the level $\pm 3\%$ in the decay rate.
 A future  calculation for arbitrary 
$m_c$ is
supposed to cross-check the $m_c=0$ result and, at the same time, make it
redundant, because no interpolation in $m_c$ will be necessary any more.  
Good luck!

The references to $B\to X_s\gamma$ NLO and NNLO calculations are collected in 
Table~\ref{TAB4}. Its  NNLO column has been constructed with an invaluable help from Mikolaj Misiak. In several of these papers also the less important decay 
$B\rightarrow X_s {\rm gluon}$ has been analyzed. We refer in particular 
to \cite{Ciuchini:1993fk,Adel:1993ah,Greub:1997hf,Bobeth:1999ww,Greub:2000an,
Greub:2000sy} and later reviews \cite{Antonelli:2009ws,Hurth:2010tk}.
Also NNLO calculation for $B\to X_s\gamma\gamma$ 
has been performed in \cite{Asatrian:2011ta,Asatrian:2014mwa}.

Most recent very advanced calculations by Misiak and collaborators can be found
in
\cite{Misiak:2015xwa,Czakon:2015exa,Rehman:2016kyv,Misiak:2017woa,Misiak:2017qsn,Misiak:2018cec,Misiak:2019ccp,Misiak:2020vlo,Hurth:2023paz}.

These are truly impressive calculations and achievements.
 While I did not
participate in this NNLO calculation, I am very satisfied that 8 members 
of the Munich club and my collaborators took part in this grand project. 
These are Mikolaj Misiak, Paolo Gambino, 
Andrzej Czarnecki, J\"org Urban, and four of my former PhD students: 
Christoph Bobeth, Martin Gorbahn, 
Ulrich Haisch and Thorsten Ewerth. 

\subsection{$B\to X_d\gamma$}
The perturbative QCD corrections to the inclusive decay $B \rightarrow X_ d \gamma$ have been analyzed in 
\cite{Ali:1998rr,Hurth:2003dk,Hurth:2003pn} and their structure is 
totally analogous to the case of the $b\to s\gamma$ transition up to 
obvious changes in flavour indices. 
However, as $\lambda_u = V_{ub} V^*_{ud}$ for $b \to d
\gamma$ is not small with respect to $\lambda_t = V_{tb} V^*_{td}$ and
$\lambda_c = V_{cb} V^*_{cd}$, one also has to take into account the terms
 proportional to $\lambda_u$.

\subsection{Exclusive Radiative Decays}
The effective Hamiltonians for inclusive radiative decays apply of course 
to exclusive decays as well. Here the additional complications are hadronic 
uncertainties. As these are beyond the scope of this writing I just refer 
to selected papers \cite{Bosch:2001gv,Beneke:2004dp,Beneke:2001at,Ali:2004hn,Bosch:2004nd,Becher:2005fg,Ball:2006eu}
and later reviews in \cite{Antonelli:2009ws,Hurth:2010tk}.

\boldmath
\section{\boldmath{$K_L\to\pi^0\ell^+\ell^-$} at NLO}\label{sec:7}
\unboldmath
\setcounter{equation}{0}
\subsection{Effective Hamiltonian}
In this and the following section we will discuss two well known decays:
$K_{\rm L}\to\pi^0\ell^+\ell^-$
and $B \to X_s \ell^+\ell^-$ with $\ell=e,\mu$.
The reason for collecting these decays close to each other is
related to the fact that the results for the first one are very helpful 
in reducing the work necessary to obtain the QCD  corrections to the second 
decay.

Thus  the effective Hamiltonian for $K_{L}\to\pi^0 \ell^+\ell^- $ 
given in (\ref{eq:HeffKpe}) includes in addition to the
operators $Q_1 .... Q_6$ the semi-leptonic operators $Q_{7V}$ 
and $Q_{7A}$ defined in (\ref{q7v7a}).
On the other hand, the effective Hamiltonian
for $B\to X_s \ell^+ \ell^- $ given in (\ref{Heff2_at_mu}) 
can be considered as the generalization of the effective Hamiltonian
for $B\to X_s\gamma$ in (\ref{Heff_at_mu})  to include the semi-leptonic operators $Q_{9V}$ and
$Q_{10A}$ defined in (\ref{Q9V}).  Evidently the operators in (\ref{q7v7a}) and 
 (\ref{Q9V}) behave identically under QCD interactions that are flavour blind. 
Only the renormalization scales differ.

The effective Hamiltonian for $K_L\to\pi^0 \ell^+\ell^-$ at scales 
$ \mu < m_c$ is then given as follows:
\begin{equation}\label{eq:HeffKpe}
{\cal H}_{\rm eff}(K_L\to\pi^0 \ell^+\ell^-) = 
\frac{G_{\rm F}}{\sqrt{2}} V_{us}^* V_{ud}
 \left[\sum_{i=1}^{6,7V} \left[ z_i(\mu)+\tau y_i(\mu)\right] Q_i 
+\tau y_{7A}(M_W)Q_{7A}\right]\,,
\end{equation}
where the four-quark operators $Q_1,...Q_6$ are familiar by now and the new operators $Q_{7V}$ and $Q_{7A}$  and $\tau$ are given by 
\begin{equation}\label{q7v7a} 
Q_{7V}=(\bar sd)_{V-A}(\bar \ell\ell)_V\,,  \qquad
Q_{7A}=(\bar sd)_{V-A}(\bar \ell\ell)_A \,,\qquad
{\tau=-\frac{V^*_{ts}V_{td}}{V^*_{us}V_{ud}}},
\end{equation}
where $(\bar \ell\ell)_V\equiv \bar \ell\gamma_\mu \ell$ and $(\bar\ell\ell)_A\equiv \bar \ell\gamma_\mu\gamma_5 \ell$.

There are three contributions to this decay: CP conserving, directly 
CP-violating and indirectly CP-violating. In 1993 when Markus Lautenbacher, 
Mikolaj Misiak, Manfred M\"unz and myself started looking 
at this decay it was not clear which of these components was the dominant 
one although there was some hope that the theoretically cleanest component, 
the directly CP-violating one, was the dominant one. 
The hope was partially based on the fact that 
at lowest order in
electroweak interactions (single photon, single Z-boson or double
W-boson exchange), this decay takes place only if the CP symmetry is
violated \cite{Gilman:1979ud}.
The CP conserving contribution to the amplitude
comes from a two photon exchange which,
although higher order in $\alpha$, could in principle be sizable.

Extensive work over  15 years on the non-perturbative CP 
conserving component and indirectly CP-violating one shows that within the
SM the latter component dominates followed by the interference between 
the two CP-violating components. The non-perturbative part calculated 
by means of chiral perturbation method turns out to be smaller than these two
components. Important  discussions on both  $K_{\rm L}\to\pi^0 e^+e^-$ and 
$K_{\rm L}\to\pi^0 \mu^+\mu^-$ can be found in 
\cite{Buchalla:2003sj,Isidori:2004rb,Friot:2004yr,Mescia:2006jd}.
\subsection{The Structure of NLO Corrections}
Back to 1993. Our goal was to calculate the coefficients  $y_{7V}$ and $y_{7A}$ 
that are relevant for the directly CP-violating component at the NLO 
level.
In order to see what was known at that time let us 
introduce $\tilde{y}_i$ through
\begin{equation}
y_i = \frac{\alpha}{2 \pi} \tilde{y}_i\,,
\label{eq:ytilde}
\end{equation}
with $\tilde {y}_i$ having the following structure according to PBE
\begin{equation}\label{y7vpbe}
\tilde{y}_{7V} =
P_0 + \frac{Y_0(x_t)}{\sin^2\theta_{\rm W}} - 4 Z_0(x_t),
\end{equation}

\begin{equation}\label{y7apbe}
\tilde{y}_{7A}=-\frac{1}{\sin^2\theta_{\rm W}} Y_0(x_t)\,,
\end{equation}
with $Y_0$ and  $Z_0$  being two one-loop master functions in the SM.

The next-to-leading QCD corrections to the coefficients above enter
only $P_0$. The top dependent terms where known from the work of Fred Gilman 
and collaborators although they have written them in a different manner as Fred 
was involved in this field before PBE was introduced. Moreover $P_0$ at 
LO was known as well.

The structure of QCD corrections to this decay belongs to class 7 introduced 
in Section 2 which 
is a bit special. In this language what was known in 1993 was 
the leading order represented by the  $1/\alpha_s$ term in $P_0$ and 
the top mass dependent part of the NLO represented 
by  $Y_0$ and  $Z_0$. 
\begin{figure}[ht]
\vspace{0.15in}
\centerline{
\epsfysize=4in
\epsffile{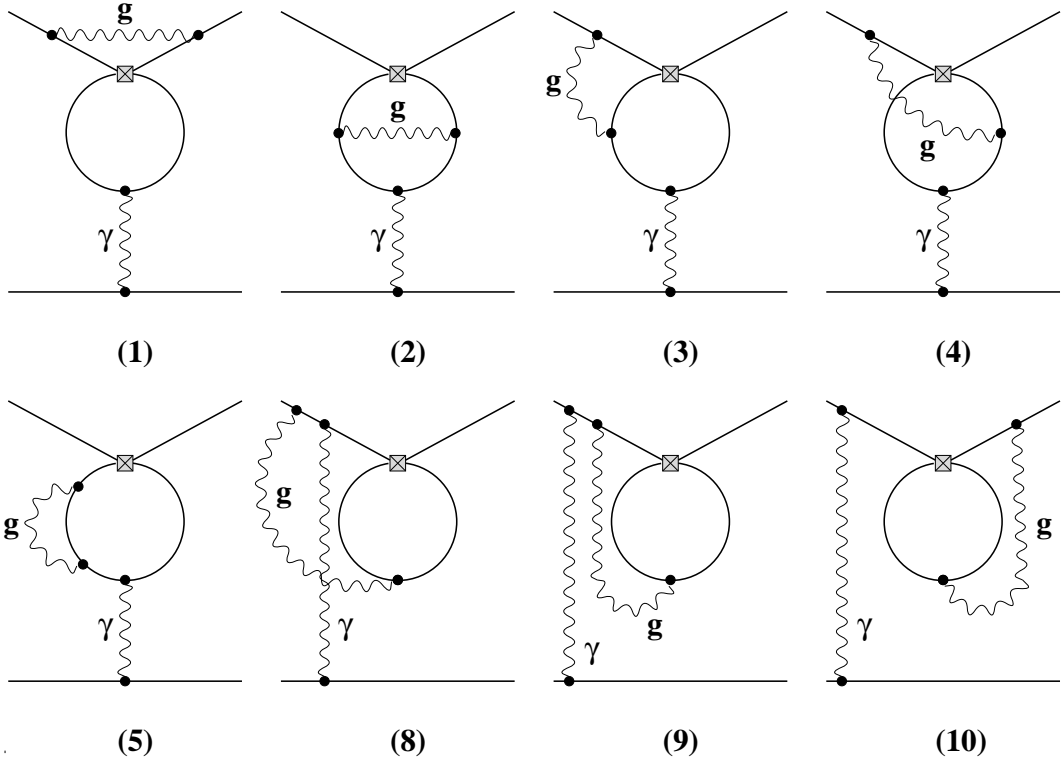}
\vspace{0.15in}
}
\caption[]{
Two--loop penguin diagrams contributing to $\gamma^{(1)}_{i7}$,
$i=1,\ldots,6$. Square vertices stand for two types of penguin
insertions. Possible
left-right reflected diagrams are not shown. The numbering of the
diagrams corresponds to the notation in Figs.~\ref{fig:5} and \ref{fig:5ew}.
\label{diag:5}}
\end{figure}

Our goal was to calculate the remaining, top mass independent, NLO corrections 
to $P_0$ that are of $\ord(1)$ in the $\alpha_s$ expansion. The most difficult 
part in this calculation is the evaluation of the two-loop anomalous 
dimension matrix involving  the
operators $Q_1 .... Q_6$ and the semi-leptonic operator $Q_{7V}$ 
\footnote{$Q_{7A}$ has no anomalous dimensions and in addition 
does not mix with the remaining operators. Therefore the only scale uncertainty in its Wilson coefficient originates in $m_t(\mu_t)$ in $Y_0(x_t)$. This 
uncertainty is practically removed through the inclusion of QCD corrections 
to this function that we have made in the context of the NLO calculation for 
$K_L\to\mu^+\mu^-$ in Section 5. In $K_L\to\pi^0 \ell^+\ell^-$ these corrections appear 
first at the NNLO level.}.
The $6\times6$ submatrix was already known from the work of Munich and 
Rome groups and only mixing between these six operators and  $Q_{7V}$ 
had to be calculated at the two-loop level: as the later operator has no
anomalous dimension only {\it six} elements of this matrix had to be calculated.
The relevant diagrams are given in Fig.~\ref{diag:5}.
After all the hard calculations that we performed before, this one 
turned out to be a relatively 
easy one and after a  month of work the two-loop $7\times 7$ matrix 
relevant 
for this decay and subsequently $P_0$ at NLO was known. 

This calculation
reduced a number of  ambiguities present in 
leading order analyses \cite{Dib:1988md,Flynn:1988ve} and enhanced $P_0$ by 
roughly $30\%$. 
The inclusion of NLO QCD
effects made also a  meaningful use of $\alpha_{\overline{MS}}$ 
in this decay possible. Our paper appeared in February 1994 \cite{Buras:1994qa}.
The two-loop mixing of all the four-quark operators with $Q_{7V}$ agreed with
  an earlier solo calculation of one of us \cite{Misiak:1992bc}, once all the convention
  differences had been taken into account. Appendix B of our common
  paper~\cite{Buras:1994qa} contains a detailed description of this issue.
  Important information on the NLO analysis of $B\to X_s \ell^+\ell^-$ in
  \cite{Misiak:1992bc} is contained there. I will comment on that in the next
  section.

\begin{table}[thb]
\caption{$K_L\to\pi^0\ell^+\ell^-$  and $B\to X_s\ell^+\ell^-$ at NLO and NNLO. For 
$B\to K^*(\rho)\ell^+\ell^-$ see text.}
\label{TAB5}
\begin{center}
\begin{tabular}{|l|l|l|}
\hline
\bf \phantom{XXXXX} Decay &  {\bf NLO} & {\bf NNLO}  \\
\hline
$K_{\rm L} \rightarrow \pi^0e^+e^-$         & \cite{Buras:1994qa} & \\
$B\rightarrow X_s \ell^+\ell^-$      & & \\
$C_i(M_W)$   &\cite{Misiak:1992bc,Buras:1994dj}  & \cite{Bobeth:1999mk}\\
$\hat\gamma({\rm Mixing})$ & \cite{Misiak:1992bc,Buras:1994dj} & \cite{Gambino:2003zm}  \\
Matrix Elements & \cite{Misiak:1992bc,Buras:1994dj} &  
\cite{Asatryan:2001zw,Asatryan:2002iy,Ghinculov:2003qd,Ghinculov:2002pe,Asatrian:2002va,Asatrian:2003yk}  \\
&&  \cite{Greub:2008cy,Bobeth:2003at,Huber:2005ig,Huber:2007vv,Huber:2008ak,Bell:2010mg} \\
$B\to X_d \ell^+\ell^-$ &\cite{Misiak:1992bc,Buras:1994dj} & \cite{Asatrian:2003vq,Seidel:2004jh}\\
$B\to K^*\mu^+\mu^-$  & \cite{Charles:1998dr,Ali:1999mm,Beneke:2000wa} &\cite{Beneke:2001at,Beneke:2004dp}\\
\hline
\end{tabular}
\end{center}
\end{table}

\section{\boldmath{$B\to X_s\ell^+\ell^-$} at NLO and NNLO}\label{sec:8}
\setcounter{equation}{0}
\subsection{Effective Hamiltonian}         
The rare decays $B \to X_s \ell^+ \ell^-$ have been the subject of 
many theoretical studies
in the framework of the SM and its extensions. Useful  reviews with a very 
good collection of references can be found in \cite{Antonelli:2009ws,Hurth:2010tk}. We set $\ell=\mu$
but the same discussion up to some kinematical factors applies to electrons.

The relevant effective Hamiltonian  at scales $\mu=O(m_b)$ is
 given by
\begin{equation} \label{Heff2_at_mu}
{\cal H}_{\rm eff}(b\to s \mu^+\mu^-) =
{\cal H}_{\rm eff}(b\to s\gamma)  - \frac{G_{\rm F}}{\sqrt{2}} V_{ts}^* V_{tb}
\left[ C_{9V}(\mu) Q_{9V}+
C_{10A}(M_W) Q_{10A}    \right]\,,
\end{equation}
where we have neglected the term proportional to $V_{us}^*V_{ub}$
and ${\cal H}_{\rm eff}(b\to s\gamma)$ is given in (\ref{Heff_at_mu}).
In addition to the operators relevant for $B\to X_s\gamma$,
there are two new operators:
\begin{equation}\label{Q9V}
Q_{9V}    = (\bar{s} b)_{V-A}  (\bar{\mu}\mu)_V\,,         
\qquad
Q_{10A}  =  (\bar{s} b)_{V-A}  (\bar{\mu}\mu)_A\,.
\end{equation}

\subsection{$B\to X_s \ell^+\ell^-$ at NLO}
As far as the last two operators are concerned, the structure of QCD 
corrections is very similar to the  case of $K_{\rm L}\to\pi^0 \ell^+\ell^-$
and one can
use  our calculations for $K_L\to\pi^0\ell^+\ell^-$ stopping the RG evolution from 
$M_W$ already 
at $\mu=\ord(m_b)$. In addition the matrix element of $Q_{9V}$ that 
also involves the mixing with $Q_1-Q_6$ operators had to be evaluated 
at the NLO level. The latter one-loop calculation was possibly the main 
new achievement in this paper which I did only with Manfred M\"unz, once
we realized during a lunch at the TUM-mensa in Garching that this paper 
could be completed rather fast \cite{Buras:1994dj}. 
At the same time, 
Mikolaj Misiak followed the instructions from Appendix B of our common 
paper~\cite{Buras:1994qa} to remove a convention mismatch in his earlier
NLO analysis of $B\to X_s\ell^+\ell^-$~\cite{Misiak:1992bc}. It led him to a 
phenomenological formula for the so-called effective coefficients that 
parametrize the decay rate in question. The formula was published in an
erratum to~\cite{Misiak:1992bc}. We compared our results prior to publication and
found perfect agreement.

While the work on $K_L\to\pi^0\ell^+\ell^-$
was harder, our results on $B\to X_s\ell^+\ell^-$ became the standard reference 
on NLO QCD corrections to these decays, dominatly because  the data 
on $B\to X_s\ell^+\ell^-$ improved dramatically in the last 20 years, while 
the decays $K_L\to\pi^0\ell^+\ell^-$ have not been observed yet. In fact 
\cite{Buras:1994dj} is one of the most cited papers from the MNLC and
my most cited paper written with a single author. 

\subsection{$B\to X_s \ell^+\ell^-$  at NNLO}
 Around the year 2000 various groups calculated NNLO corrections to 
$B\to X_s\ell^+\ell^-$ and $B\to K^*\ell^+\ell^-$ putting in particular emphasis on 
the reduction of various scale uncertainties present in our NLO calculations. 
An important role in these efforts played 
the forward-backward asymmetry and the point in the invariant dimuon mass 
$s_0$ at which  this asymmetry vanishes. The calculation of $s_0$ is 
theoretically rather clean for both decays, the feature pointed out 
by Burdman \cite{Burdman:1998mk} in the context of $B\to K^*\ell^+\ell^-$. A meaningful discussion of 
$s_0$ begins first at the NLO level. From this point of view the NNLO 
calculations in question amount to NLO corrections to $s_0$ reducing 
considerably the scale dependence of the LO result that we could have 
discussed already in 1994 but somehow did not.

A compact summary of the NNLO calculations has been presented 
 by Hurth and Nakao \cite{Hurth:2010tk}. As I was not 
involved in these efforts I will not describe them here in detail but 
in order to be complete in references as much as possible I will very 
briefly summarize 
what has been done for the perturbative contributions to the decays 
in question. The review in \cite{Hurth:2010tk} turned out 
to be very useful.

First of all let us stress that for the NNLO calculations of $B\to X_s\ell^+\ell^-$ 
many parts of the NLO calculations of $B\to X_s\gamma$ can be used. 
This is evident when one compares the formal structure of QCD corrections 
given in (\ref{E6}) for Class 6 relevant for  $B\to X_s\gamma$ with the 
structure of QCD corrections given in (\ref{E7}) for Class 7 to which 
 $B\to X_s\ell^+\ell^-$ belongs. Indeed, whereas $\ord(\alpha_s)$ corrections 
in the latter decay represent the NNLO terms, they represent the NLO terms in 
the case of $B\to X_s\gamma$. Therefore below I will only list the calculations 
that are specific to  $B\to X_s\ell^+\ell^-$ and were not done in the context 
of $B\to X_s\gamma$. Again we divide the calculations in three steps as in 
 previous sections:

{\bf Step 1:}

In \cite{Bobeth:1999mk} $\ord(\alpha_s)$ corrections to the 
Wilson coefficient of $Q_{9V}$ have been completed. More explicitly such 
corrections to the penguin function $Z$ were still missing, while those 
to the function $Y$ were already known from the calculations for 
$B_s\to\mu^+\mu^-$ described before. In this manner the large 
matching scale uncertainty of $16\%$ at the NLO
level has been practically eliminated. Note that the coefficient of 
$Q_{10A}$ at this order, represented by the function $Y$, was also known as 
already mentioned in a footnote few pages before.

{\bf Step 2:}

The $\ord(\alpha_s)$ corrections to the term $P_0$ have been calculated 
by  Gambino, Gorbahn and Haisch \cite{Gambino:2003zm} 
who first generalized my two-loop calculation of the 
mixing between $(Q_1-Q_6)$ operators with the semi-leptonic operator $Q_{9V}$, 
done in collaboration with Lautenbacher, Misiak and M\"unz ($\hat\gamma_{se}^{(1)}$ in 
(\ref{ggew})) to the next order by calculating $\hat\gamma_{se}^{(2)}$. 
At this order also three loop mixing in the $(Q_1-Q_6)$ sector is required. 
It has been 
calculated by Gorbahn and Haisch \cite{Gorbahn:2004my} and entered already 
$B\to X_s\gamma$ at the NNLO level.

{\bf Step 3:}

Let me note that until this point the calculations discussed in this section were 
fully in the domain of MNLC. However, in Step 3 at the NNLO level other 
groups dominated the NNLO calculations, in particular Christoph Greub and 
his powerful army. 
In fact  the four-quark matrix elements including
the corresponding bremsstrahlung contributions have been  calculated for the
low-$q^2$ region in
\cite{Asatryan:2001zw,Asatryan:2002iy,Ghinculov:2003qd},
bremsstrahlung contribution for the forward-backward asymmetry in $B
\to X_s \ell^+\ell^-$ 
in \cite{Ghinculov:2002pe,Asatrian:2002va,Asatrian:2003yk}, and the
four-quark matrix elements in the high-$q^2$ region in
\cite{Ghinculov:2003qd,Greub:2008cy}.  The
two-loop matrix element of the operator $Q_{9V}$ has been  estimated using
the corresponding result in the decay mode $B \to X_u \ell \nu$ and
also Pade approximation methods \cite{Bobeth:2003at}. {The role of collinear photons has been investigated in  \cite{Huber:2008ak}.}
For QED corrections we refer to \cite{Bobeth:2003at,Huber:2005ig,Huber:2007vv,Huber:2008ak,Huber:2015sra}. The SCET methods 
at NNLO have been used in \cite{Bell:2010mg}. Phenomenology of $B\to X_s \ell^+\ell^-$ for the Belle II era has been presented in \cite{Huber:2020vup}.

 For the study of the decay $B\to X_d \ell^+\ell^-$ we refer to 
\cite{Asatrian:2003vq,Seidel:2004jh}. The analysis is very similar but 
this time CKM suppressed terms have to be kept as in the case of 
$B\to X_d\gamma$ decay. Here NLO QED corrections to all angular observables have been calculated in \cite{Huber:2019iqf}.

Finally I would like to mention
a paper on NNLO corrections to $B\to X_s\ell^+\ell^-$ in 
the MSSM. This is the work done in collaboration 
with my PhD students Christoph Bobeth and Thorsten 
Ewerth \cite{Bobeth:2004jz}. The main motivation for this work was the reduction of scale
uncertainties in $s_0$ which differs  a bit from the one in the SM and in order  to feel this difference perturbative uncertainties have to be under 
control. I should 
emphasize that in contrast to the rest of my papers described  above, 
my contribution to this paper was minor.  
Indeed  Christoph and Thorsten should be fully credited for this work. 

\subsection{$B\to K \ell^+\ell^-$ and $B\to K^*(\rho)\ell^+\ell^-$}
 The NLO and NNLO QCD corrections discussed for inclusive decays 
can be of course  
 also used for corresponding exclusive decays that are easier to measure.
 Again as in the case of $B\to K^*(\rho)\gamma$ formfactor uncertainties 
 matter. In particular the $B\to K^*\mu^+\mu^-$ decay has been investigated
  by many authors of whom we can cite only a few. 
Among the older papers let me just mention 
\cite{Charles:1998dr,Ali:1999mm,Beneke:2000wa}. In particular in
1999, Ali et al.\ calculated the dilepton mass spectrum
and $A_{\rm FB}$ in the SM and various SUSY scenarios using na\"\i\/ve
factorization and QCD sum rules on the light cone \cite{Ali:1999mm}. Later
it was shown by Beneke et al.\ \cite{Beneke:2001at,Beneke:2004dp}  
that $B\to K^*\mu^+\mu^-$ admits a systematic theoretical
description using QCD factorization in the heavy quark limit
$m_b\to\infty$. This limit is relevant for small invariant dilepton masses and 
reduces the number of independent
form factors from 7 to 2. Spectator effects, neglected in na\"\i\/ve
factorization, also become calculable. 
In \cite{Ali:2006ew}, a
calculation of $B\to K^*\mu^+\mu^-$ using soft-collinear theory
(SCET) was presented.

There are other aspects related to 
formfactors but this topic is outside the scope of this review and I 
refer only to selected papers. The  older papers \cite{Bobeth:2008ij,Egede:2008uy, Altmannshofer:2008dz,Bharucha:2010im} discuss several
formfactor issues in detail and define
various symmetries and asymmetries in the SM and  study them in selected New Physics models.
 In these papers a definite 
progress on the calculation of formfactors has been achieved. 
A lot of information can 
also be found in the reviews \cite{Antonelli:2009ws,Hurth:2010tk}.
Finally an important paper discussing theoretical aspects of  $B\to K \ell^+\ell^-$ 
and
$B\to K^*\ell^+\ell^-$ at large $q^2$, in particular in connection with hadron-quark 
duality is the paper by Beylich, Buchalla and Feldmann
\cite{Beylich:2011aq}.

I should mention at this stage that starting with 2013 there was a great activity in this field due to the so-called anomalies in these decays,
departures from SM expectations in some observables in $B_d\to K(K^*)\mu^+\mu^-$ 
found by  LHCb. It is not possible to review this topic here and I refer 
to \cite{Alguero:2022wkd} and references therein. In this contex
the knowledge of formfactors has been significantly improved
\cite{Straub:2015ica,Gubernari:2018wyi,Parrott:2022dnu,Parrott:2022rgu,Parrott:2022smq}. 

The next pages of this review are dedicated to the NLO and NNLO QCD corrections 
 done in the QCD factorization (QCDF) approach to two-body B decays. This means 
Gerhard Buchalla, one of its fathers, is entering  the scene and I can collect the energy for 
the rest of the review. The next section was written first by Gerhard in 2011
and  updated here by me in consultation with him and the second father of QCDF,
Martin Beneke.

%%%%%%%%%%%%%%%%%%%%%%%%%%%%%%%%%%%%%%%%%%%%%%%%%%%%%%%%%%%%%%%%%
%   QCD factorization
%%%%%%%%%%%%%%%%%%%%%%%%%%%%%%%%%%%%%%%%%%%%%%%%%%%%%%%%%%%%%%%%%
\section{QCD Factorization for Exclusive \boldmath{$B$} Decays (by G. Buchalla)}
\label{sec:qcdf}

\subsection{Introduction}

\label{sec:intro}

The formulation of factorization theorems for exclusive hadronic
$B$-meson decays in 1999 made an entire new class of processes
accessible to systematic calculations of higher-order corrections
in QCD \cite{Beneke:1999br,Beneke:2000ry}. 
These processes include $B$ decays into a pair of light
mesons, the prototype of which is $B\to \pi\pi$, but also rare and 
radiative decays, such as $B\to K^*\gamma$ or $B\to K^*\ell^+\ell^-$.
In the heavy-quark limit, that is up to relative corrections of
order $\Lambda_{\rm QCD}/m_b$, the problem of computing exclusive 
hadronic decay
amplitudes simplifies considerably. In this limit the decay amplitudes
can be written as hard-scattering kernels, which are {\it process dependent}
but perturbatively calculable, multiplied by hadronic quantities such
as $B\to\pi$ form factors, meson decay constants and light-cone wave 
functions, which are nonperturbative but {\it process-independent.}  
The decomposition into calculable hard contributions and universal 
hadronic quantities is in full analogy with the factorization of 
short-distance and long-distance terms that is the basis of almost 
any application of QCD to high-energy processes. 
Correspondingly the framework is refered to as QCD factorization for 
exclusive hadronic $B$ decays, or QCD factorization for short.
In the present section we review the factorization formula and give an
overview of the NLO and NNLO calculations performed for exclusive $B$ 
decays. We conclude with a brief outlook on highlights of QCD
  factorization, special applications to precision flavour physics,
and recent developments.

\subsection{Factorization formula}
\label{sec:fform}

The matrix element of an operator $Q_i$ in the effective weak 
Hamiltonian for the decay of a $\bar B$ meson into a pair of
light mesons $M_1M_2$ is given by \cite{Beneke:1999br,Beneke:2000ry}
\begin{eqnarray}
\label{fff}
\langle M_1 M_2|Q_i|\bar{B}\rangle &=&
F^{B\to M_1}(m_2^2)\,\int_0^1 du\,T_{i}^I(u)\,\Phi_{M_2}(u)
\,\,+\,\,(M_1\leftrightarrow M_2)\nonumber\\
&&\hspace*{-2cm}
+\,\int_0^1 d\xi du dv \,T_i^{II}(\xi,u,v)\,
\Phi_B(\xi)\,\Phi_{M_1}(v)\,\Phi_{M_2}(u), 
\end{eqnarray}
up to power corrections of order $\Lambda_{\rm QCD}/m_b$.
$F^{B\to M_{1,2}}(m_{2,1}^2)$ are $B\to M_{1,2}$ form factors,
where $m_{1,2}$ denote the light-meson masses,
and $\Phi_M$ is the light-cone distribution amplitude for the
quark-antiquark Fock state of meson $M$. Here the light-cone distribution
amplitudes are understood to include the decay constant $f_M$ of meson $M$
in their normalization. These quantities define the nonperturbative input
needed for the computation of the decay amplitudes in QCD factorization.
They are simpler than the full matrix element on the l.h.s. of (\ref{fff})
and universal in the sense that they appear as well in many other processes, 
which are different from $\bar B\to M_1M_2$.  
The $T_{i}^I(u)$ and $T_i^{II}(\xi,u,v)$ are the hard-scattering functions,
which are process-specific and depend in particular on the operator $Q_i$.
They are calculable by standard methods in perturbative QCD.

The formula (\ref{fff}) exhibits the factorization of the
short-distance kernels $T_i$ and the long-distance hadronic quantities
$F^{B\to M}$ and $\Phi_M$. The factorization of the latter takes, in general, 
the form of
a convolution over the parton momentum fractions $\xi$, $u$, $v\in [0,1]$. 
A graphical representation of (\ref{fff}) is given in Fig.~\ref{fig:fform},
where index $j$ accounts for the possibility of more than a
single $B\to M_1$ form factor. 
%%%%%%%%%%%%%%%%%%%%%%%%%%%%%%%%%%%%%%%%%%%%%%%%%%%%%%%%%%%%%%%%%%%
\begin{figure}[t]
 \begin{center}
  \vspace{-3cm}
  \includegraphics[width=\textwidth]{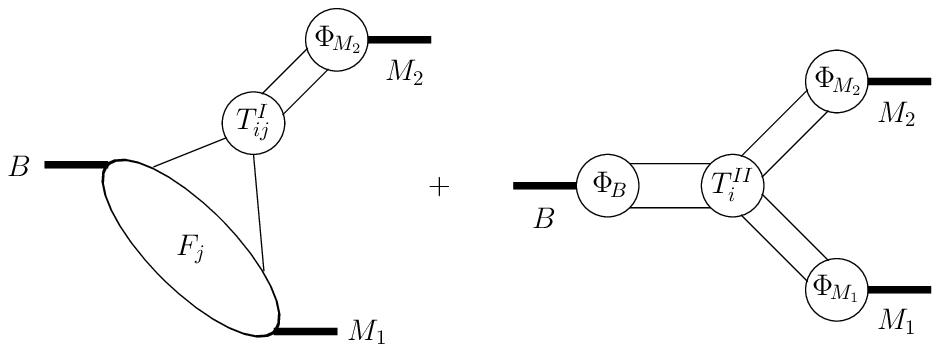}
  \vspace{-15.5cm}
 \end{center}
\caption{\label{fig:fform} Factorization formula.}
\end{figure}
%%%%%%%%%%%%%%%%%%%%%%%%%%%%%%%%%%%%%%%%%%%%%%%%%%%%%%%%%%%%%%%%%%%
The second term ($\sim T^{II}$) is distinguished from the first
($\sim T^I$) by the participation of the $B$-meson spectator quark
in the hard interaction, indicated by the spectator line entering
the kernel $T^{II}$. The spectator interaction requires the exchange
of a hard gluon. $T^{II}$ starts therefore at order $\alpha_s$,
whereas $T^{I}$ is of order unity, schematically
\begin{equation}\label{tiexp}
T^I=T^{I}_{(0)} +\frac{\alpha_s}{4\pi}T^{I}_{(1)}+
\left(\frac{\alpha_s}{4\pi}\right)^2 T^{I}_{(2)}+\ldots
\, ,\qquad\quad 
%\end{equation} 
%\begin{equation}\label{tiiexp}
T^{II}=\frac{\alpha_s}{4\pi}T^{II}_{(1)}+
\left(\frac{\alpha_s}{4\pi}\right)^2 T^{II}_{(2)}+\ldots
\end{equation}

The above description relies on the fact that in the two-body
decay of the $B$ meson the final-state particles are necessarily
very energetic, with light-like four-momenta, in the rest frame of the $B$. 
A meson emitted from the hard interaction, such as $M_2$ from
$T^I$ in Fig.~\ref{fig:fform}, is then described by its light-cone
distribution amplitude. At leading power in $\Lambda_{\rm QCD}/m_b$
the amplitude is determined by the contribution from the light-cone 
wave function of leading twist, which corresponds to the
simplest, two-particle Fock state. Higher Fock states give
power-suppressed contributions and are therefore absent in the heavy-quark
limit. For example, an additional energetic gluon, collinear to the
light-like quark and anti-quark in meson $M_2$ will generate an
additional, far off-shell propagator when attached to the hard
process $T$, which results in a power suppression. The properties
of the light-cone wave functions, which vanish at the endpoints ($u=0$,~$1$),
also imply the suppression of highly asymmetric configurations 
where one parton carries almost the entire meson momentum and the
other parton is soft.  

To leading order in QCD, at ${\cal O}(\alpha^0_s)$, the factorized matrix 
element in (\ref{fff}) reduces to a particularly simple result.
The second term $T^{II}$ is absent at this order and $T^I(u)$ becomes a
$u$-independent constant.
Taking the matrix element of operator $Q_1=(\bar ub)_{V-A}(\bar du)_{V-A}$
for $\bar B\to\pi^+\pi^-$ as an example\footnote{In the present section 
the numbering of operators $Q_{1,2}$ and their coefficients is reversed 
with respect to the notation of the previous sections.}, 
the factorization formula then states that
\begin{equation}\label{qcdflo}
\langle\pi^+\pi^-|(\bar ub)_{V-A}(\bar du)_{V-A}|\bar B\rangle =
\langle\pi^+|(\bar ub)_{V-A}|\bar B\rangle\,
\langle\pi^-|(\bar du)_{V-A}|0\rangle =
i F^{B\to\pi}(0) f_\pi m^2_B\,.
\end{equation}
This corresponds to the prescription of factorizing the matrix
element of the 4-quark operator into a product of matrix elements
of bilinear quark currents. Such an ansatz, which has a long history
in phenomenological applications \cite{Fakirov:1977ta}, thus receives
its proper justification in the context of QCD factorization.
The approximation in (\ref{qcdflo}) means that the emission of the
$\pi^-$ is independent of the remaining $\bar B\to\pi^+$ transition.
The intuitive argument for this, namely that the energetic and highly
collinear, colour-singlet $\bar u d$ pair forming $\pi^-$ has little
interaction with the rest of the process, has been described long time ago by Bjorken in
\cite{Bjorken:1988kk}. The factorization theorem (\ref{fff}) is the
formal implementation of this idea and it allows us to compute
corrections systematically.

Factorization also works for decays of the type $\bar B\to D^+\pi^-$
with a heavy and a light meson in the final state, if it is the light meson
that is emitted from the hard interaction (meson $M_2$ in
Fig.~\ref{fig:fform}). In this case spectator scattering is
power suppressed and the factorization theorem in the heavy-quark limit
takes the form
\begin{equation}\label{ffhl}
\langle D M_2|Q_i|\bar{B}\rangle =
 F^{B\to D}(m_2^2)\,\int_0^1 du\,T_{i}^I(u)\,\Phi_{M_2}(u)\,.
\end{equation}
The expression in (\ref{ffhl}) had already been used
in \cite{Politzer:1991au} to compute the order-$\alpha_s$
corrections to the ratio of the $\bar B\to D\pi^-$
and $\bar B\to D^*\pi^-$ decay rates, prior to the systematic 
development of QCD factorization.

The factorization theorem can be formulated using soft-collinear
effective theory (SCET) \cite{Bauer:2000yr,Bauer:2001yt}.
This formalism is useful for proving factorization \cite{Bauer:2001cu}
and for disentangling the hard and hard-collinear scale in explicit terms.
QCD factorization and SCET are theoretical concepts that are fully
compatible with each other, but they refer to different aspects
of the problem of $B$-decay matrix elements. 
In some sense the relation between QCD factorization and SCET
is similar to the relation between the heavy-quark expansion
(HQE) and heavy-quark effective theory (HQET) in their application
to {\it inclusive\/} $B$ decays.
QCDF \cite{Beneke:1999br,Beneke:2000ry} refers to
the separation of the matrix elements into simpler long-distance
quantities and calculable hard interactions, where the long-distance
form factors are defined in full QCD. SCET, on the other hand, is a general
effective field theory formulation for the relevant QCD modes
(hard, hard-collinear, collinear, soft) and allows a further
separation of scales, for instance in the transition form factors. 
However, working with form factors in full QCD often seems preferable
in practice. An excellent review of SCET can be found in \cite{Beneke:2015wfa}.
Some aspects are also discussed in \cite{Petrov:2016azi}.

\subsection{NLO calculations}
\label{sec:nlo}

Next-to-leading-logarithmic accuracy (NLO) in the decay amplitudes
requires the calculation of the ${\cal O}(\alpha_s)$ terms
in the factorization formula (\ref{fff}). 
The relevant diagrams for the kernels $T^{I,II}$ are shown in 
Fig.~\ref{fig:figlett1}. They provide a concrete illustration of 
the schematic picture in Fig.~\ref{fig:fform}.
%%%%%%%%%%%%%%%%%%%%%%%%%%%%%%%%%%%%%%%%%%%%%%%%%%%%%%%%%%%%%%%%%%%
\begin{figure}
[t]
%[htbp]
\begin{center}
  \vspace{-2.5cm}
\hspace*{1.5cm}\includegraphics[width=\textwidth]{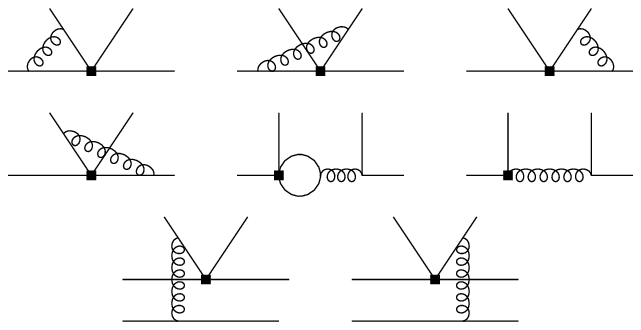}\hspace*{-2cm}
  \vspace{-14.5cm}
\end{center}
\caption{Order $\alpha_s$ corrections to the hard
scattering kernels $T^I_i$ (first two rows) and $T^{II}_i$
(last row). In the case of $T^I_i$, the spectator quark does
not participate in the hard interaction and is not drawn.}
\label{fig:figlett1}
\end{figure}
%%%%%%%%%%%%%%%%%%%%%%%%%%%%%%%%%%%%%%%%%%%%%%%%%%%%%%%%%%%%%%%%%%%
As an example, we quote the contribution from current-current
operators $Q_{1,2}$ to the $\bar B\to\pi^+\pi^-$ amplitude.
(The penguin contributions are given in \cite{Beneke:1999br,Beneke:2001ev}.) 
Up to power corrections and to NLO precision this amplitude reads 
\begin{equation}\label{bppnlo}
\langle\pi^+\pi^-|{\cal H}_{\rm eff}|\bar B\rangle =
i \frac{G_F}{\sqrt{2}}V_{ub}V^*_{ud} F^{B\to\pi}(0) f_\pi m^2_B
\, \left[a_{1,I} + a_{1,II}\right]+\ldots
\end{equation}
where
\begin{equation}\label{a1i}
a_{1,I}=C_1+\frac{C_2}{N}+\frac{C_2}{N} \frac{C_F\alpha_s}{4\pi}
\left[-12\ln\frac{\mu}{m_b}-18+\int_0^1 du\ 3
\left(\frac{1-2u}{1-u}\ln u-i\pi\right)\Phi_\pi(u) \right]\,,
\end{equation}
and ($\bar u\equiv 1-u$, $\bar v\equiv 1-v$)
\begin{equation}\label{a1ii}
a_{1,II}=\frac{C_2}{N} \frac{C_F\pi\alpha_s}{N}
\frac{f_B f_\pi}{m^2_B F^{B\to\pi}(0)}
\int_0^1\frac{d\xi}{\xi}\Phi_B(\xi) \,
\int_0^1\frac{du}{\bar u}\Phi_\pi(u)\, \int_0^1\frac{dv}{\bar v}\Phi_\pi(v)\,.
\end{equation}
Here $C_F=(N^2-1)/(2N)$ and $N$ is the number of colours.

The contribution in (\ref{a1i}) represents the $T^I$-part of
the matrix element. Coefficients $C_{1,2}$ and $\alpha_s$
are evaluated at a scale $\mu={\cal O}(m_b)$. As it must be the case,
the scale and scheme dependence of the $\alpha_s$-correction
cancels against the corresponding dependence of the NLO
coefficients up to terms of order $\alpha^2_s$. The constant $-18$
refers to the NDR scheme as defined in \cite{Buras:1992tc}.
Whereas form factor and decay constant are scheme and scale independent,
the pion distribution amplitude $\Phi_\pi$ has such a dependence.
Since it enters only at ${\cal O}(\alpha^2_s)$ in $a_{1,I}$ this is
irrelevant at NLO.

Hard-gluon exchange between the two final-state pions leads to
a perturbative rescattering phase and thus to an imaginary
part in (\ref{a1i}). Together with the phase from penguin loops
this gives the formally leading contribution to the rescattering
phase in the heavy-quark limit. The numerical value of the
imaginary part has to be taken with caution because it has to
compete with $\Lambda_{\rm QCD}/m_b$ power corrections, which are hard to
quantify. In any case, the rescattering phase is predicted to
be suppressed, either by $\alpha_s$ or by $\Lambda_{\rm QCD}/m_b$. 

The spectator-scattering term $a_{1,II}$ in (\ref{a1ii})
is an additional, qualitatively different contribution,
which first arises at order $\alpha_s$. The scale dependent
quantities $C_2$ and $\alpha_s$ are evaluated at a scale
$\mu_h=\sqrt{\Lambda \mu}$, representing the typical virtuality of the 
semi-hard (or, more precisely, hard-collinear) gluon in this process.

The factorized structure of the amplitude in (\ref{bppnlo}) is in close 
analogy with the amplitude for other weak processes, for instance 
$B^0$-$\bar B^0$ mixing. All of them consist of certain 
long-distance quantities multiplying calculable short-distance functions. 
A technical complication
specific to (\ref{bppnlo}) is the presence of meson distribution
amplitudes, whose factorization involves an integration over
parton momentum fractions. 

The first calculation of hadronic two-body $B$-decay amplitudes
complete to NLO in QCD was performed in \cite{Beneke:1999br}
for the three $\bar B\to\pi\pi$ channels $\bar B_d\to\pi^+\pi^-$, 
$\bar B_d\to\pi^0\pi^0$ and $B^-\to\pi^-\pi^0$. 
The class of heavy-light final states $\bar B_d\to D^{(*)+}L^-$,
with light meson $L^-=\pi^-$, $\rho^-$, $K^{(*)-}$, $a^-_1,\ldots$, 
was analyzed in detail at NLO in \cite{Beneke:2000ry}
(see also \cite{Chay:2000wp}).
A more recent discussions of phenomenological applications of these modes
can be found in \cite{Fleischer:2010ca}.
The NLO calculations were subsequently extended to all
$B\to\pi\pi$ and $B\to\pi K$ channels \cite{Beneke:2001ev}
and eventually to all decays $B\to PP$ and $B\to PV$, where $P$ $(V)$
is a light pseudoscalar (vector) meson \cite{Beneke:2003zv}.
Decays into flavour-singlet mesons of the type $B\to K^{(*)}\eta^{(')}$ 
and their special properties were treated in \cite{Beneke:2002jn}.
The decays $B\to VV$ are more complicated
due to the existence of different helicity amplitudes for the pair of
vector mesons. Only decays into light vector mesons with longitudinal 
polarization are strictly calculable in QCD factorization.
Early papers on this subject are \cite{Cheng:2001aa,Kagan:2004uw}.
Comprehensive studies have been given in
\cite{Beneke:2006hg,Cheng:2008gxa,Bartsch:2008ps}, where
\cite{Cheng:2008gxa} also considers final states with axial vector mesons.

The methods of QCD factorization can also be employed for rare and
radiative $B$ decays. In this case the dominant part of the
amplitude comes from bilinear quark currents, whose matrix elements
are directly given by form factors. However, the nonleptonic Hamiltonian
contributes to the transition as well and requires a nontrivial application 
of the factorization formula. The NLO results for the exclusive
decays $B\to K^*\gamma$ and $B\to\rho\gamma$ were obtained in   
\cite{Beneke:2001at,Bosch:2001gv,Ali:2001ez}.
This calculation involves the NLO Wilson coefficients for
$b\to s\gamma$ \cite{Chetyrkin:1996vx}, which depend on three-loop 
anomalous dimensions, and two-loop virtual corrections to the
matrix elements of local operators \cite{Greub:1996tg,Buras:2001mq}.
The work of \cite{Beneke:2001at} included the
generalization to $B\to K^*\ell^+\ell^-$ at moderate values of the
dilepton mass $q^2$.

\subsection{NNLO calculations}
\label{sec:nnlo}

Important progress has been achieved in extending perturbative 
calculations in QCD factorization for charmless two-body $B$ decays 
to the NNLO, including effects of order $\alpha^2_s$.
In many cases this level of accuracy is probably below the
size of uncertainties from other sources, in particular from power
corrections. However, the explicit knowledge of NNLO corrections
is of conceptual interest as it extends the factorization formula
to the next nontrivial level in perturbation theory.
In addition, there are quantities for which the NNLO effects are
likely to be also numerically important. These are cases where
a contribution is absent at ${\cal O}(1)$ and thus the
${\cal O}(\alpha_s)$ term, a NLO contribution in the general
counting scheme, is effectively the lowest order. Examples are
strong phases relevant for direct CP violation,
hard spectator scattering, or the color-suppressed amplitude coefficient 
$a_2$, which is accidentally small at leading and next-to-leading order 
and therefore rather sensitive to NNLO effects.

We briefly summarize the first classes of NNLO corrections that have been computed in the past.
These are, first, the
${\cal O}(\alpha^2_s)$ one-loop hard-spectator 
interactions for current-current operators \cite{Beneke:2005vv,Kivel:2006xc,Pilipp:2007mg} and for penguin contributions \cite{Beneke:2006mk,Jain:2007dy}.
It should be mentioned that here the hard-spectator interactions of 
${\cal O}(\alpha^2_s)$ are counted as NNLO terms since they enter the decay 
amplitudes at the NNLO level. A different convention has been used
in \cite{Beneke:2005vv,Beneke:2006mk}, where the same corrections are
refered to as NLO due to the fact that hard-spectator interactions first
arise at ${\cal O}(\alpha_s)$.

Second, the two-loop vertex corrections ($T^I$) have been addressed
for the first time in \cite{Bell:2007tv}, where the imaginary
part is computed explicitly. The corresponding real part has been 
obtained in \cite{Bell:2009nk}, completing the NNLO vertex corrections
for current-current operators. These results have been
confirmed in \cite{Beneke:2009ek}.
The first phenomenological analysis of exclusive $B$ decays
at NNLO has been presented in \cite{Bell:2009fm} for the tree-dominated 
$B\to\pi\pi$, $\pi\rho$ and $\rho\rho$ decays.

The numerical impact of NNLO effects for the coefficients $a_1(\pi\pi)$ 
and $a_2(\pi\pi)$, 
which determine the topological tree-amplitudes in $B\to\pi\pi$, 
is illustrated by the following compilation from \cite{Bell:2009nk}:
\begin{eqnarray}
a_1(\pi\pi) &=& 1.008 + [0.022+0.009 i]_{I,\alpha_s}+
[0.024+0.026 i]_{I,\alpha^2_s}
\nonumber\\
&& - [0.012]_{II,\alpha_s}-[0.014+0.011 i]_{II,\alpha^2_s}-[0.007]_{P}
\nonumber\\
&=& 1.019^{+0.017}_{-0.021}+\left(0.025^{+0.019}_{-0.015}\right)i\,,
\label{a1nnlo}
\end{eqnarray}
\begin{eqnarray}
a_2(\pi\pi) &=& 0.224 -[0.174+0.075 i]_{I,\alpha_s}-
[0.030+0.048 i]_{I,\alpha^2_s}
\nonumber\\
&& + [0.075]_{II,\alpha_s}+[0.032+0.019 i]_{II,\alpha^2_s}+[0.045]_{P}
\nonumber\\
&=& 0.173^{+0.088}_{-0.073}-\left(0.103^{+0.051}_{-0.054}\right)i\,.
\label{a2nnlo}
\end{eqnarray}
In both equations the first line lists the leading order result together 
with the vertex corrections ($I$) at various orders in $\alpha_s$.
Similarly, the second line displays the amplitude from hard-spectator 
interactions ($II$). It includes a model-dependent
estimate of power corrections ($P$) from twist-3 contributions to the
light-cone wave function of the pion. The third line shows the
full result together with the total uncertainty. The detailed input
can be found in \cite{Bell:2009nk}.

From (\ref{a1nnlo}) we see that the calculation is well under control
and the uncertainties are small. Note that the NLO correction 
is suppressed due to small Wilson coefficients.
In (\ref{a2nnlo}) the cancellation of leading and next-to-leading
vertex contributions is clearly visible. This implies the dominance of
the hard spectator term and the {relatively} large impact of NNLO
effects and power corrections. In addition, the first inverse moment of
the $B$-meson distribution amplitude $\Phi_B$, which determines
hard-spectator scattering as can  be seen in (\ref{a1ii}), is not well known. 
An accurate prediction of the $B\to\pi^0\pi^0$ branching
ratio, very sensitive to $a_2$, is therefore difficult. The measured  
number appears to be higher than theoretical estimates. The agreement
is better for $B\to\rho^0\rho^0$ \cite{Bell:2009fm}.

Another class of $B$ decays are those with 
  heavy-light final states.
  Prototypes of this class are the decays
  $\bar B_{(s)}\to D^{(*)+}_{(s)} L^-$, with $L$ a light meson, 
  $L=\pi$, $\rho$, $K^{(*)}$ or $a_1(1260)$, since in these cases only the
  dominant color-allowed tree amplitude $a_1(D^+L^-)$ contributes at
  leading power. 
  The factorization formula (\ref{ffhl}) is simplified here
  and offers interesting tests of the QCD dynamics in these processes.
  The QCD corrections to the amplitudes $a_1(D^+L^-)$ have been computed
  at NNLO in~\cite{Huber:2016xod}.
  The NNLO predictions for the branching ratios from factorization
  at leading power tend to be larger than experimental measurements. 
  Phenomenological implications have been discussed in \cite{Huber:2016xod}
  and more recently in \cite{Bordone:2020gao,Cai:2021mlt}.

We conclude with a few remarks concerning rare and radiative decays.
The radiative decays $B\to V\gamma$ were discussed at NNLO
in \cite{Ali:2007sj}. 
Conceptual aspects of the factorization formula for $B\to V\gamma$
at higher orders in $\alpha_s$ have been treated in
\cite{Becher:2005fg,DescotesGenon:2004hd,Chay:2003kb}.
The rare decays $B\to K^* l^+l^-$, $B\to\rho l^+l^-$ with order-$\alpha_s$
corrections \cite{Beneke:2001at,Beneke:2004dp}, 
mentioned in \ref{sec:nlo} above, have formally
the structure of NNLO processes, requiring the dominant Wilson coefficient
$C_9$ at NNLO \cite{Gambino:2003zm}.

\subsection{Outlook}
\label{sec:outlook}

QCD factorization can be applied to a large number of exclusive
$B$ decays and the associated phenomenology is very rich. We will not
go into any detail here, but restrict ourselves to a few remarks. 
The literature quoted throughout the present section contains
much further information on these topics.

Not all observables accessible in principle to a factorization
calculation are equally useful in practice. Accurate estimates of
power corrections are still beyond our control and effects of typically
$10$ -- $20\%$ are to be expected. This makes it difficult to compute
direct CP asymmetries because these are sensitive to the relatively small
strong phases, of which only the perturbative (though formally leading)
part is calculable. However, there are many cases where the level of
precision attainable with QCD factorization provides the basis for
accurate predictions and flavour-physics tests.  

A first example are suitable ratios of hadronic and semileptonic
rates, e.g.
\begin{equation}\label{ftest}
\frac{\Gamma(B^-\to\pi^-\pi^0)}{
\left. d\Gamma(\bar B_d\to\pi^+l^-\bar\nu)/dq^2\right|_{q^2=0}}=
3\pi^2 f^2_\pi |V_{ud}|^2 |a_1+a_2|^2\,,
\end{equation}
which are known as factorization tests. While not directly relevant
for flavour physics, they test QCD factorization independently of
uncertainties from form factors and $V_{ub}$. At present the precision
is still experimentally limited.

The parameter $S(\rho^+_L\rho^-_L)$ of mixing-induced CP violation in 
$\bar B_d\to\rho^+_L\rho^-_L$ is a rather clean quantity. The penguin amplitude
is numerically small (even smaller than for $\bar B_d\to\pi^+\pi^-$) and can
be computed in QCD factorization with little absolute uncertainty.
CKM phases may be extracted to within a few degrees 
\cite{Beneke:2006hg,Bartsch:2008ps}.
Additional methods to constrain the penguin contribution also benefit
from QCD factorization. Using $\bar B_d\to\bar K^{*0}_L K^{*0}_L$ and
$V$-spin symmetry should ultimately allow a precise extraction of
the CKM angle $\gamma$ with a theory error of 
$\pm 1^\circ$ \cite{Bartsch:2008ps}.

More generally, QCD factorization can be employed to estimate the
size of $SU(3)$ breaking in approaches that rely on flavour symmeties
to determine CKM quantities from CP violation in hadronic 
$B$ decays \cite{Fleischer:2008uj}.

Promising observables are exclusive rare and radiative decays such as
$B\to K^*\gamma$, $B\to\rho\gamma$ or $B\to K^*l^+l^-$.
They are dominated by form-factor terms, similar to
semileptonic modes, but also receive (rather moderate) hadronic
contributions. To those the framework of factorization can be
successfully applied.

The framework of QCD factorization in $B$ decays has been extended to
  include QED effects, both for two-body charmless \cite{Beneke:2020vnb}
  and heavy-light final states \cite{Beneke:2021jhp}.
  In spite of the complications for electrically charged final states,
  the formalism is similar to QCD factorization, provided the definitions
  of the light-cone distribution amplitudes and form factors are
  appropriately generalized \cite{Beneke:2021pkl,Beneke:2022msp}.  
  QED effects play a nontrivial role also in rare and radiative decays
  of $B$ mesons. Detailed investigations of this topic have been
  performed for the rare processes
  $B_{d,s}\to\mu^+\mu^-$~\cite{Beneke:2017vpq,Beneke:2019slt,Feldmann:2022ixt},
  $B\to \gamma\ell\nu_\ell$~\cite{Beneke:2018wjp}
  and $B\to \ell\nu_\ell$~\cite{Cornella:2022ubo}.
  An interesting application of SCET to decays of new very heavy resonances has
been presented in \cite{Alte:2018nbn,Alte:2019iug}.

In the upcoming era of precision experiments with $B$ mesons, 
QCD factorization in the heavy quark limit, at NLO and beyond, 
provides us with an important tool to control theory predictions 
at a level adequate for discoveries in flavour physics.  Summaries
of the theory status in non-leptonic heavy meson decays can be found in 
\cite{Feldmann:2014iha,Beneke:2015wfa,Bell:2015koa,Bell:2019qya,Bell:2020qus}.

\section{Electric Dipole Moments}\label{sec:EDMs}
\subsection{Preliminaries}
Even though the experimental sensitivities have improved a lot, no EDM of a fundamental particle has been observed so far and we have only upper bounds on them
to our disposal. Yet, as they test CP violation in flavour conserving processes
and are very strongly suppressed in the SM, they play a very important role
in the search for NP. In fact a measurement of
a non-vanishing EDM of any particle including nucleons, nuclei, atoms and 
molecules will be a clear signal of NP similar to lepton flavour 
violating processes. But whereas the latter are in most cases theoretically 
clean, the case of EDMs is very different. In addition to 
 the short distance dynamics present in the Wilson coefficients of 
 contributing operators,  not only hadronic physics as in other processes
 discussed by us,
but often also nuclear physics including nuclear many-body calculations and 
atomic physics is responsible for the final value of a given EDM. Therefore the identification of NP responsible through a measurement of a non-vanishing EDM is much more challenging than is the case of 
remaining observables discussed by us.

A useful review about EDMs can be found in \cite{Engel:2013lsa} which updates the review in \cite{Pospelov:2005pr}. See also \cite{Batell:2012ge,Chupp:2014gka} and in particular in the case of EDMs of diamagnetic atoms the review in \cite{Yamanaka:2017mef}. I can recommend all of them as well as 
\cite{Jung:2013hka,Dekens:2018bci}. Here we will only list references
in which RG analyses in the context of the SMEFT have been performed.
In particular we will list the papers in which one-loop and two-loop anomalous
dimensions of the relevant operators have been calculated.
In this context a very transparent presentation is given in \cite{Cirigliano:2016nyn}, where references to the relevant papers can be found. Possibly, also
Section~17.3 in my book could be useful for the first reading.

\subsection{One-Loop and Two-Loop Anomalous Dimensions}
The one-loop anomalous dimensions can be found
\cite{Degrassi:2005zd,Hisano:2012cc,Elias-Miro:2013gya,Elias-Miro:2013mua,Grojean:2013kd,Alonso:2013hga,Bhattacharya:2015rsa}. They not only allow to study the QCD RG evolution of non-leptonic operators but also the impact of Yukawas on this evolution.

The NLO and NNLO  mixing of dipole operators has already been discussed
in connection with $B\to X_s\gamma$ decay. Here the papers
\cite{Misiak:1994zw} and \cite{Gorbahn:2005sa} should be mentioned.
More recent NLO studies of dipole operators in connection with electric dipole moments in the context of the SMEFT can be found in
\cite{Panico:2018hal,Brod:2018lbf,Brod:2018pli} and very recently in
\cite{Brod:2022bww}. In particular I can recommend the latter paper in which
some errors in the previous literature have been corrected.
These analyses  play a significant role in constraining CP-violating phases of Yukawa couplings. In particular the correlations between EDM and LHC data
are a powerful tool to look for NP. Finally, a model independent analysis
of the magnetic and electric dipole moments of the muon and electron in the
framework of the WET and the SMEFT can be recommended.

\section{Standard Model Effective Field Theory}\label{sec:SMEFT}
The WET and also 
the SMEFT play these days
very important roles in the tests of the SM and of
the NP beyond it.
In order to increase
the precision of these tests it is necessary to go beyond the LO analyses both in the WET and also in
the SMEFT. To this end, as already discussed extensively above, it is
mandatory to include first in the renormalization group (RG) analyses in these
theories the one-loop matching contributions, both between these two theories
as well as when passing thresholds at which heavy particles are integrated out.
But this is not the whole story. To complete a NLO analysis and remove
various renormalization scheme (RS) dependences in the one-loop matching also
two-loop anomalous dimensions of all operators in the WET and SMEFT have
to be included. This is a big {challenge} because of the large number of operators
involved in both theories.

The present status of these efforts in the case of non-leptonic {meson} $\Delta F=1$ decays and $\Delta F=2$ quark mixing processes is as follows:
\begin{itemize}
\item
  The matchings in question are known by now both at tree-level \cite{Jenkins:2017jig} and one-loop
  level \cite{Dekens:2019ept}. Previous partial results can be found, for example, in \cite{Aebischer:2015fzz,Bobeth:2017xry,Hurth:2019ula, Endo:2018gdn,Grzadkowski:2008mf}.
\item
  The one-loop ADMs  relevant for the RG in WET \cite{Jenkins:2017dyc,Aebischer:2017gaw} and SMEFT \cite{Jenkins:2013zja,Jenkins:2013wua,Alonso:2013hga} are
  also known.
\item
  The two-loop QCD ADMs relevant for RG evolutions for both    $\Delta F=2$ and $\Delta F=1$ non-leptonic transitions in WET are also known \cite{Buras:2000if,Aebischer:2021raf,Aebischer:2020dsw}.
\item
  The two-loop QCD ADMs relevant for RG evolutions of   $\Delta F=2$ transitions in SMEFT are also known \cite{Aebischer:2022anv} and the ones for $\Delta F=1$ transitions should be known soon.
\item
  On-shell methods for the computation of the one-loop and two-loop ADMs in the
  SMEFT have been developed in \cite{EliasMiro:2020tdv,Bern:2020ikv,Machado:2022ozb}. They allow a good insight into the flavour structure of the ADMs.
  \end{itemize}

One complication in the RG evolution from NP scale $\muNP$ down to hadronic scale $\muLow$ is the fact that within SMEFT the most convenient is the Warsaw basis \cite{Grzadkowski:2010es} and in this  basis  anomalous dimensions of SMEFT operators have been calculated. But in the process of matching of the SMEFT on to the  WET most useful
is the so-called JMS basis \cite{Jenkins:2017jig}. Finally for QCD evolution in the WET the most useful
is the BMU basis \cite{Buras:2000if} that we discussed in Section~\ref{BMU}.
The Wilson coefficients in the BMU basis at $\muLow$ are given then in terms
of the SMEFT ones at $\muNP$ as follows \cite{Aebischer:2021raf}
\begin{align}
  \label{eq:fullEvol}
  \vec{C}_\text{BMU}(\muLow) &
  = \hat U_\text{BMU}(\muLow, \muEW) \;
    \hat M_\text{JMS}(\muEW)\, \hat K(\muEW) \;
    \hat U_\text{SMEFT}(\muEW,\muNP) \;
    \vec{\mathcal{C}}_\text{SMEFT}(\muNP) .
\end{align}
  Here, the matrix $\hat M_\text{JMS}$ summarizes the JMS$\to$BMU basis transformation
in the WET. The matrix $\hat K$ summarizes matching relations
between the WET in the JMS basis and the SMEFT in the Warsaw basis.
At the NLO level, in the process of basis changes, Fierz transformations
have to be performed which brings in evanescent operators. Detailed discussion
of this issue can be found in \cite{Dekens:2019ept,Aebischer:2021raf,Aebischer:2022tvz,Aebischer:2022aze,Aebischer:2022rxf,Fuentes-Martin:2022vvu}.
Explicitly
\begin{align}
  \label{eq:SMEFTWET}
  \vec{\mathcal{C}}_\text{BMU}(\muEW) &
  = \hat M_\text{JMS}(\muEW) \; \vec{\mathcal{C}}_\text{JMS}(\muEW),
&
  \vec{\mathcal{C}}_\text{JMS}(\muEW) &
  = \hat K(\muEW) \; \vec{\mathcal{C}}_\text{SMEFT}(\muEW)\,
\end{align}
with
\begin{align}
  \hat M_\text{JMS}(\muEW) &
  = \hat M^{(0)} +\frac{\alS(\muEW)}{4\pi} \hat M^{(1)},
&
  \hat K(\muEW) &
  = \hat K^{(0)} +\frac{\alS(\muEW)}{4\pi} \hat K^{(1)}.
\end{align}

The two RG evolution matrices have the familiar structure 
\begin{align}
  \label{eq:UBMU}
  \hat U_\text{BMU}(\muLow,\, \muEW) &
  = \left[1 + \hat J_\text{BMU}\frac{\alS(\muLow)}{4\pi} \right]
    \hat U_\text{BMU}^{(0)}(\muLow,\, \muEW)
    \left[1 - \hat J_\text{BMU} \frac{\alS(\muEW)}{4\pi} \right],
\\
  \label{eq:USMEFT}
  \hat U_\text{SMEFT}(\muEW,\, \muNP) &
  = \left[1 + \hat J_\text{SMEFT} \frac{\alS(\muEW)}{4\pi} \right]
    \hat U_\text{SMEFT}^{(0)}(\muEW,\, \muNP)
    \left[1 - \hat J_\text{SMEFT} \frac{\alS(\muNP)}{4\pi} \right]
\end{align}
and 
\begin{align}
  \vec{\mathcal{C}}_\text{SMEFT}(\muNP) &
  = \vec{\mathcal{C}}^{(0)}_\text{SMEFT} +
    \frac{\alS(\muNP)}{4\pi} \vec{\mathcal{C}}_\text{SMEFT}^{(1)}\,.
\end{align}
The formulae for all these matrices can be found in \cite{Aebischer:2021raf}.
In Section 5 of that paper detailed discussion of the cancellation of renormalization scheme dependences between various factors in (\ref{eq:fullEvol}) can
be found as well as their numerical evaluation. There one can also find the
impact of the top-Yukawa couplings on $\hat U_\text{SMEFT}(\muEW,\, \muNP)$.

The large number of operators motivated many groups to develop various 
computer codes for tree level and one-loop matchings as well RG running.
Most recent review of these tools can be found in \cite{Dawson:2022ewj}.
\section{Summary}\label{sec:12}
Our story is approaching the end and we reached the summit from which the 
full field of QCD and QED corrections to weak decays can be seen. In fact,
35 years after the Ringberg Workshop  NLO and NNLO QCD and 
QED corrections to the most important decays are known. 
{While the members of the MNLC conquered most of the 
NLO and NNLO summits, 
several other researchers outside MNLC contributed in an important manner to
these efforts as one can find investigating numerous tables in our review.}

These efforts
 increased 
significantly 
the accuracy of predictions of the SM for the full field of weak decays. 
This is an important step towards 
the indirect searches for New Physics through flavour violating and CP-violating 
processes. Personally I do not think that the calculations of  still higher 
orders
of perturbation theory are really required and we should wait for the data in order 
to see what kind of new physics will be identified directly at the LHC and 
indirectly through flavour violating processes and more generally through 
high precision experiments. With the technology developed in the last 35
years, calculations of higher order QCD corrections in the extensions of the 
SM selected by Nature should be straightforward even if often tedious. 
Most of these calculations have been by now automatized as exemplified 
by  \cite{Rosiek:2010ug},  where further references can be found.

On the other hand, still significant progress in non-perturbative calculations 
is required in order to increase the power of tests of the SM through FCNC 
processes, in particular in the case of non-leptonic $K$, $D$ and $B$ decays. 
While in the case of $B$-decays, QCDF presented in Section~\ref{sec:qcdf} allowed 
to make an important progress, the case of non-leptonic $K$-decays is a 
different story.
In this context 
I am very curious whether direct CP violation in $K_L\to\pi\pi$ decays 
represented by the ratio $\epe$ is well described by the SM or not. Even 
35 years after the seminal Ringberg 1988 workshop it was not possible to 
obtain the prediction for $\epe$ on which various competing groups would
agree with each other. This is the subject of another story, this time devoted
exclusively to $\epe$ and the $\Delta I=1/2$ rule \cite{Buras:2020wyv,Buras:2022cyc}.
 We should hope that in this 
 decade the answer will be provided by lattice groups.

At the 1988 Ringberg Workshop, a first estimate of the $K^0-\bar K^0$ hadronic matrix element by the European Lattice Collaboration was presented with $\hat B_K=0.87\pm0.20$ \cite{Buras:1989er} and the progress in the following two decades was slow. But
in the last decade much faster development took place in calculating
this parameter. Indeed during the last  years an impressive 
progress in calculating  $\hat B_K$ has been achieved by means of unquenched 
lattice simulations. In particular in the previous decade, lower numerical values like 
$\hat B_K=0.724\pm0.008\pm0.028$ \cite{Antonio:2007pb} and $\hat B_K=0.749\pm0.027$  \cite{Aoki:2010pe} appeared in the literature. By now the  accuracy 
has been significantly increased so that the most recent update from FLAG reads
 $\hat{B}_K =0.7625(97)$~\cite{FlavourLatticeAveragingGroupFLAG:2021npn} while the new UTfit analysis is based on $\hat B_K=0.756(16)$ \cite{Bona:2022pjv}.

Interestingly these new values are very close to $\hat B_K=0.75$ obtained 
in the large-N limit of QCD \cite{Gaiser:1980gx,Buras:1985yx}. Including 1/N corrections Bardeen,  G\'erard and myself
\cite{Bardeen:1987vg,Gerard:1990dx} found some indications for
$\hat B_K\le 0.75$. A more precise 
analysis of  G\'erard \cite{Gerard:2010jt} put this result on firm footing. 
Thus afterall our large-N result for $\hat B_K$ presented already at the 1988 
Ringberg 
workshop and briefly mentioned at the beginning of this writing 
has been confirmed by much more precise lattice simulations more than 
28 years later. 
In this context also the paper by 
Bijnens and Prades \cite{Bijnens:1995br} should be mentioned. See \cite{Gerard:2010jt} for more details.

Motivated by the progress made by lattice groups we have improved in 2014
our old calculation of $\hat B_K$ within large $N$ approach through the 
inclusion of lowest-lying vector meson contributions in addition to the pseudoscalar ones to hadronic matrix elements of current-current operators and the calculation of the corresponding Wilson coefficients in a momentum scheme at the NLO
\cite{Buras:2014maa}. This  
improved significantly the matching between quark-gluon short distance contributions and long distance meson contributions over our result in 1987. We find
\be\label{BBGBK}
\hat B_K=0.73\pm 0.02,  \qquad  ({\rm Dual~QCD,}~2014).
\ee
in an excellent agreement with the lattice QCD values quoted above. 
However, we are aware of the fact that while lattice 
calculations will have one day a good control over their errors, this is not quite 
the case here. On the other hand, we could provide the explanation 
why $\hat B_K$ is found within $2\%$ from its large $N$ 
value.

{\bf Acknowledgements}

I would like to thank all the members of the MNLC for fantastic time we 
had together conquering the NLO and NNLO summits of weak decays over so 
many years. In fact 
after 35 years I can confidently state that my team was much bigger 
than the one of colonel Hunt and whereas he reached only the south {col}
on the way to 
the Mount Everest summit, I was lucky to stand on many NLO and 
few NNLO summits. But similarly to 
his case the sponsors played also in our case a very important role. 
In addition to the Physics Department of the Technical University Munich, 
the thanks 
go primarly to 
the German `Bundesministerium f\"ur Bildung und Forschung' (BMBF) and 
 Deutsche Forschung Gemeinschaft (DFG) which were both very helpful 
in supporting us for many years.
Some support came also from the MPI for Physics and the German-Israeli 
Foundation. In the last years Martin
Gorbahn, Joachim Brod and myself were also strongly supported by  the Cluster of Excellence Universe in Munich and the TUM-IAS. Also the advanced ERC grant
2011-2016 helped me a lot after my retirement in April 2012.
Finally, financial support from the Excellence Cluster ORIGINS,
funded by the Deutsche Forschungsgemeinschaft (DFG, German Research
Foundation), 
Excellence Strategy, EXC-2094, 390783311 is acknowledged.

Very special thanks go to Gerhard Buchalla for writing Section 9 of this 
review. I also thank Paolo Gambino and Matthias Jamin for some help in finding 
references.
Finally, but very importantly, I would like to thank Guido Martinelli for 
the discussion at the seminal supper in 1988 in the Ringberg castle and 
the competition which definitely kept us working for several years. 
Among the members of his 
team my particular thanks go to Marco Ciuchini and Luca Silvestrini. 

A number of my friends read V1  and subsequent  versions of this NLO story pointing out misprints and making comments on the full text. These are Jason Aebischer, Gerhard Buchalla, Joachim Brod, Svjetlana Fajfer, Robert Fleischer, Paolo Gambino,
Jean-Marc G\'erard, Jennifer Girrbach, Tobias Huber, Jacky Kumar, Alexander Lenz, Mikolaj Misiak, Ulrich Nierste, Matthias 
Jamin, Cecilia Tarantino, Andreas Weiler and Peter Weisz. I appreciate all
these comments and thank them a lot. {Finally, I would like to thank one of the referees in Physics Reports for very helpful comments.}

\addcontentsline{toc}{section}{References}

\small

\bibliographystyle{JHEP}
\bibliography{bookv4}

\end{document}